%% file: ANA-STDM-2017-33-PAPER.tex
\pdfoutput=1
\newcommand*{\ATLASLATEXPATH}{}
\documentclass[PAPER,atlasdraft=false, cernpreprint=true, texlive=2016, UKenglish]{\ATLASLATEXPATH atlasdoc}
 
\usepackage[subfigure]{\ATLASLATEXPATH atlaspackage}
 
\usepackage{\ATLASLATEXPATH atlasbiblatex}
\usepackage{\ATLASLATEXPATH atlasjetetmiss}
 
\usepackage{\ATLASLATEXPATH atlasphysics}
\usepackage{\ATLASLATEXPATH atlasmisc}
 
\graphicspath{{logos/}{figures/}}
 
\usepackage{ANA-STDM-2017-33-PAPER-defs}

\AtlasTitle{A measurement of soft-drop jet observables in $pp$ collisions with the ATLAS detector at $\sqrt{s} = 13 ~\text{Te\kern -0.1em V}$}
 
\PreprintIdNumber{CERN-EP-2019-269}
\AtlasJournal{Phys. Rev. D.}
\AtlasJournalRef{Phys. Rev. D 101 (2020) 052007}
\AtlasDOI{10.1103/PhysRevD.101.052007}

\AtlasVersion{3.0}
\AtlasAbstract{
Jet substructure quantities are measured using jets groomed with the soft-drop grooming procedure in dijet events from 32.9 fb${}^{-1}$ of $pp$ collisions collected with the ATLAS detector at $\sqrt{s}=13 ~\text{Te\kern -0.1em V}$.
These observables are sensitive to a wide range of QCD phenomena.
Some observables, such as the jet mass and opening angle between the two subjets which pass the soft-drop condition, can be described by a high-order (resummed) series in the strong coupling constant $\alpha_{\text{S}}$.
Other observables, such as the momentum sharing between the two subjets, are nearly independent of $\alpha_{\text{S}}$.
These observables can be constructed using all interacting particles or using only charged particles reconstructed in the inner tracking detectors.
Track-based versions of these observables are not collinear safe, but are measured more precisely, and universal non-perturbative functions can absorb the collinear singularities.
The unfolded data are directly compared with QCD calculations and hadron-level Monte Carlo simulations.
The measurements are performed in different pseudorapidity regions, which are then used to extract quark and gluon jet shapes using the predicted quark and gluon fractions in each region.  All of the parton shower and analytical calculations provide an excellent description of the data in most regions of phase space.
}
 
\author{The ATLAS Collaboration}
 
\AtlasRefCode{STDM-2017-33}

\hypersetup{pdftitle={ATLAS document},pdfauthor={The ATLAS Collaboration}}

\begin{document}
 
\maketitle
 
\tableofcontents
 
\section{Introduction}
 
Jets are collimated sprays of particles that are initiated by high-energy quarks and gluons.
Grooming techniques systematically remove soft and wide-angle radiation, making the structure of the jet robust against contamination from multiple simultaneous proton--proton interactions (pileup) as well as against final-state radiation and the underlying event.
This internal structure of a jet has been successfully used to tag the origin of jets in precision measurements and searches at the Large Hadron Collider (LHC)~\cite{Asquith:2018igt,Larkoski:2017jix}.
While grooming has been a powerful tool for applications of jet substructure techniques, it also provides a unique opportunity for the study of the strong force itself.
If groomed in a suitable way, the radiation pattern inside the resulting jet can be predicted from first principles in QCD.  The differential cross sections as a function of key observables such as the groomed jet mass have been computed beyond leading-logarithmic accuracy~\cite{Frye:2016aiz,Frye:2016okc,Marzani:2017mva,Marzani:2017kqd,Kang:2018vgn,Kang:2018jwa} as an expansion in the strong coupling constant $\alphas$ along with logarithms of ratios of physical scales.  New `Sudakov safe' observables~\cite{Larkoski:2015lea} that are the ratio of attributes that are both infrared-safe and collinear-safe cannot be expressed as an expansion in $\alphas$, but can be described with a series in fractional powers of $\alphas$.  For particular grooming configurations, observables such as the ratio of subjet energies can be \textit{independent} of $\alphas$~\cite{Larkoski:2015lea}.  These nonstandard and universal behaviors are now being tested with precision at the LHC and the Relativistic Heavy Ion Collider (RHIC).
 
While many grooming procedures suppress difficult-to-model soft and wide-angle radiation, only one grooming algorithm has been successfully used for calculations beyond the formal precision of the parton shower (leading logarithm).  This \textit{soft-drop} grooming procedure~\cite{Larkoski:2014wba} is a generalization of the modified mass drop procedure~\cite{Dasgupta:2013ihk} and is formally insensitive to \textit{nonglobal logarithmic corrections}~\cite{Dasgupta:2001sh}: resummation terms resulting from radiation which leaves the jet cone and then produces radiation that reenters the jet. Soft-drop jet observables have been calculated to next-to-leading-logarithm (NLL) and next-to-next-to-leading-logarithm (NNLL) accuracy.   The soft-drop jet mass has recently been measured in dijet events~\cite{STDM-2017-04,Sirunyan:2018xdh}.
In the region where the calculations are expected to be accurate, the agreement with the data is excellent, and nonperturbative effects~\cite{Hoang:2019ceu} have become the most important theoretical source of uncertainty instead of higher-order effects.
 
This analysis goes beyond the jet mass by adding other soft-drop jet observables that are connected with the grooming procedure.  Furthermore, in addition to measuring observables reconstructed using all interacting particles, charged-particle observables are measured using tracks. These track-based observables can be probed with better experimental precision compared to the calorimeter-based observables.  Charged-particle observables are not formally collinear-safe, but universal nonperturbative functions, like parton distribution functions, can absorb the relevant singularities and allow for precise predictions~\cite{Krohn:2012fg,Waalewijn:2012sv,Chang:2013rca,Elder:2017bkd}.
Finally, the differences between these distributions in regions with different quark/gluon composition is used to understand how the behavior and sensitivity of the different observables depends on the origin of the jet.
Previous measurements of groomed jet observables have been conducted at the LHC by CMS~\cite{CMS-TOP-17-013,Sirunyan:2018xdh}, ATLAS~\cite{STDM-2017-04,STDM-2017-34}, and ALICE~\cite{Acharya:2019djg}, and at RHIC by the STAR Collaboration~\cite{KunnawalkamElayavalli:2019csg} and additional studies at detector level have been performed using CMS data~\cite{Sirunyan:2018gct,Sirunyan:2017bsd,Larkoski:2017bvj,Tripathee:2017ybi}.

\section{Soft-drop procedure}
\label{sec:softdropprocedure}

The soft-drop grooming algorithm proceeds as follows.
After a jet is clustered using any algorithm, its constituents are then reclustered using the Cambridge/Aachen (C/A) algorithm~\cite{Dokshitzer:1997in,Wobisch:1998wt},
which iteratively clusters the closest constituents in rapidity and azimuth.
This typically produces a jet with the same constituents as the original jet, but with a modified jet clustering history, which is sensitive to the angle-ordered nature of parton shower evolution.
Then, the last step of the C/A clustering algorithm is undone, breaking the jet $j$ into the last two subjets, $j_1$ and $j_2$, which were clustered together.
These two subjets are then used to evaluate the soft-drop condition:
\begin{align*}
\label{eq:softdrop}
\frac{\text{min}(p_\text{T,$j_1$},p_\text{T,$j_2$})}{p_\text{T,$j_1$}+p_\text{T,$j_2$}}> z_\text{cut}\left(\frac{\Delta R_{12}}{R}\right)^\beta,
\end{align*}
\noindent where $p_\text{T,$j_i$}$ is the transverse momentum of subjet $j_i$, and $\Delta R_{12}$ is the distance between the two subjets in $y$--$\phi$ space.\footnote{\AtlasCoordFootnote}  The parameters $z_\text{cut}$ and $\beta$ are algorithm parameters explained in greater detail below, and $R$ is the jet radius parameter.
If $j_1$ and $j_2$ fail the soft-drop condition, then the subjet with the lower $p_\text{T}$ is removed, and the one with the higher $p_\text{T}$ is relabeled as $j$ and the procedure is iterated. If the soft-drop condition is satisfied, then the algorithm is stopped, and the resulting jet $j$ is the soft-dropped jet.
If no pairs of subjets in the declustering satisfy the soft-drop condition, then the resulting jet is the zero vector.
 
The parameters $z_\text{cut}$ and $\beta$ determine the sensitivity of the algorithm to soft and wide-angle radiation.  As $\beta\rightarrow \infty$ (and $z_\text{cut}<1$),
the soft-drop condition is always satisfied, and no grooming is applied.
Decreasing $\beta$ preferentially removes wide-angle radiation and increasing $z_\text{cut}$ preferentially removes soft radiation.  The theoretical calculations are performed for a range in $\beta$ and assume $z_\text{cut}$ is small enough so that it does not introduce large logarithms (which was explicitly checked in Refs.~\cite{Marzani:2017mva,Marzani:2017kqd}).  This measurement adopts the same choice as the available theoretical calculations: $z_\text{cut}=0.1$ and $\beta\geq 0$.  Several $\beta$ values are tested to probe different scales of angular strulcture inside the jets.
 
This paper measures three closely related substructure observables, which are calculated from jets after they have been groomed with the soft-drop algorithm.
These are the jet mass, the $p_\text{T}$ balance $z_g$ (which is the left-hand side of Eq.(~\ref{eq:softdrop})) of the splitting which passes the soft-drop condition,
and $r_g$, which is the opening angle $R_{12}$ of this splitting in Eq.(~\ref{eq:softdrop}).
These three observables -- the jet mass, $z_g$ and $r_g$ -- are described in greater detail in Section~\ref{sec:objectdef}.
These observables are approximately related by $m^2/p_\text{T}^2\sim z_g r_g^2$, and each probes different aspects of the structure of the jet.

\section{ATLAS detector}
 
The ATLAS detector~\cite{PERF-2007-01} at the LHC covers nearly the entire solid angle around the collision point.
It consists of an inner tracking detector surrounded by a thin superconducting solenoid, electromagnetic and hadronic calorimeters,
and a muon spectrometer incorporating three large superconducting toroidal magnets.
 
The inner-detector system (ID) is immersed in a \SI{2}{\tesla} axial magnetic field
and provides charged-particle tracking in the range \(|\eta| < 2.5\).
The high-granularity silicon pixel detector, the innermost layer of the tracking detector, covers the vertex region and typically provides four measurements per track,
the first hit being typically recorded in the insertable B-layer that was installed before Run~2~\cite{ATLAS-TDR-2010-19,PIX-2018-001}.
It is followed by the silicon microstrip tracker, which usually provides eight measurements per track.
These silicon detectors are complemented by the transition radiation tracker,
which enables radially extended track reconstruction up to \(|\eta| = 2.0\).
 
The calorimeter system covers the pseudorapidity range \(|\eta| < 4.9\).
Within the region \(|\eta|< 3.2\), electromagnetic calorimetry is provided by barrel and
endcap high-granularity lead/liquid-argon (LAr) detectors,
with an additional thin LAr presampler covering \(|\eta| < 1.8\),
to correct for energy loss in material upstream of the detectors.
Hadronic calorimetry is provided by the steel/scintillator-tile detector,
segmented into three barrel structures within \(|\eta| < 1.7\), and two copper/LAr hadronic endcap calorimeters which cover $1.5 < |\eta| < 3.2$.
The solid angle coverage is completed with forward copper/LAr and tungsten/LAr calorimeter modules covering $3.1 < |\eta| < 4.9$, which are
optimized for electromagnetic and hadronic measurements respectively.
 
Interesting events are selected for recording by the first-level trigger system implemented in custom hardware,
followed by selections made by algorithms implemented in software in the high-level trigger~\cite{TRIG-2016-01}.
The first-level trigger makes decisions at the \SI{40}{\MHz} bunch crossing rate to keep the accepted-event rate below \SI{100}{\kHz},
which the high-level trigger further reduces in order to record events to disk at about \SI{1}{\kHz}.
 
\section{Datasets}
\label{sec:MCandata}
 
These measurements use the dataset of $pp$ collisions recorded by the ATLAS detector in 2016, corresponding to an integrated luminosity of
$32.9$ \ifb~\cite{ATLAS-CONF-2019-021,Avoni_2018}~at a center-of-mass energy of $\sqrt{s}=13$~\TeV. Events are only considered if they were collected during stable beam conditions and satisfy all data quality requirements~\cite{DAPR-2018-01}.
Due to the high instantaneous luminosity and the large total inelastic proton--proton ($pp$) cross section, on average there are about 25 simultaneous (pileup) collisions
in each bunch crossing.
 
The measurements presented in this paper use a variety of Monte Carlo (MC) event generator samples to estimate the impact of detector efficiency and resolution as well as for comparison with the unfolded data.
Dijet events were generated at leading order (LO) with \PYTHIA 8.186~\cite{Sjostrand:2006za,Sjostrand:2007gs}, with the $2 \rightarrow 2$ matrix element convolved with the
NNPDF2.3LO parton distribution function (PDF) set~\cite{Ball:2012cx} and using the A14 set of multiple-parton-interaction and parton-shower parameters~\cite{ATL-PHYS-PUB-2014-021}.
\PYTHIA8 uses a $\pT$-ordered parton shower model.  Additional dijet events were generated using different generators, in order to study the impact of modeling uncertainties.
\SHERPA 2.1~\cite{Gleisberg:2008ta,Bothmann:2019yzt} was used to generate events using multileg $2\rightarrow 2$ and $2\rightarrow 3$ matrix elements,
which were matched to parton showers following the CKKW prescription~\cite{Catani:2001cc}. These \SHERPA events were generated using the CT10nlo PDF set~\cite{Lai:2010vv}
and the default \SHERPA set of tuned parameters. \HERWIGpp 2.7~\cite{Bahr:2008pv,Corcella:2000bw} was used to provide a sample of events with an angle-ordered parton shower model.
These events were generated with the $2\rightarrow 2$ matrix element, convolved with the CTEQ6L1 PDF set~\cite{Pumplin:2002vw}
and configured with the UE-EE-5 set of tuned parameters~\cite{Seymour:2013qka}.
 
All generator events were passed through a full simulation of the ATLAS detector~\cite{SOFT-2010-01} implemented in \GEANT 4~\cite{Agostinelli2003250}, which describes the interactions of particles with the detector and the subsequent digitization of analog signals. The effects of pileup were simulated with unbiased $pp$ collisions using the \PYTHIA 8.186 generator with the A2~\cite{ATL-PHYS-PUB-2012-003} set of tuned parameters and the MSTW2008LO~\cite{Martin:2009iq} PDF set; these events were overlaid on the nominal dijet events. These events are then reweighted such that the distribution of the average number of interactions per bunch crossing matches that seen in data.
 
\section{Event selection and object reconstruction}
 
Since the data are unfolded to particle level, it is necessary to define both the particle-level and detector-level objects used in the measurement.
The former are chosen to be as close as possible to the latter in order to minimize the model dependence caused by an extrapolation
from the phase space measured at detector level to the phase space measured at particle level.
Section~\ref{sec:events} describes the particle-level and detector-level event selection criteria.
Following this, Section~\ref{sec:objectdef} describes the particle-level and detector-level jet reconstruction procedure
for both the calorimeter-based (all-particle) observables and the track-based (charged-particle) observables.
 
\subsection{Jet and event selection}
\label{sec:events}
 
Detector-level events are required to have at least one primary vertex reconstructed from at least two tracks with $p_\textrm{T}$ greater than 400 MeV.
The primary hard-scattering vertex of the event is chosen to be the one with the highest $\sum_\text{tracks} p_\textrm{T}^{2}$. The inputs to the jet clustering algorithm are locally calibrated topological calorimeter-cell clusters~\cite{PERF-2014-07}.
 
Jets are clustered with $\Fastjet$~\cite{Fastjet} using the anti-$k_t$~\cite{Cacciari:2008gp} algorithm with radius parameter $R=0.8$.
A series of simulation-based calibration factors are applied to ensure that the detector-level jet $\pT$ is the same as the particle-level value on average~\cite{PERF-2016-04}.
Each event is required to have at least two reconstructed jets, where the transverse momentum of the leading jet, $p_\text{T}^\text{lead}$, is greater than 300 ~\GeV.
The jet selection is applied to ungroomed jets, which ensures that the same jets are studied for all grooming configurations.
In order to enhance the dijet topology and allow an interpretation of quark or gluon origin of the jets in the event,
the leading two jets are required to be well-balanced: $\pT^\text{lead}/\pT^\text{sublead} < 1.5$. Both jets are required to have $|\eta| < 1.5$, and only jets with a nonzero mass are retained.
 
Events are selected using single-jet triggers.
Due to the large cross section for jet production, most of the jet triggers are prescaled. Therefore events which pass these triggers are randomly discarded with some
fixed probability. The lowest-\pt-threshold unprescaled $R = 0.4$ single-jet trigger in 2016 is fully efficient for $R=0.8$ dijet events where the leading-jet $\pT$ is greater than 600~\GeV.
In events where the leading jet has 300~\GeV~ $< p_\text{T} < 600$~\GeV, a prescaled trigger is used with an average prescale value of 1000 (the inverse of the probability to be recorded).
While this results in a lower effective luminosity, it provides access to the lower $\pT$ region.

The inputs to particle-level jets are stable particles ($c\tau > 10$ mm) excluding muons and neutrinos.
These jets are clustered using the same radius parameter as the detector-level jets and have the same $\eta$ and $\pt$ cuts as for the detector-level selection.
 
\subsection{Inputs for jet substructure}
\label{sec:objectdef}
 
Two types of jet substructure observables are measured: calorimeter-based observables, which correspond to observables reconstructed from all particles inside the jet at particle level,
and track-based observables, which correspond to observables reconstructed from charged particles.
Track-based observables are theoretically more complicated to describe, but are experimentally cleaner to measure due to the precise angular measurement from the ID.
For both the calorimeter-based and track-based measurements, the jet selection is performed on the calorimeter-based jets,
while the soft-drop grooming is applied to the cluster inputs and the track inputs respectively (Section~\ref{sec:softdropprocedure}).
The jets after the application of this algorithm are often referred to as groomed, and the constituents of these jets are used to compute the jet substructure observables.
It is noted that since the event selection is applied to un-groomed jets, some selected jets are left with one constituent after grooming, resulting in jets with a mass of zero.
 
For the calorimeter-based observables, the same constituents are used to calculate the observables as are used to create the jets
described in Section~\ref{sec:events} for both detector level and particle level.
For detector-level track-based observables, the soft-drop procedure is applied to tracks matched to the ungroomed jet via ghost association~\cite{Cacciari:2008gn},
and jet substructure observables are calculated using the groomed tracks.  These tracks are selected with a $p_\text{T}>500$ MeV requirement
and assigned to the primary vertex in accord with the track-to-vertex matching.
Tracks not included in vertex reconstruction are assigned to the primary vertex if it has the smallest $|\Delta z_0\sin\theta|$ compared to any other reconstructed vertex, up to a maximum distance of 3.0 mm.
Tracks not matched to the primary vertex are not considered.  At particle level, these track-based observables are built using the charged-particle constituents of the particle-level jets, excluding muons.

Both the leading and subleading jet are used in this measurement.
In order to expose differences between quark and gluon jets, the more forward and more central of the two jets are distinguished and measured separately.
Between the leading and subleading jets, the one with the smaller $|\eta|$ will be referred to as the ``central'' jet, and the other one as the ``forward'' jet.
For a fixed jet \pt at high rapidity where the high-$x$ contribution is more important, jets are more often quark-initiated due to the large contribution of valence quarks.

\section{Observables}
 
Three substructure observables are calculated from the two jets groomed with the soft-drop algorithm (using the C/A algorithm with $R=0.8$ to recluster the jets), including the jet mass, $z_g$, and $r_g$.
These three observables completely characterize the splitting from the soft-drop condition, and they are all measured using both the calorimeter and tracker inputs.

\paragraph{Jet mass:}
One of the most basic and important jet substructure observables is the jet mass:
\begin{align}
m^2=\left(\sum_{i\in\text{jet}} E_i\right)^2-\left(\sum_{i\in\text{jet}} \vec{p}_i\right)^2,
\end{align}
 
where $i$ refers to the constituents of the jet. The measurement is performed for a dimensionless version of the jet mass: the relative mass $\rho\equiv \log(m^2/\pt^2)$,
where $m$ is groomed and $\pt$ is ungroomed (groomed jet \pt is not infrared- and collinear-safe~\cite{Marzani:2017mva}).  The calorimeter-cluster inputs are treated as massless and tracks are assigned the pion mass.
Since the probability distribution of $\rho$ is approximately linear in the resummation regime
($\Lambda_\text{QCD}/\pt\lesssim m/ \pt \lesssim z_\text{cut}$, where $\Lambda_\text{QCD}$ is the energy scale of hadronization)~\cite{Frye:2016aiz,Frye:2016okc,Marzani:2017mva,Marzani:2017kqd,Kang:2018vgn,Kang:2018jwa}, the binning for $\rho$ is evenly spaced.  For $\rho$, the distributions are normalized to the integrated cross section, $\sigma_{\mathrm{resum}}$, measured in the resummation region, $-3.7 < \rho < -1.7$.
By changing $\beta$, the distribution shifts to higher values as fewer constituents are removed from the jets during grooming.
 
An example of the distribution of $\rho$ in simulation at detector level (particle level) for the calorimeter-based (all particles) definition is shown in Figure~\ref{fig:observs:mass}(a)
for the more central of the two jets and for $\beta=0$.
For this observable, particularly in the lower-relative-mass region, there are nontrivial detector effects which occur due to the calorimeter granularity,
resulting in a distribution with different shapes at particle and detector levels.
As expected, the distribution of $\log(m^2/\pt^2)$ is approximately
linear for $\beta=0$ in the resummation regime.
 
One way to reduce the impact of these detector corrections is to consider track-based (charged-particle) observables.
An example of the track-based (charged-particle-based) $\rho$ is shown in Figure~\ref{fig:observs:mass}(b), where tracks (charged particles) are used for both the mass and the \pt.
As in the calorimeter case, the mass is calculated using the groomed jet, while the \pT is calculated using the ungroomed constituents, but no calibration is applied to the ungroomed jet since no such calibration exists for track-based inputs.
Although the particle-level distributions only include charged particles, the distributions are similar to those shown in Figure~\ref{fig:observs:mass}(a),
but in this case the impact of the detector corrections is significantly smaller.
 
\begin{figure}[h!]
\begin{center}
\subfigure[$\rho$ distribution, $\beta = 0$, calorimeter-based]{\includegraphics[width=0.48\textwidth,page=1]{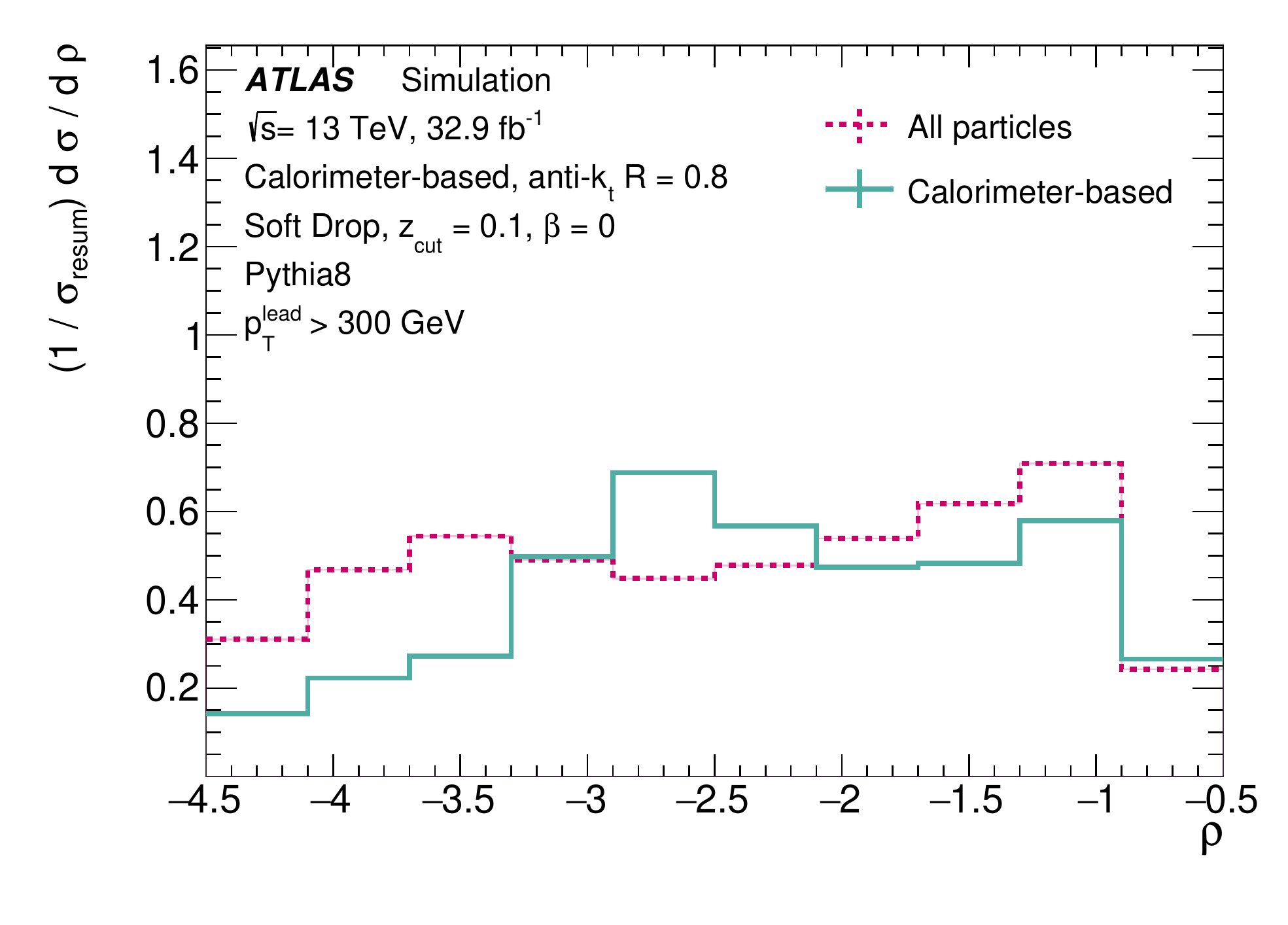}}
\subfigure[$\rho$ distribution, $\beta = 0$, track-based]{\includegraphics[width=0.48\textwidth,page=1]{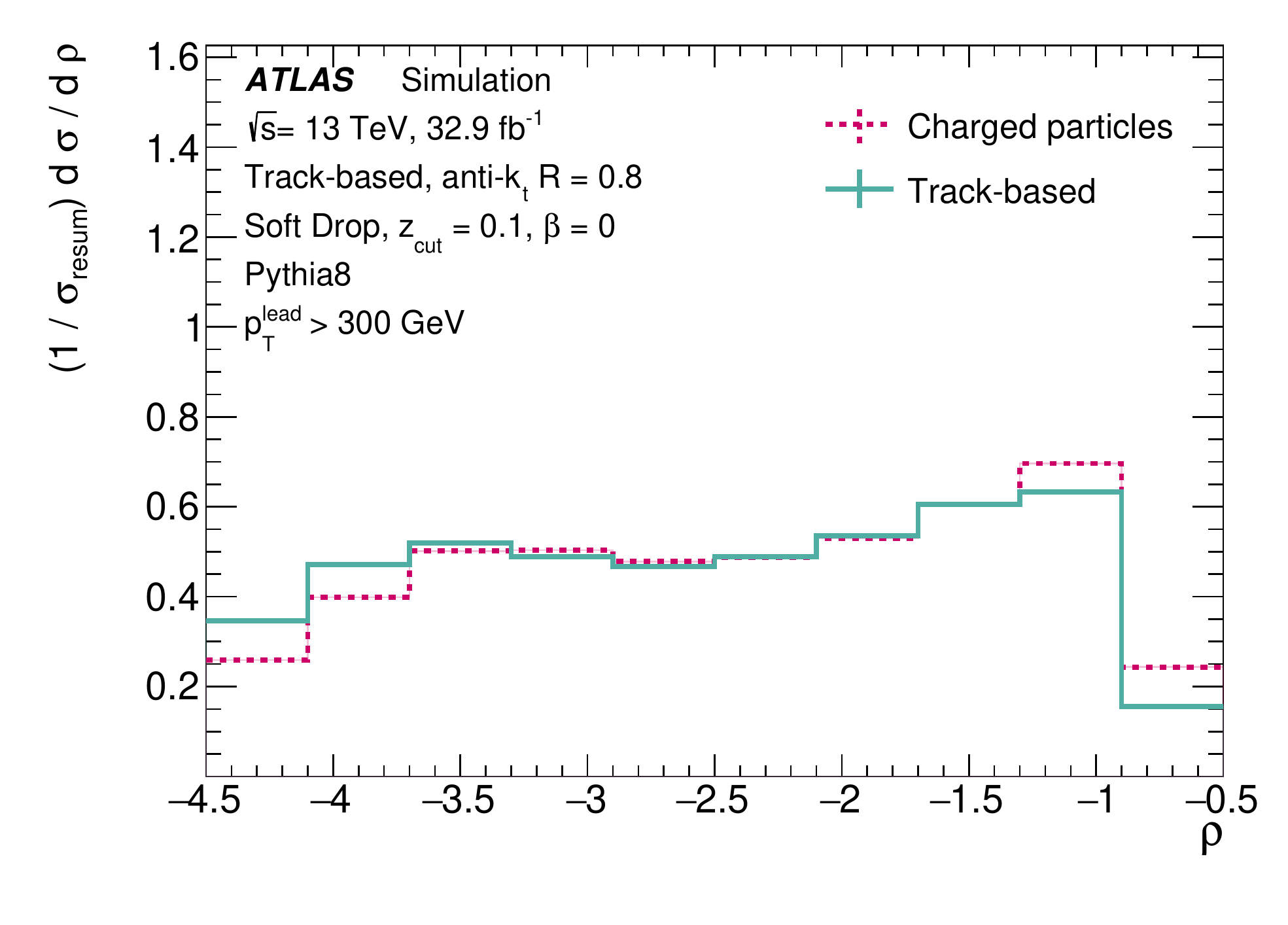}}
\end{center}
\caption{The distribution in simulation of $\rho$ at detector level and particle level for the more central of the two jets
for $\beta=0$ for (a) calorimeter-based (all particles), and (b) track-based (charged particles). The statistical uncertainties are drawn, but are too small to be visible.}
\label{fig:observs:mass}
\end{figure}
 
\paragraph{$z_g$:}
An important quantity when describing the hard splitting scale that defines the mass is $z_g$, which is
$\text{min}(p_{\mathrm{T, j_{1}}}, p_{\mathrm{T,{j_2}}}) / (p_{\mathrm{T, j_{1}}}+p_{\mathrm{T, {j_2}}})$
for the splitting that satisfies the soft-drop condition.  If no such splitting occurs, then the jet is not included in the measurement.  Symmetric splittings are characterized by $z_g\sim 0.5$.
Figure~\ref{fig:observs:zg} shows an example of the normalized distribution in simulation of $z_g$ at detector level (particle level) with $\beta=0$ for both
the calorimeter-based (all particles) and track-based (charged particles) definitions. For $\beta=0$ and $z_{\mathrm{cut}}=0.1$, $z_g$ must be greater than $0.1$ in order to pass the soft-drop condition,
and therefore bins with $z_g$ values less than 0.1 are not shown (this is not the case for $\beta > 0$).
As in the case with the mass, the distributions of the charged-particles and all-particles versions of $z_g$  are similar. Detector effects for the calorimeter-based $z_g$ are smaller than for the relative mass, because $z_g$ is less sensitive to the angular distribution of energy within the jet.
 
The binning is evenly spaced in $z_g$ and the distributions are normalized to the integrated cross section $\sigma$.
 
\begin{figure}[h!]
\begin{center}
\subfigure[$z_g$ distribution, $\beta = 0$, calorimeter-based]{\includegraphics[width=0.48\textwidth,page=1]{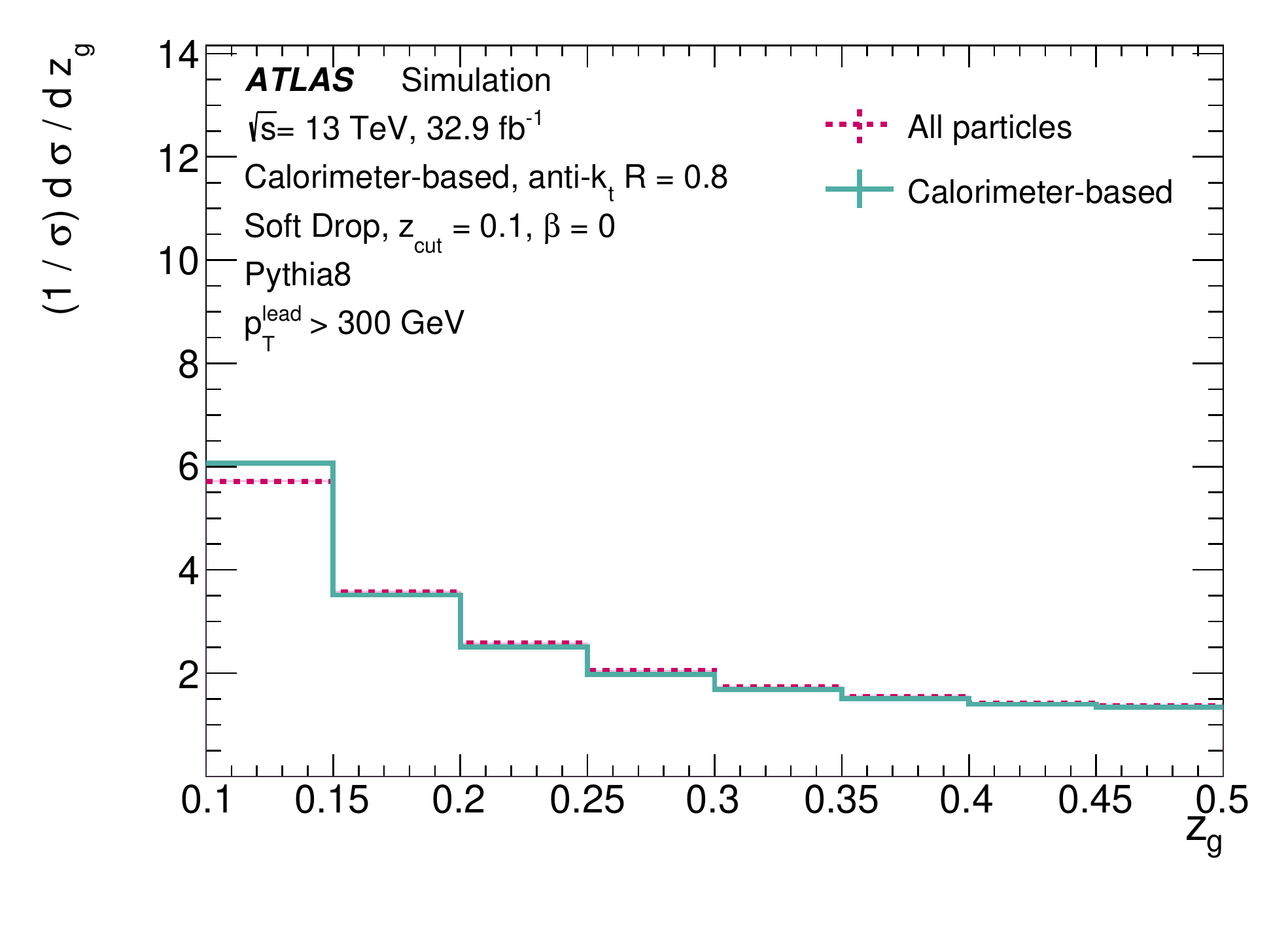}}
\subfigure[$z_g$ distribution, $\beta = 0$, track-based]{\includegraphics[width=0.48\textwidth,page=1]{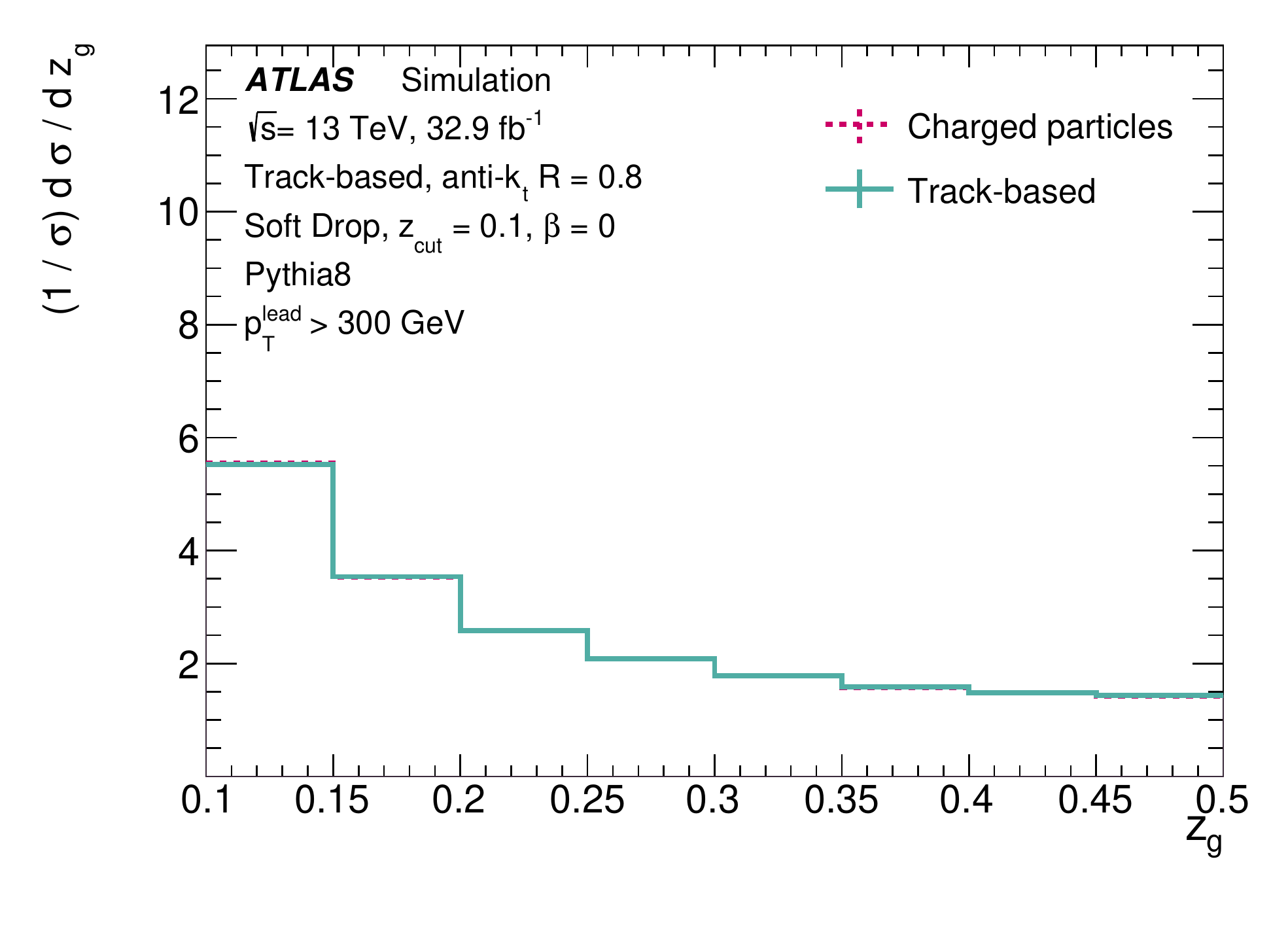}}
\end{center}
\caption{The distribution in simulation of $z_g$ at detector level and particle level
for $\beta=0$ for (a) calorimeter-based (all particles), and (b) track-based (charged particles). The statistical uncertainties are drawn, but are too small to be visible.}
\label{fig:observs:zg}
\end{figure}

\paragraph{$r_g$:}
The opening angle $\Delta R_{12}$ between the two subjets that pass the soft-drop condition is $r_g$. This angle is smaller than the jet radius by definition.
Although $r_g$ is highly correlated with the relative mass and $z_g$, it is useful for explicitly exposing the angular distribution.
Figure~\ref{fig:observs:rg} shows an example of the normalized  calorimeter-based (all particles)
and track-based (charged particles) $r_g$ distributions.
As expected, there are large detector effects for the calorimeter-based case, especially at low angles.
Due to the correlation between mass and $r_g$, the distribution shapes and detector effects look similar to the ones shown in Figure~\ref{fig:observs:mass}.
 
The binning for $r_g$ is logarithmically spaced.  The distributions are normalized to the integrated cross section $\sigma$.  Similar to $\rho$, increasing $\beta$ shifts the distribution to higher values as there is less grooming.
 
\begin{figure}[h!]
\begin{center}
\subfigure[$r_g$ distribution, $\beta = 0$, calorimeter-based]{\includegraphics[width=0.48\textwidth,page=1]{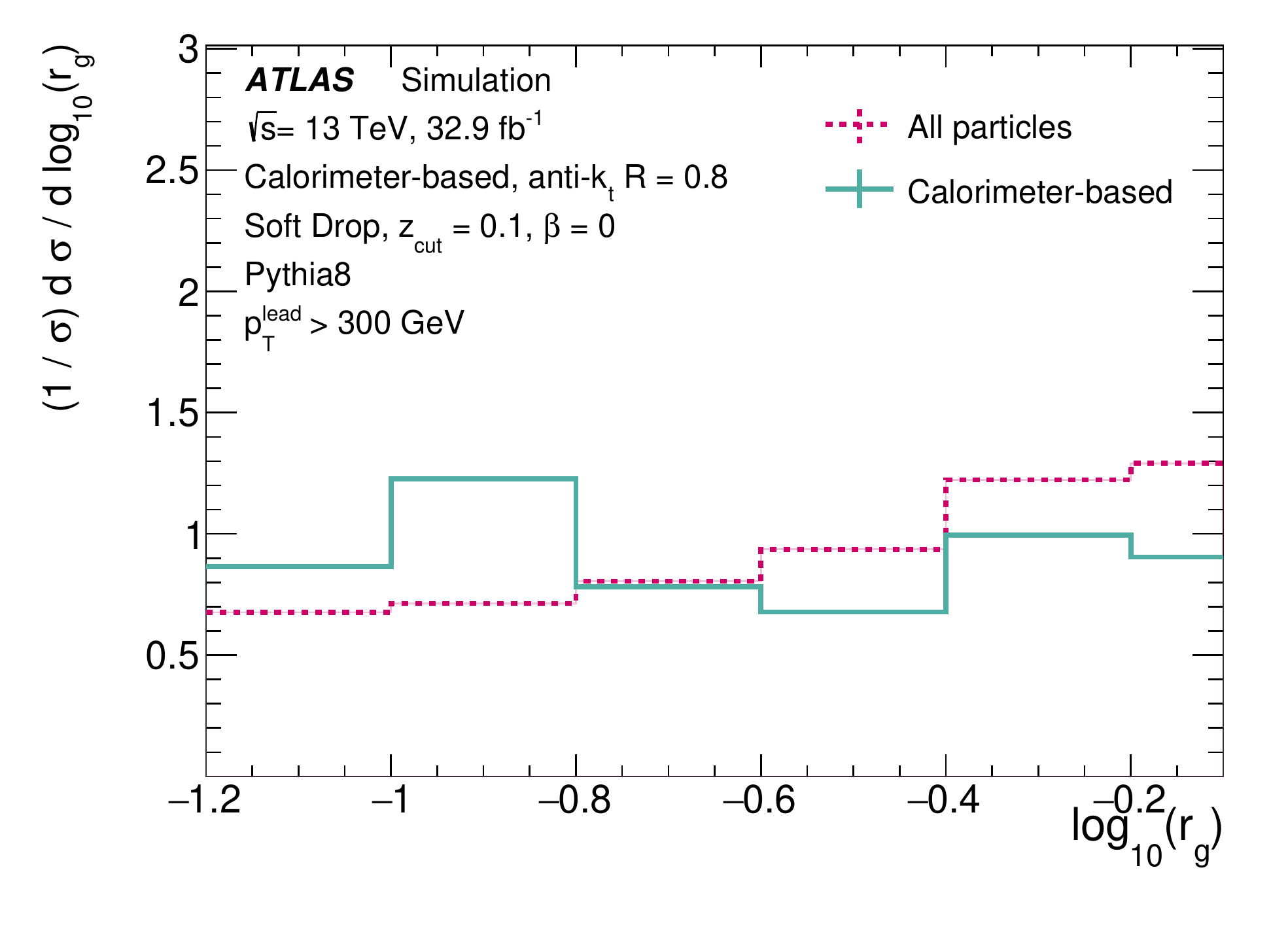}}
\subfigure[$r_g$ distribution, $\beta = 0$, track-based]{\includegraphics[width=0.48\textwidth,page=1]{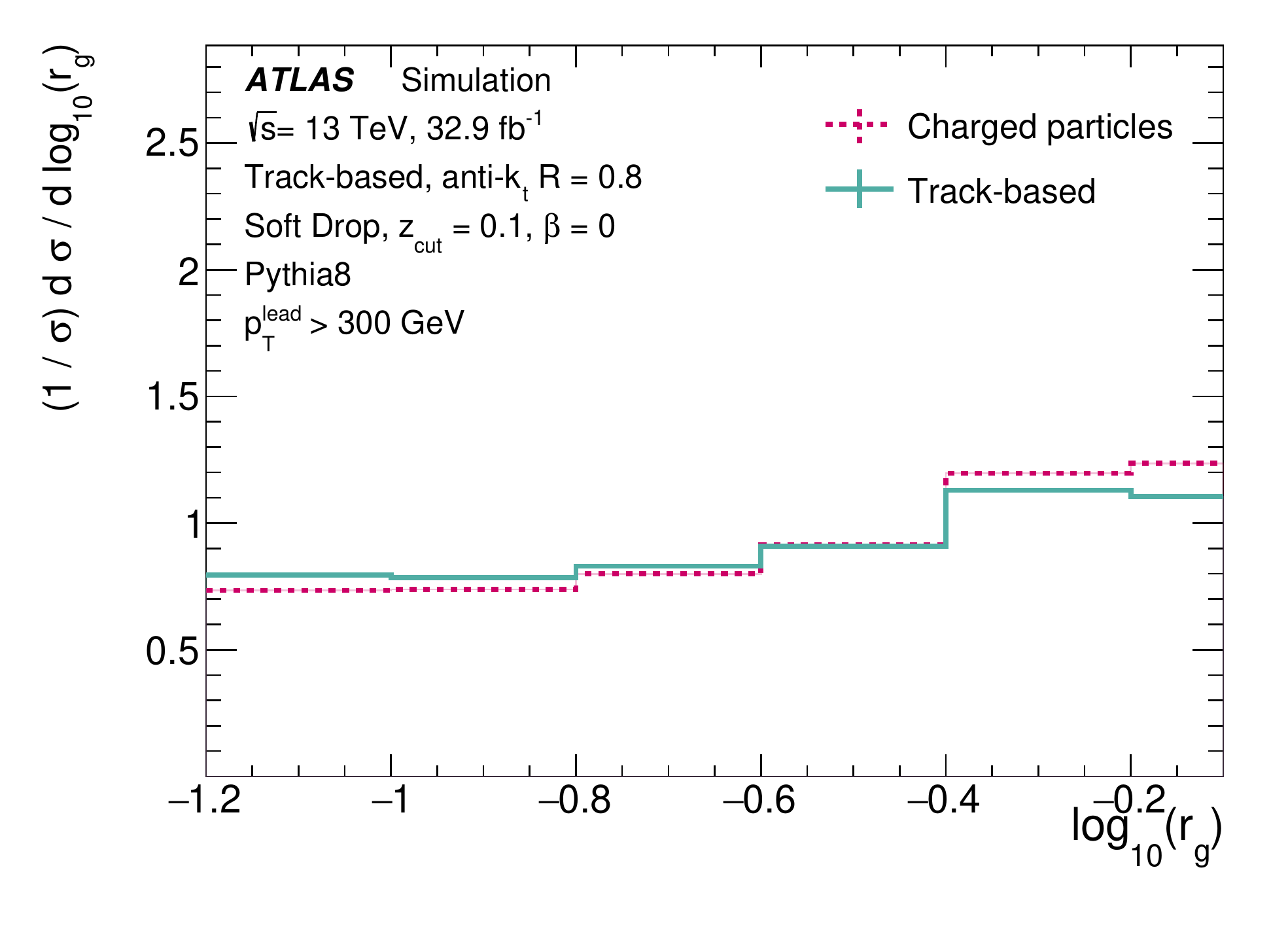}}
\end{center}
\caption{The distribution in simulation of $r_g$ at detector level and particle level for the more central of the dijet system
for $\beta=0$ for (a) calorimeter-based (all particles), and (b) track-based (charged particles). The statistical uncertainties are drawn, but are too small to be visible.}
\label{fig:observs:rg}
\end{figure}

\section{Unfolding}

The substructure observables are reconstructed in bins of the transverse momentum of the jet, and the double-differential distributions are unfolded using \PYTHIA 8.186.
An iterative Bayesian technique~\cite{DAgostini:1994zf} is used with one (four) iterations for track-based (calorimeter-based) observables. These values were chosen
to minimize the total uncertainty, and are implemented in the RooUnfold framework~\cite{Adye:2011gm}.

The probability distributions of obtaining a particle-level value given a detector-level observation, $\Pr(\text{particle-level}|\text{detector-level})$, in \PYTHIA8 for $\beta=0$ are presented for all three observables for the calorimeter-based and track-based definitions in Figure~\ref{fig:response:m}.
While the unfolding is done simultaneously in $\pT$ and the jet observable, the unfolding matrices are shown inclusively in \pT for simplicity.
As anticipated, the unfolding matrices for the track-based observables have significantly smaller off-diagonal elements than their calorimeter-based analogs.
 
\begin{figure}[h!]
\begin{center}
\subfigure[$\rho$ distribution, $\beta = 0$, calorimeter-based]{\includegraphics[width=0.40\textwidth]{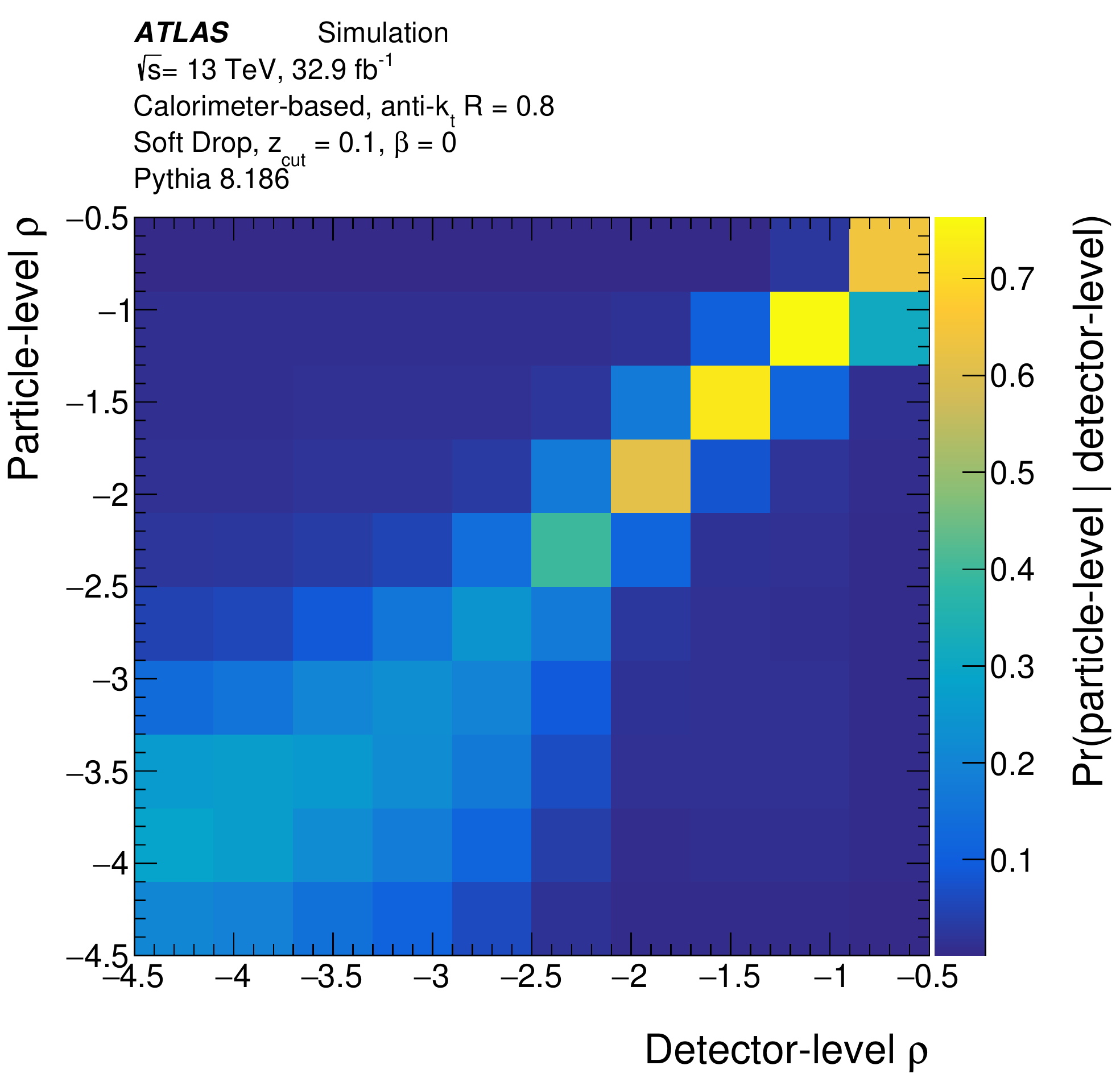}}
\subfigure[$\rho$ distribution, $\beta = 0$, track-based]{\includegraphics[width=0.40\textwidth]{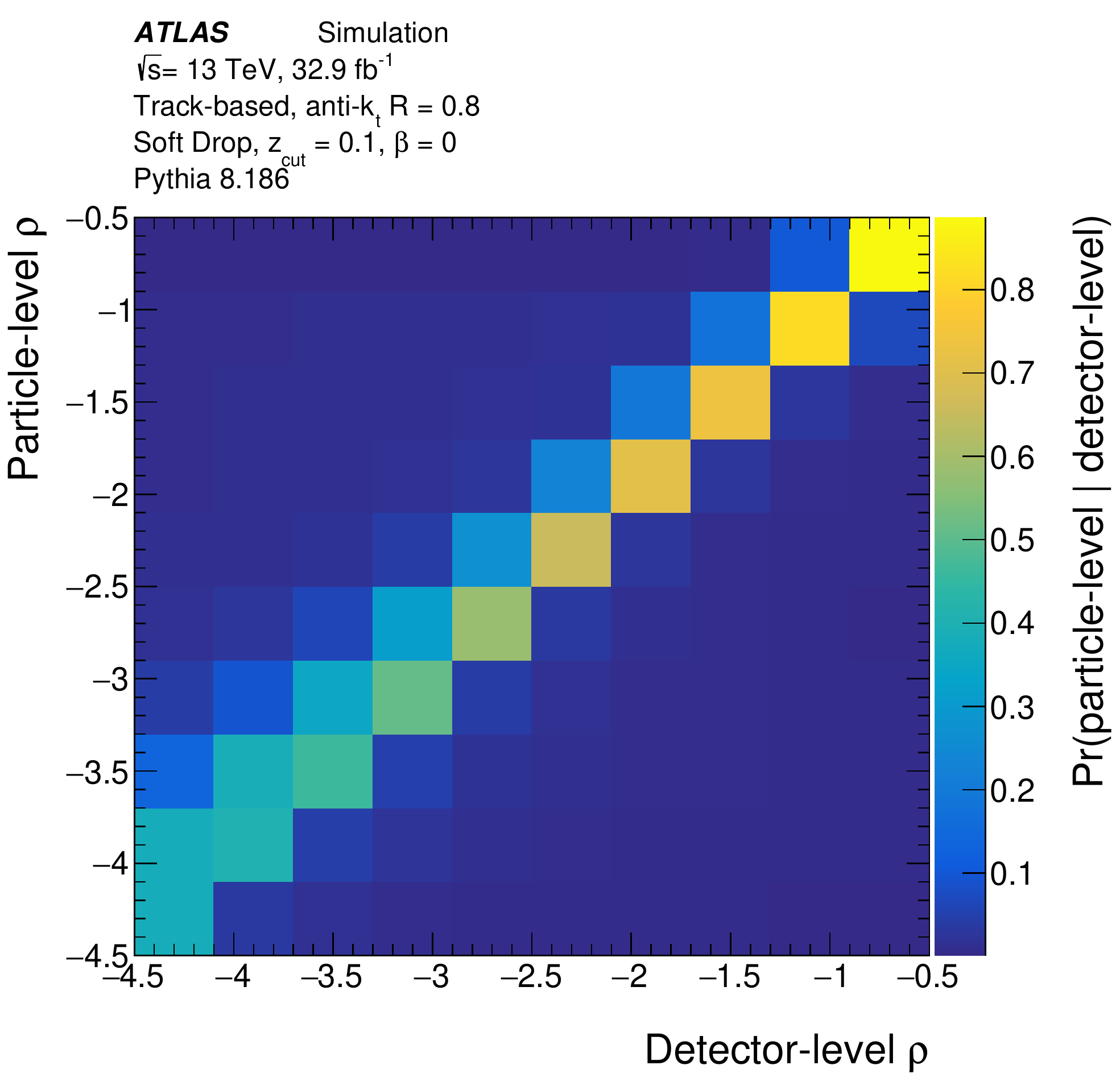}}
\subfigure[$z_g$ distribution, $\beta = 0$, calorimeter-based]{\includegraphics[width=0.40\textwidth]{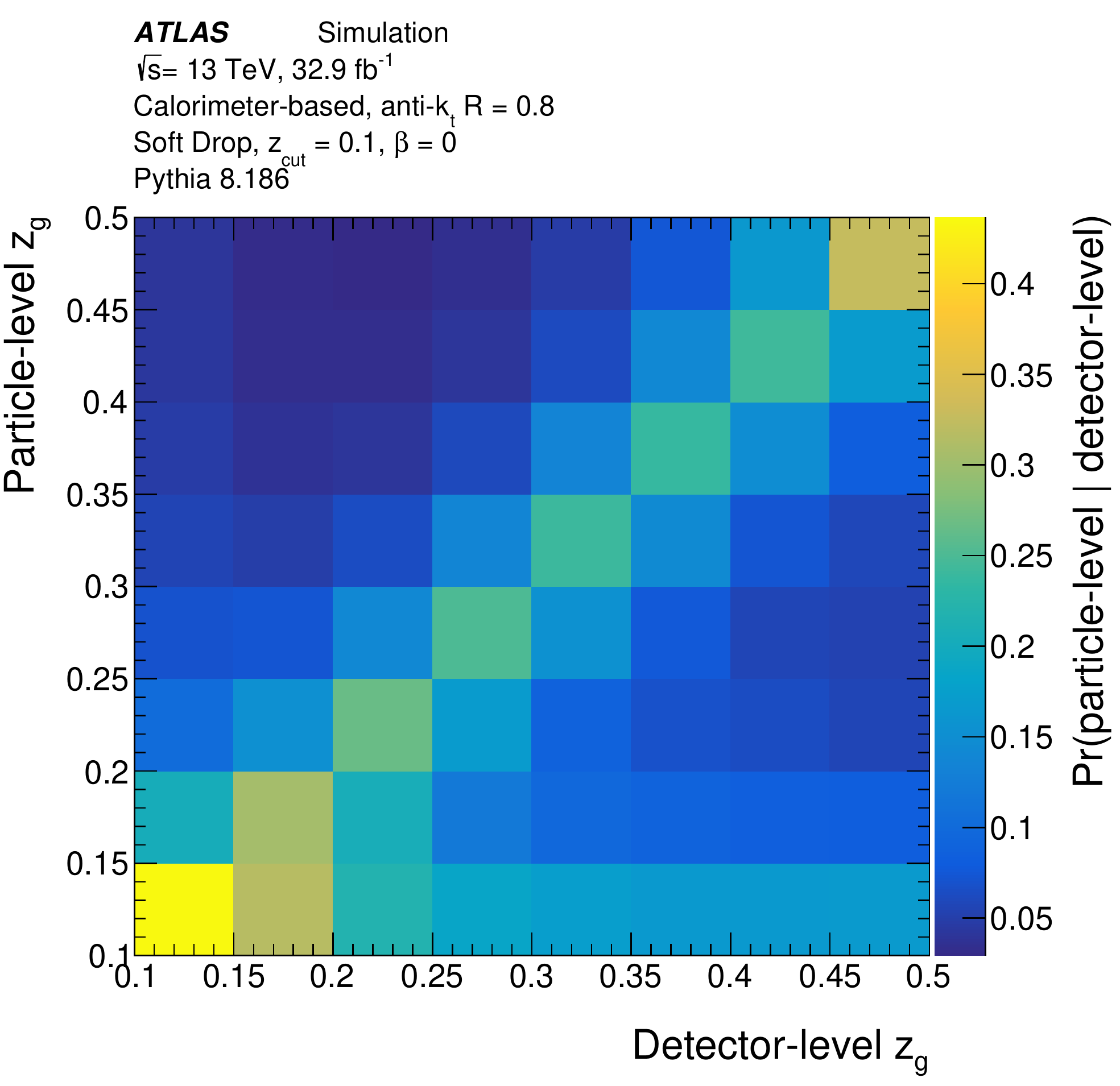}}
\subfigure[$z_g$ distribution, $\beta = 0$, track-based]{\includegraphics[width=0.40\textwidth]{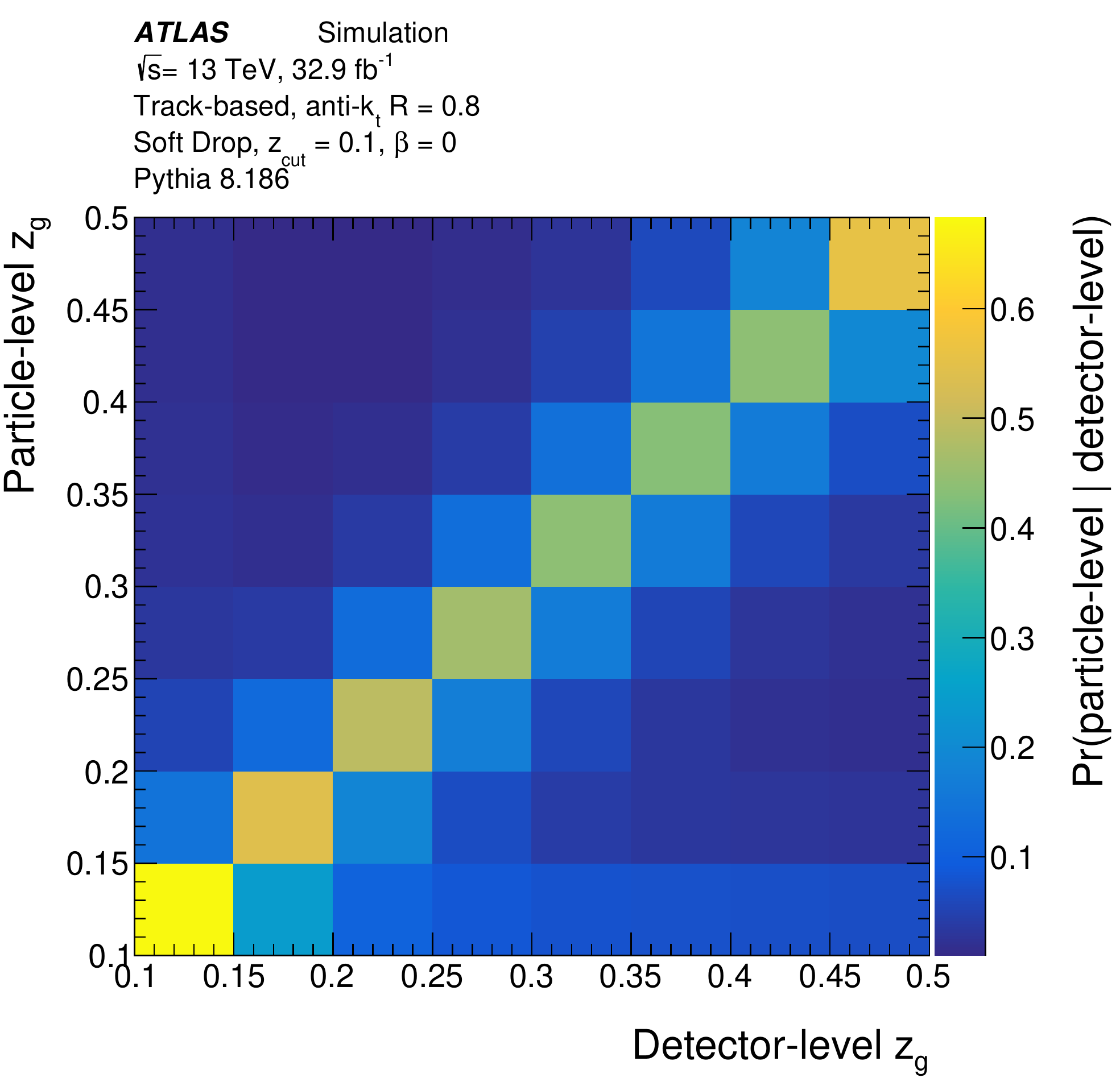}}
\subfigure[$r_g$ distribution, $\beta = 0$, calorimeter-based]{\includegraphics[width=0.40\textwidth]{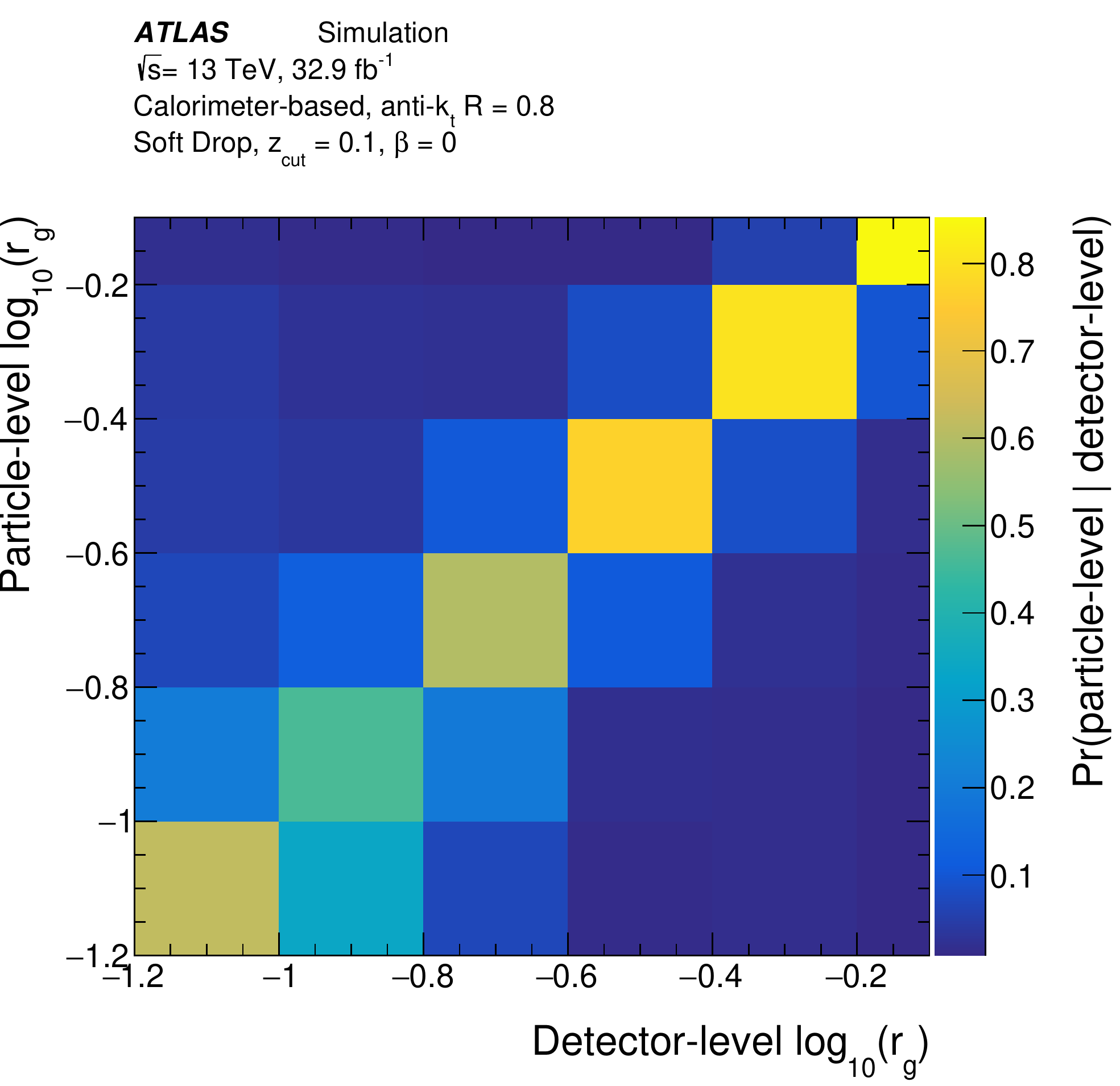}}
\subfigure[$r_g$ distribution, $\beta = 0$, track-based]{\includegraphics[width=0.40\textwidth]{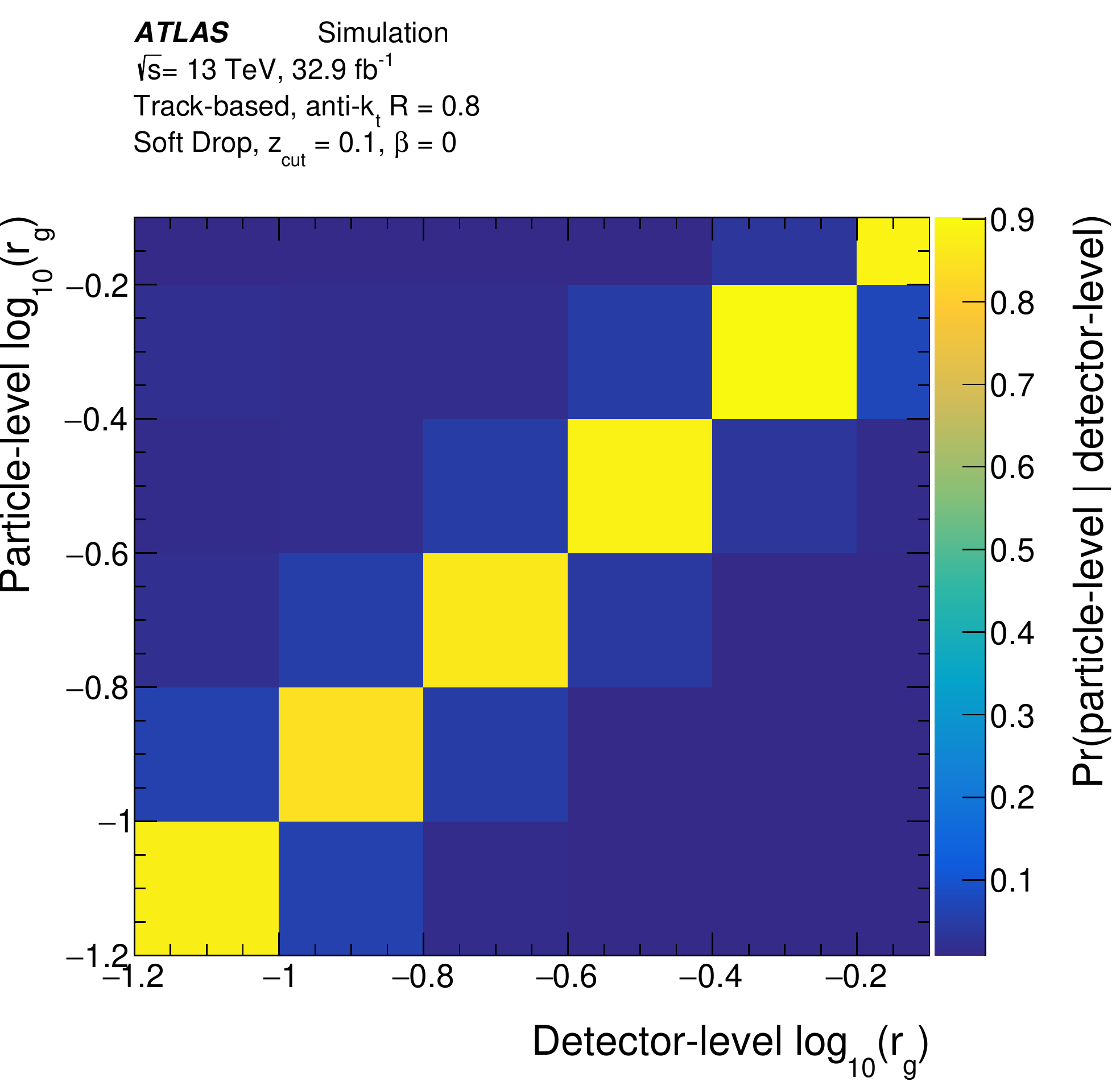}}
\end{center}
\caption{The distribution of $\Pr(\text{particle-level}|\text{detector-level})$ for the more central jet for (top) $\rho$, (middle) $z_g$, and (bottom) $r_g$ with $\beta=0$ for \PYTHIA8 for (left) calorimeter-based definition, and (right) track-based definition. }
\label{fig:response:m}
\end{figure}

\section{Uncertainties}
\label{sec:systematics}
% The next lines are included from the .//sections/systematics.tex input file
 
Several sources of statistical and systematic uncertainties are considered for this analysis.
The data and simulation statistical uncertainties are evaluated from pseudo-experiments using the bootstrap method~\cite{efron1979}.
The uncertainties from the calorimeter-cell reconstruction, track reconstruction, and MC modeling are determined by applying variations to the simulation, as detailed in Sections~\ref{sec:systs:exp},~\ref{sec:syst:trk}, and~\ref{sec:systs:modeling}, respectively. The impact of the calorimeter-cell cluster uncertainties on the jets is taken into account for both the calorimeter-based measurement as well as the track-based measurement since it impacts the selection of jets.
The varied simulation is then used to repeat the unfolding procedure and the deviation from the nominal result is used to estimate the uncertainty.
The uncertainty in the pileup modeling is determined by reweighting the pileup profile up by 10\% in MC simulation.
The uncertainty in the unfolding procedure (unfolding nonclosure) is computed using a data-driven reweighting procedure~\cite{Malaescu:2009dm}.
In this method, the particle-level spectrum is reweighted such that the reconstructed spectrum better matches the data distribution, while the response matrix is left unchanged.
The difference between the reweighted detector-level simulation after unfolding and the generator-level simulation from the same generator is then taken as an uncertainty.
All uncertainties are symmetrized unless stated otherwise.

A summary of all the uncertainties considered is given in Section~\ref{sec:systs:summary}.
 
\subsection{Calorimeter-cell cluster uncertainties}
\label{sec:systs:exp}
 
Uncertainties on the reconstruction of calorimeter cell clusters are estimated using comparisons between tracks with momentum $p$ and clusters with energy $E$ in data and in simulation.

Calorimeter cell clusters require seed cells that exceed the noise threshold; if a particle interacts with the material in front of the calorimeter
and produces many spread-out low-energy secondary particles, there may not be sufficient localized energy to seed a cluster.
The rate at which particles do not seed a cluster is studied with tracks that do not match a calorimeter cell cluster within $\Delta R<0.2$, where
tracks are extrapolated to the calorimeter layer corresponding to the energy-weighted position of the calorimeter cell cluster.
This rate is studied at 13 TeV $pp$ collisions using tracks isolated from all other track candidates by at least $\Delta R=0.4$.
The data/MC difference is then used to derive the cluster reconstruction efficiency uncertainty, which is evaluated in bins of pseudorapidity and energy~\cite{PERF-2015-05}.
To assess the impact of this uncertainty on the unfolded results, clusters are randomly removed at a rate determined by the measured difference between data and simulation -- less than 5\% for low-momentum clusters and negligible beyond 10~\GeV.

The cluster energy scale and resolution uncertainties are determined in three separate regions.
For $E<30$~\GeV, there are enough events to derive these uncertainties using the full $E/p$ distribution in data~\cite{PERF-2015-05}.
For any clusters with $30 < E < 350$~\GeV, the uncertainties are derived from the combined testbeam data~\cite{Abat:2010zza}.
Finally, for regions outside of the testbeam and $E/p$ coverage, a $\pt$- and $\eta$-independent 10\% uncertainty is assigned as a conservative estimate of the uncertainty,
as done in previous studies~\cite{PERF-2015-05}.
 
For the regions where the uncertainty is derived using $E/p$, the mean and standard deviation of the distributions are extracted in bins of $E$ and $|\eta|$.  Only tracks with at least one associated cluster are included, using the same matching criteria as for the cluster efficiency.
Depending on the fit quality, either the mean and $\sigma$ of a Gaussian fit to the data, or the distribution mean and RMS values are used.  For example, for $p\approx 25$~\GeV~and $\eta\approx 0$, the data and simulation are consistent with $\langle E/p\rangle=1$ and $\sigma(E/p)=0.22$ within 1\% for the mean and 5\% for the standard deviation.
 
To evaluate the cluster energy scale uncertainty, the cluster energy in simulation is shifted according to the difference of the $E/p$ mean value between data and MC simulation. Similarly, to evaluate the cluster energy resolution uncertainty, cluster energies are smeared according to data/MC differences in the $E/p$ distribution by one standard deviation. The effect of the energy scale and resolution uncertainties is defined as the relative difference between the nominal and modified jet substructure observable. A series of validation studies
which probe the jet energy scale, jet mass scale, and jet mass resolution were performed to ensure
that this prescription is also valid for nonisolated clusters within jets.

The cluster angular resolution is estimated using a similar method by studying the modeling of the $\Delta R$ distribution between tracks and calorimeter-cell clusters.
 
\subsection{Tracking uncertainties}
\label{sec:syst:trk}
 
Systematic uncertainties are evaluated for the track reconstruction efficiency, fake rate, and momentum scale. The efficiency is decomposed into two components: one from the uncertainty in the inner detector material (`inclusive efficiency') and one from the modeling of pixel cluster merging inside dense environments, such as inside the core of high-energy jets (`efficiency within jets').
 
The inclusive efficiency uncertainty is due to the material uncertainty, which is constrained by detector construction knowledge and photon conversions as well as hadronic interactions~\cite{PERF-2015-07}.
The total relative uncertainty on the efficiency is 0.5\% for $|\eta| < 0.1$ and grows to 2.7\% for the $2.3 < |\eta| < 2.5$ region.
The impact of this uncertainty in the measured distributions is evaluated by randomly removing tracks in simulation with a $p_\text{T}$- and $|\eta|$-dependent probability.
 
The uncertainty in the tracking efficiency in dense environments is due to the modeling of pixel cluster merging.
This is studied using the $dE/dx$ method~\cite{PERF-2015-08,ATL-PHYS-PUB-2016-007}: the rate of pixel clusters assigned to single tracks with a large charge (comparable to twice a minimum ionizing particle charge) in the core of jets is measured in data and in simulation.  The comparison between data and simulation results in an additional 0.4\% (absolute)
uncertainty that is only applied to tracks within a $\Delta R = 0.1$ of a jet.
 
Fake tracks result from random combinations of hits mostly from charged particles that happen to overlap in space.
Outside of jets, the fake rate is highly pileup dependent, as the chances for many low-$p_\text{T}$ particles to be close increases with the number of particles in the event.  However, inside jets, the density from primary charged particles is also high and can result in an increased fake rate.  The fake rate itself is much less than 1\%, but fake tracks can have a large $p_\text{T}$.  The modeling of the fake rate is studied with a dedicated measurement that enriches the rate of fake tracks by inverting various track quality criteria~\cite{ATL-PHYS-PUB-2017-016}.  The simulation reproduces the fake rate to within about 30\% of the observed rate in data.  The fake-rate uncertainty is estimated by randomly removing 30\% of fake tracks.
 
The leading source of uncertainty in the track parameters is in the $q/p_\text{T}$ ($q$ is the electric charge) from a potential sagitta distortion due to detector-misalignment weak modes~\cite{ATL-PHYS-PUB-2015-051}. This bias is corrected for, once per data-taking period, and the correction is about $0.1$/TeV except at $\phi\approx 0$ and $|\eta|\approx 2.5$ where the correction can reach 1/TeV. The impact on the measurement is smaller than that of the other tracking uncertainties.
 
\subsection{Modeling uncertainty}
\label{sec:systs:modeling}
 
Since the detector response depends on the energy and angular distribution of particles inside jets, it is sensitive to the fragmentation model used for the unfolding.  An uncertainty is estimated by repeating the unfolding procedure using \SHERPA and comparing that with the nominal unfolding that uses \PYTHIA8, and taking the full difference as the uncertainty.
The result of performing this procedure with \HERWIGpp instead of \SHERPA produces a similar uncertainty.
In addition to the direct sensitivity to the fragmentation modeling, there is also an indirect sensitivity to the quark/gluon fractions and the jet momentum distribution.  An uncertainty due to the PDFs is evaluated as the spread in the unfolded distributions from 100 NNPDF2.3LO eigenvector variations.
 
\subsection{Summary of uncertainties}
\label{sec:systs:summary}
 
Figure~\ref{results:uncert:unfold:central_inclPt} presents a summary of the total and individual uncertainties for all observables and $\beta=0$ for both the calorimeter-based and track-based measurements,
where all of the uncertainties are summed in quadrature to obtain the total uncertainty.
The uncertainties change with $\beta$, due to the differing angular sensitivity, but the overall conclusions are similar.
For the calorimeter-based $\rho$, the fragmentation modeling is the dominant uncertainty for most of the mass range, while the pileup modeling and cluster energy scale uncertainties dominate at high relative mass.
A similar description is true for the track-based $\rho$, where the fragmentation modeling is the dominant
uncertainty across the entire $\rho$ range and the effects from the unfolding nonclosure are subdominant, while the tracking uncertainties are typically negligible.
Analogous results hold for the calorimeter-based $r_g$ observable, while for the track-based $r_g$ measurement, subdominant effects are seen from the cluster energy scale, fake rate,
and data statistical uncertainty.
For calorimeter-based $z_g$, the cluster energy scale and modeling uncertainties are most important, and the uncertainties are generally smaller than for $\rho$ and $r_g$.
A similar description holds for the track-based $z_g$, whose uncertainty is dominated by the modeling and unfolding nonclosure uncertainties.

\begin{figure}[h!]
\begin{center}
\subfigure[$\rho$, $\beta = 0$, calorimeter-based]{\includegraphics[width=0.49\textwidth,page=1]{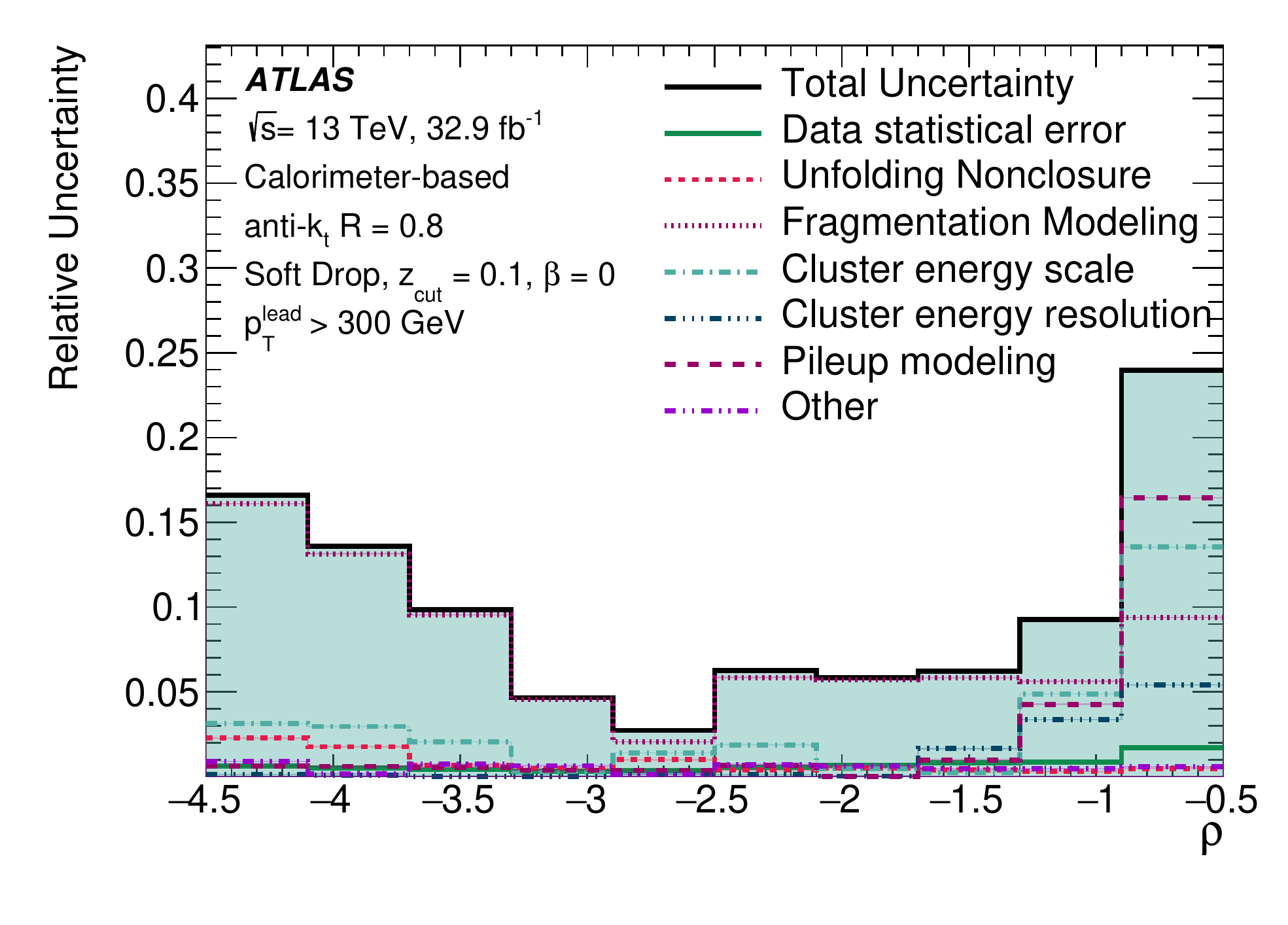}}
\subfigure[$\rho$, $\beta = 0$, track-based]{\includegraphics[width=0.49\textwidth,page=1]{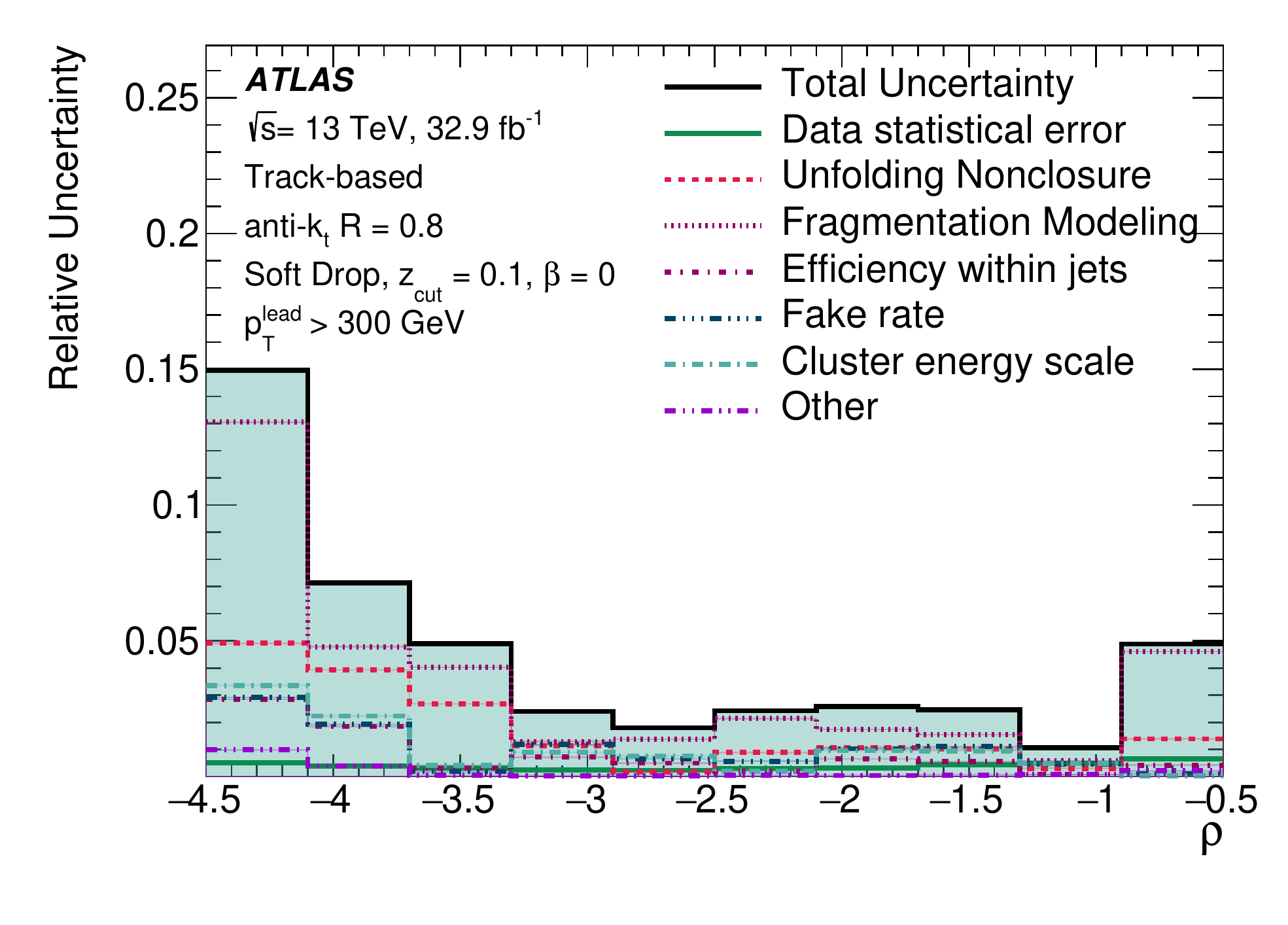}}
\subfigure[$z_g$, $\beta = 0$, calorimeter-based]{\includegraphics[width=0.49\textwidth,page=1]{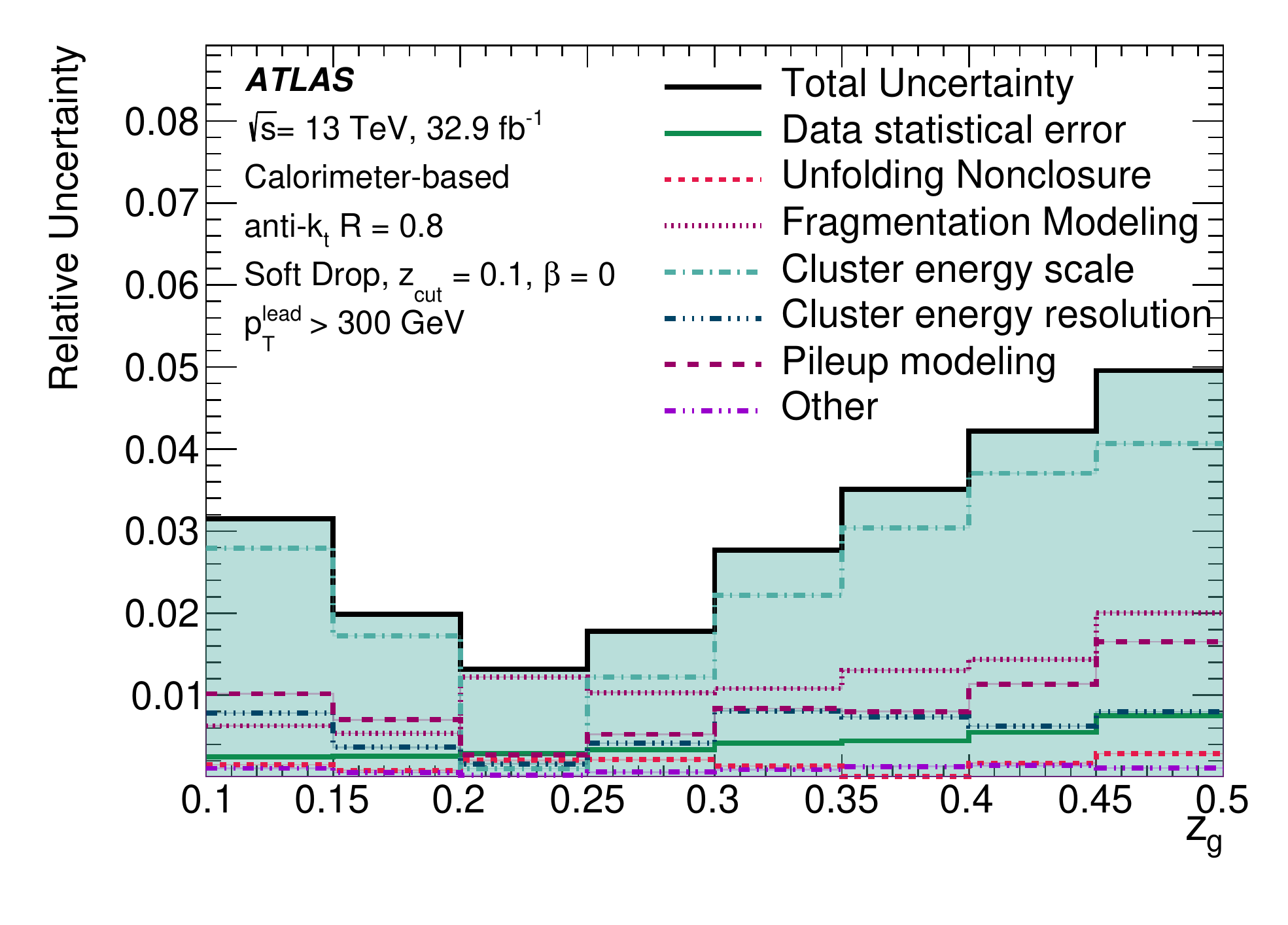}}
\subfigure[$z_g$, $\beta = 0$, track-based]{\includegraphics[width=0.49\textwidth,page=1]{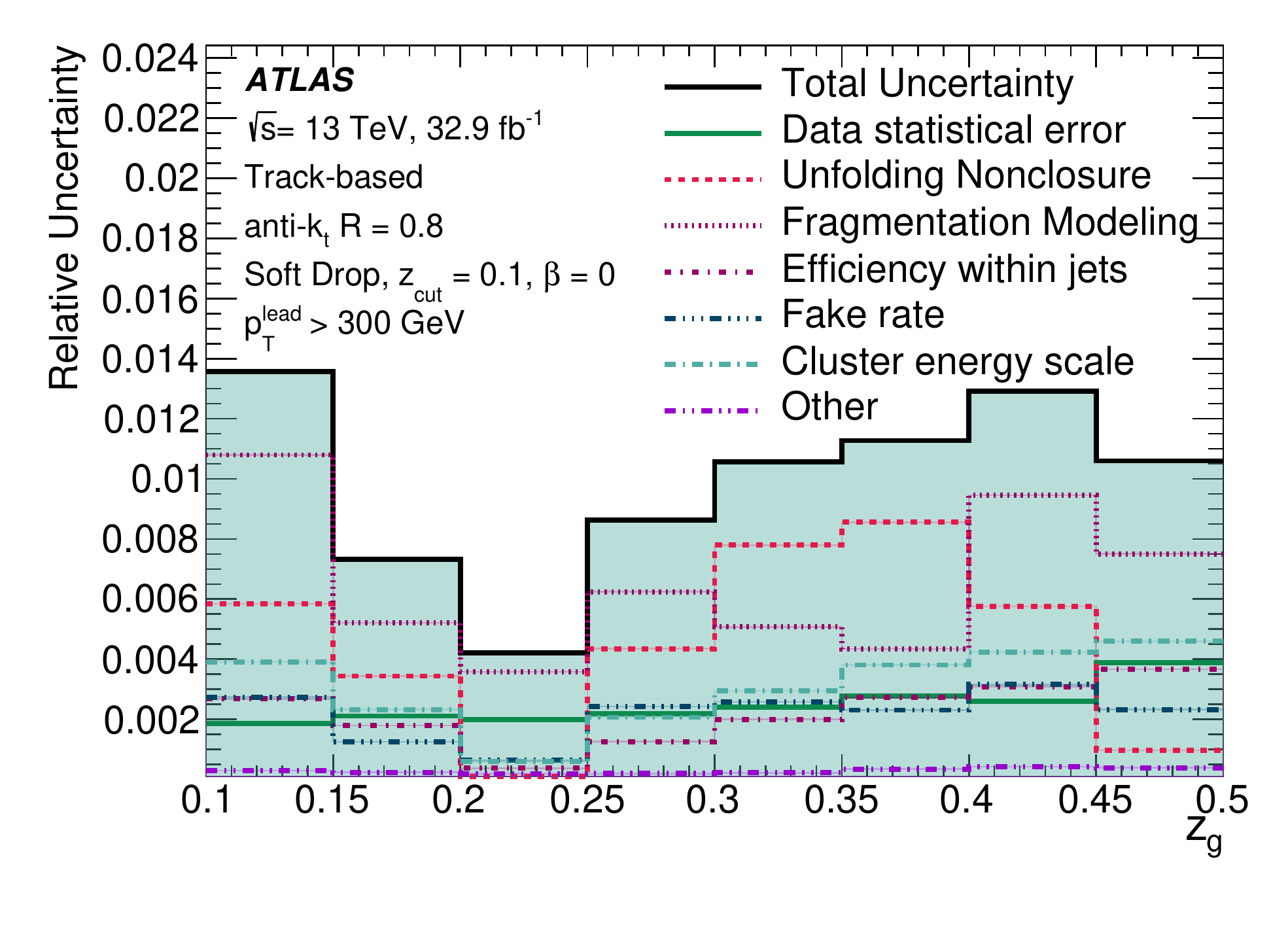}}
\subfigure[$r_g$, $\beta = 0$, calorimeter-based]{\includegraphics[width=0.49\textwidth,page=1]{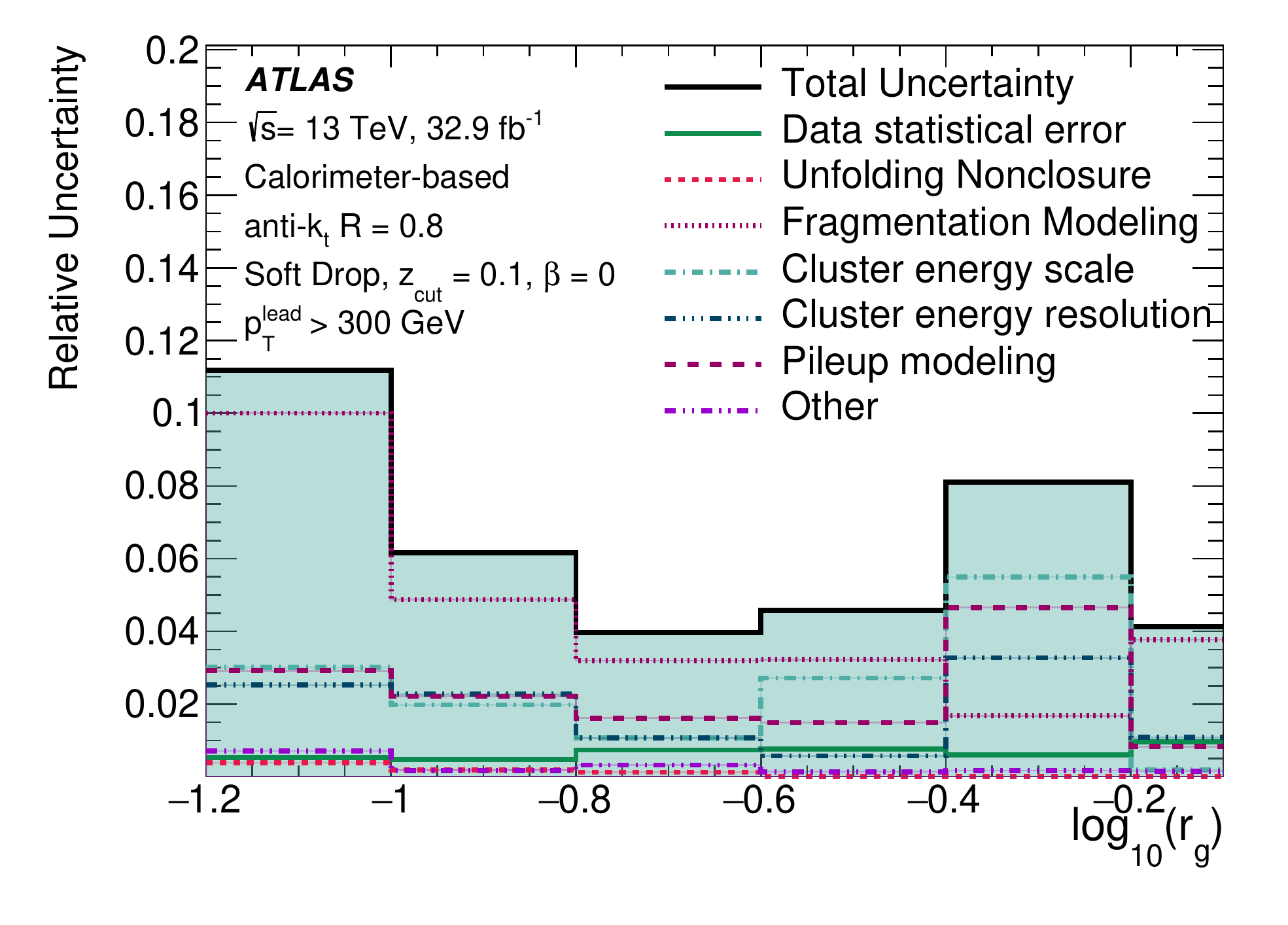}}
\subfigure[$r_g$, $\beta = 0$, track-based]{\includegraphics[width=0.49\textwidth,page=1]{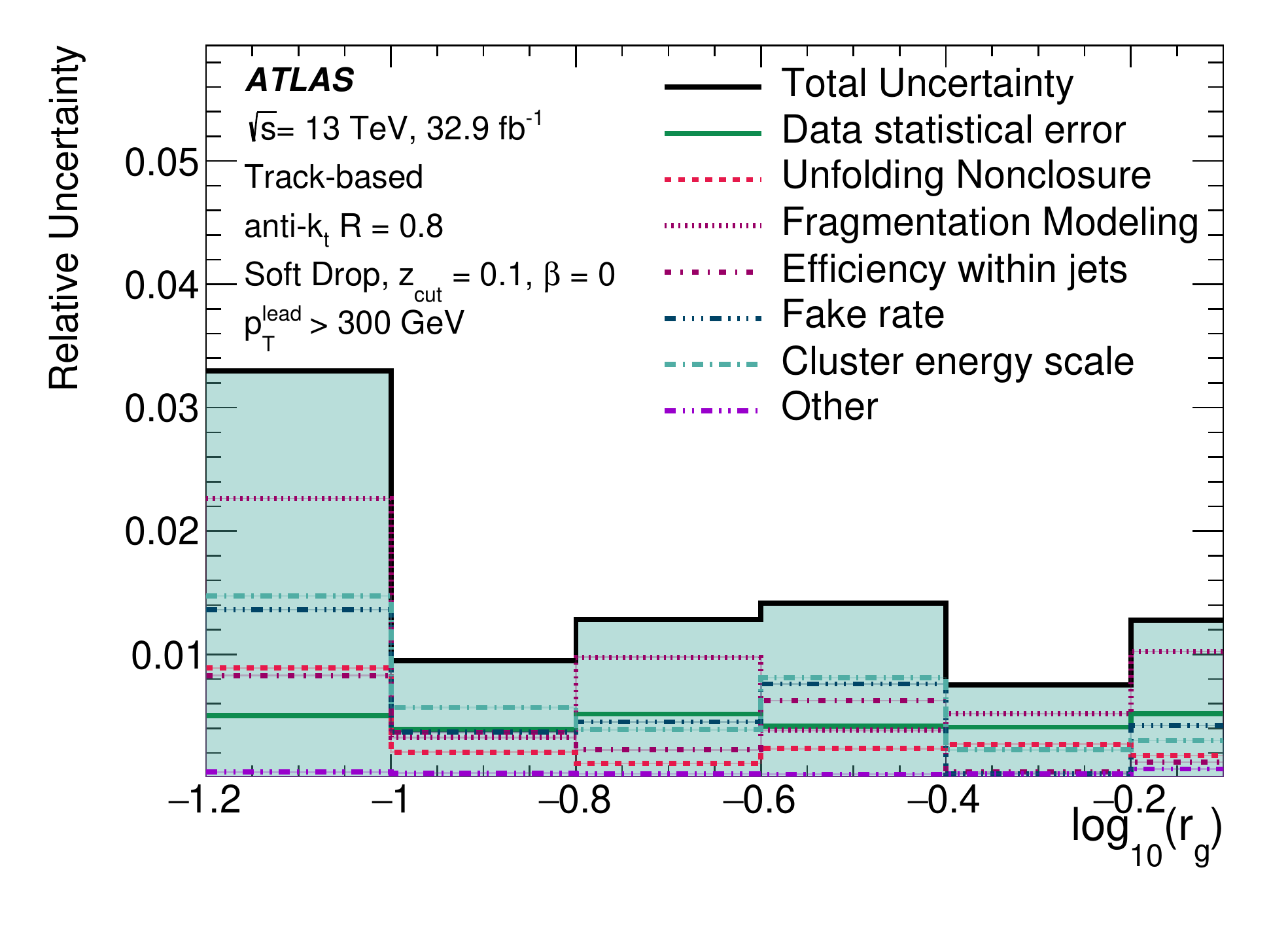}}
\end{center}
\caption{Total and individual uncertainties inclusive in \pt for $\beta = 0$ for calorimeter-based observables (left) and track-based observables (right) for $\rho$ (top), $z_g$ (middle), and $r_g$ (bottom).}
\label{results:uncert:unfold:central_inclPt}
\end{figure}

% End of text imported from the .//sections/systematics.tex input file
 
\section{Results}
% The next lines are included from the .//sections/results.tex input file
The unfolded data are presented in several different ways, in order to highlight various aspects of the measurement.
Since these distributions change slowly as a function of \pT, most of the results are shown inclusively in \pT.  Section~\ref{sec:mcComparison} provides a comparison between the unfolded data and several MC predictions,
highlighting the various regions of each measurement which are well-modeled by simulation. This is followed by a comparison
between the unfolded data and state-of-the-art analytical predictions in Section~\ref{sec:theoryComparison}.
Section~\ref{sec:trackCalo} directly compares the results of the measurements of the calorimeter- and track-based observables.
While these observables are unfolded to different particle-level definitions, this comparison highlights the similarities between the different
definitions, as well as demonstrating the improved precision in track-based measurements of observables sensitive to the angular structure of the jet.
The forward and central measurements are compared in Section~\ref{sec:forwardCentral}, and these measurements are used as input to the
extraction of the quark- and gluon-jet distributions of these observables, which are shown in Section~\ref{sec:quarkGluon}.
 
\subsection{Comparison with MC predictions}
\label{sec:mcComparison}
 
Figures~\ref{results:unfold:rho:mc}--~\ref{results:unfold:rg:mc} compare the unfolded data from both jets with the particle-level distributions
from MC generators described in Section~\ref{sec:MCandata}.
Several trends are visible in these results. For $\rho$, the MC predictions are mostly accurate within 10\% except for the lowest relative masses, which are dominated by nonperturbative physical effects. This becomes more visible for larger values of $\beta$, where more soft radiation is included within the jet,
increasing the size of the nonperturbative effects. In addition, in the high-relative-mass region, where the effects of the fixed-order calculation are relevant,
some differences between MC generators are seen. A similar trend may be seen for $r_g$, where the small-angle region shows more pronounced differences
between MC generators, since this corresponds to the region where nonperturbative effects are largest. Overall, these effects are smaller than for the relative mass.
Unlike the other two observables, $z_g$ is modeled well within about 10\% across most of the spectrum. However, there is some tension between the predictions and the
unfolded data, which is visible particularly for the track-based observables, which have better precision.
 
In general, the MC predictions show similar behavior for the calorimeter-based and track-based definitions, both in their overall distributions
and in their agreement with the unfolded data distribution. However, as the tracking measurement is more precise, the disagreement between data and MC simulation in the nonperturbative regions is more significant.  For instance, in Figure~\ref{results:unfold:zg:mc}(e)--~\ref{results:unfold:zg:mc}(f), the \HERWIGpp prediction does not agree with the unfolded distribution at high values of $z_g$
for the track-based case, but it does agree in the calorimeter-based case.

\begin{figure}[hp!]
\begin{center}
\subfigure[$\beta = 0$, calorimeter-based]{\includegraphics[width=0.45\textwidth,page=1]{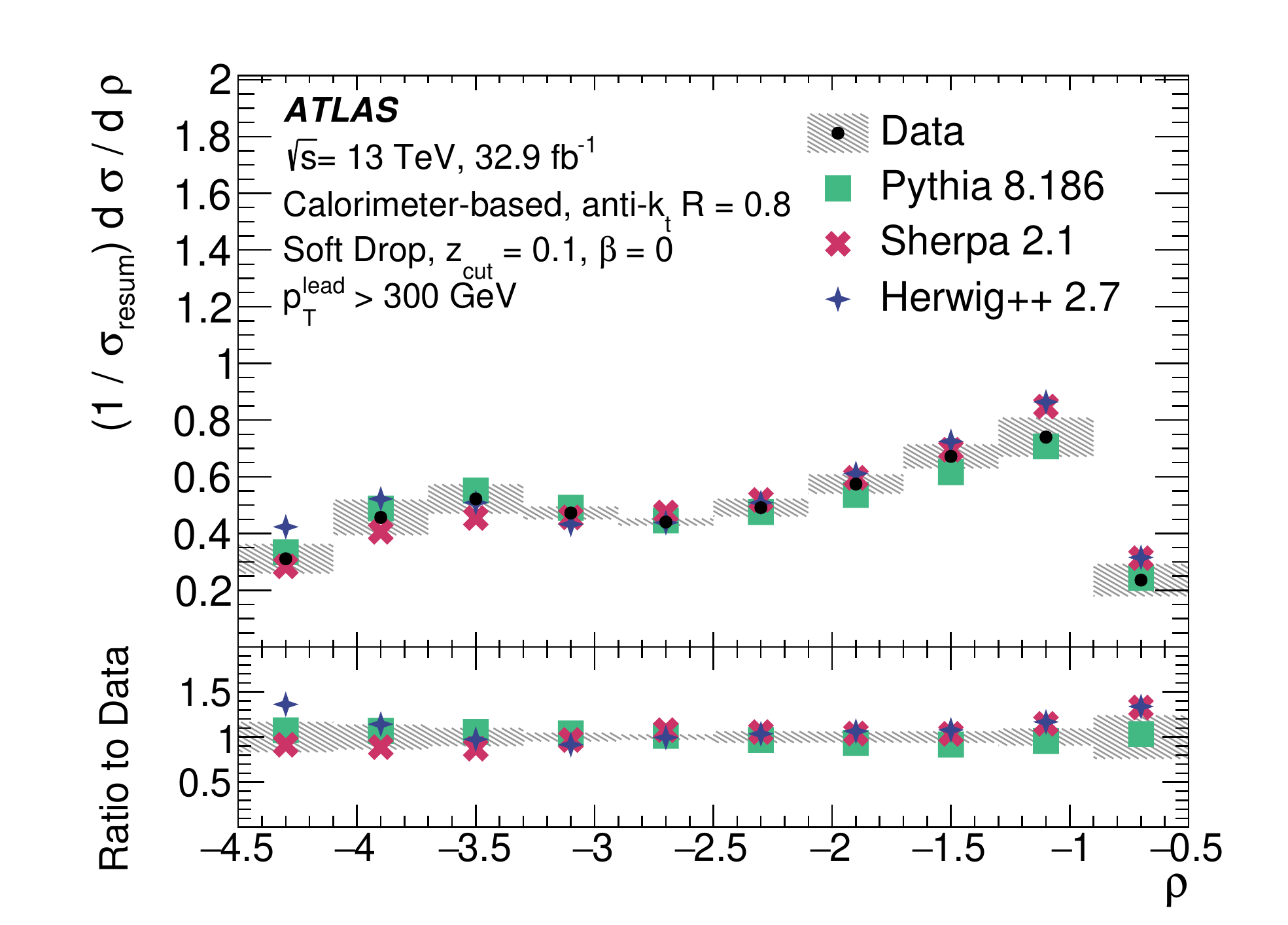}}
\subfigure[$\beta = 0$, track-based]{\includegraphics[width=0.45\textwidth,page=1]{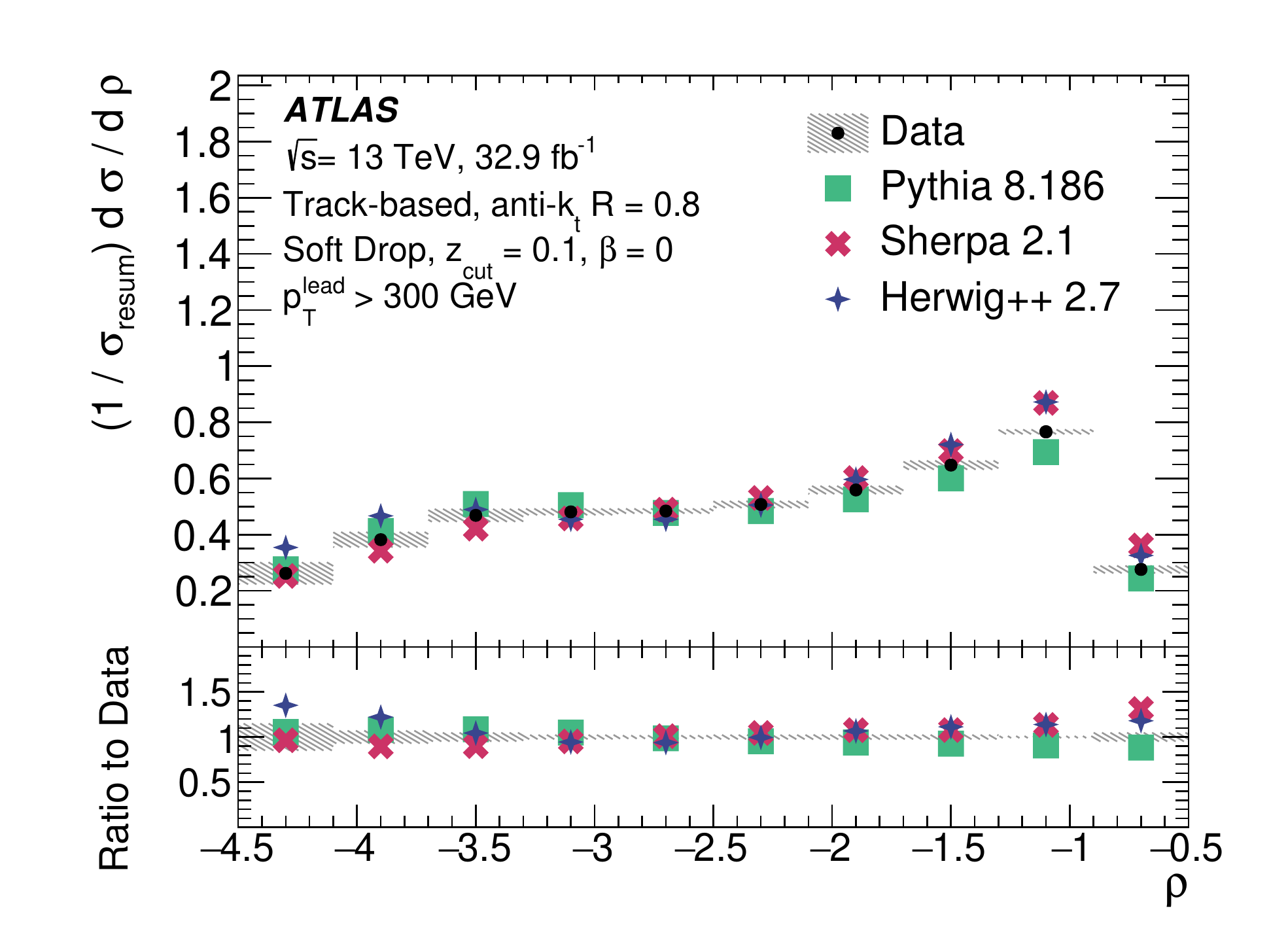}}
\subfigure[$\beta = 1$, calorimeter-based]{\includegraphics[width=0.45\textwidth,page=1]{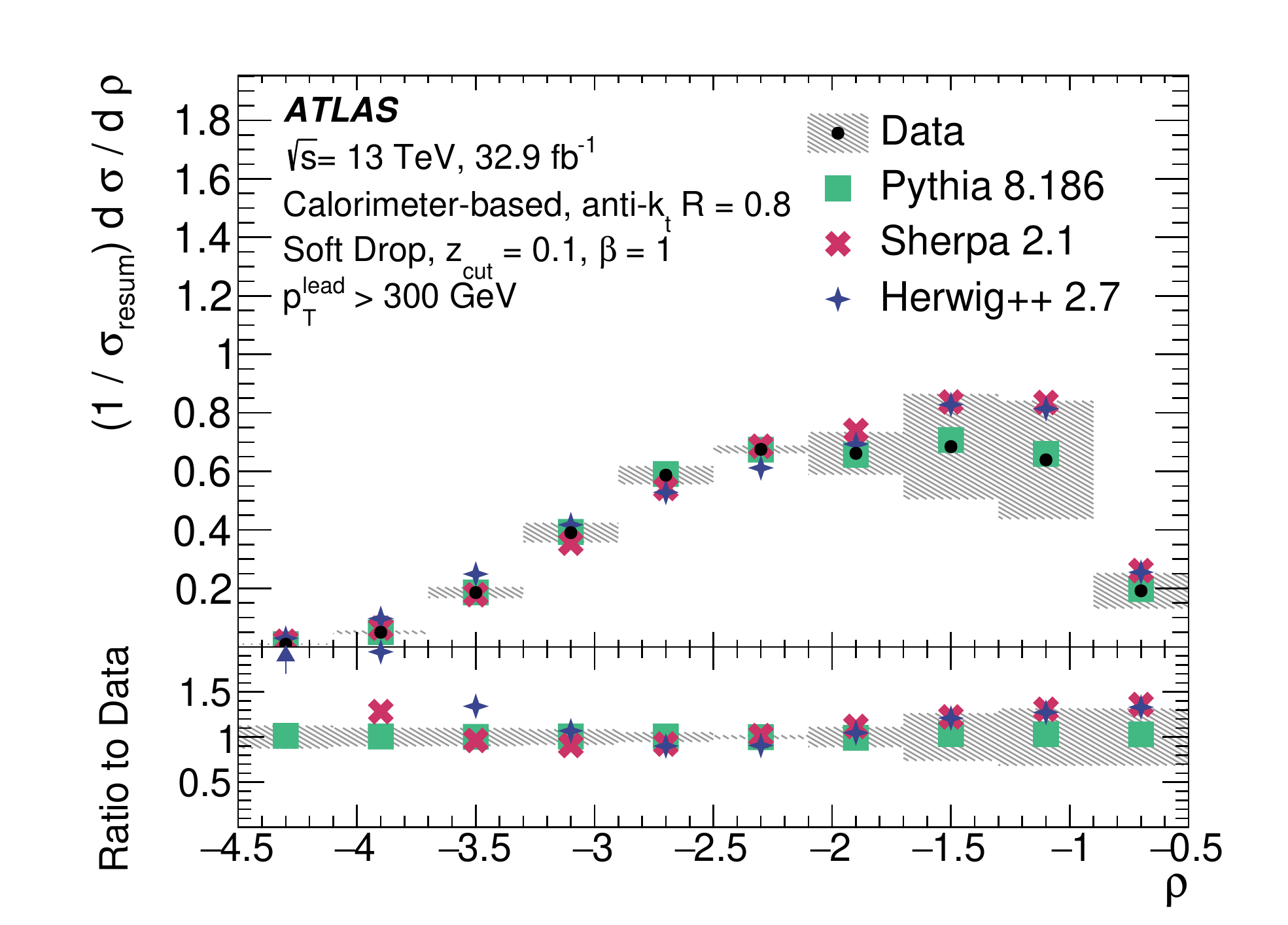}}
\subfigure[$\beta = 1$, track-based]{\includegraphics[width=0.45\textwidth,page=1]{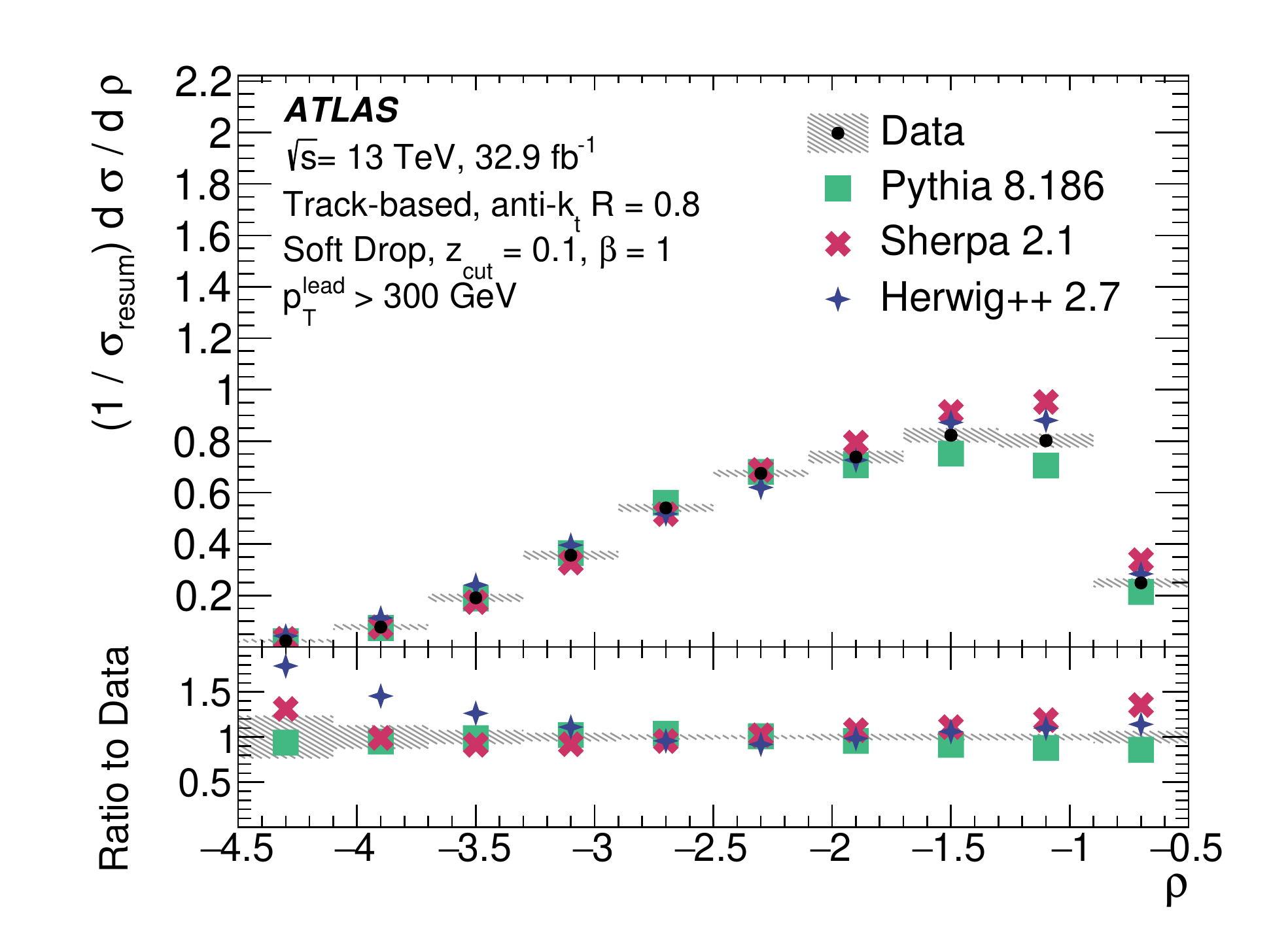}}
\subfigure[$\beta = 2$, calorimeter-based]{\includegraphics[width=0.45\textwidth,page=1]{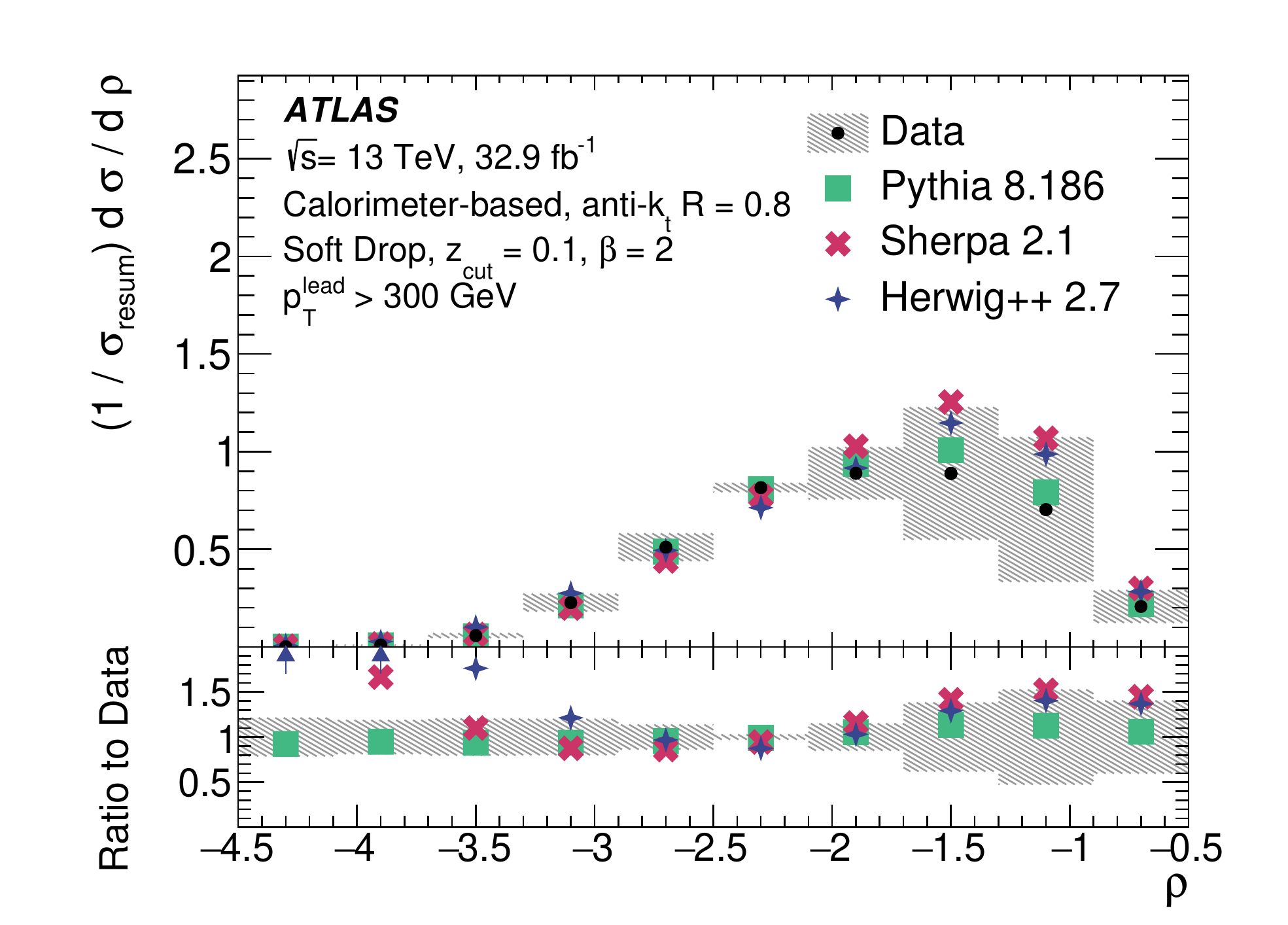}}
\subfigure[$\beta = 2$, track-based]{\includegraphics[width=0.45\textwidth,page=1]{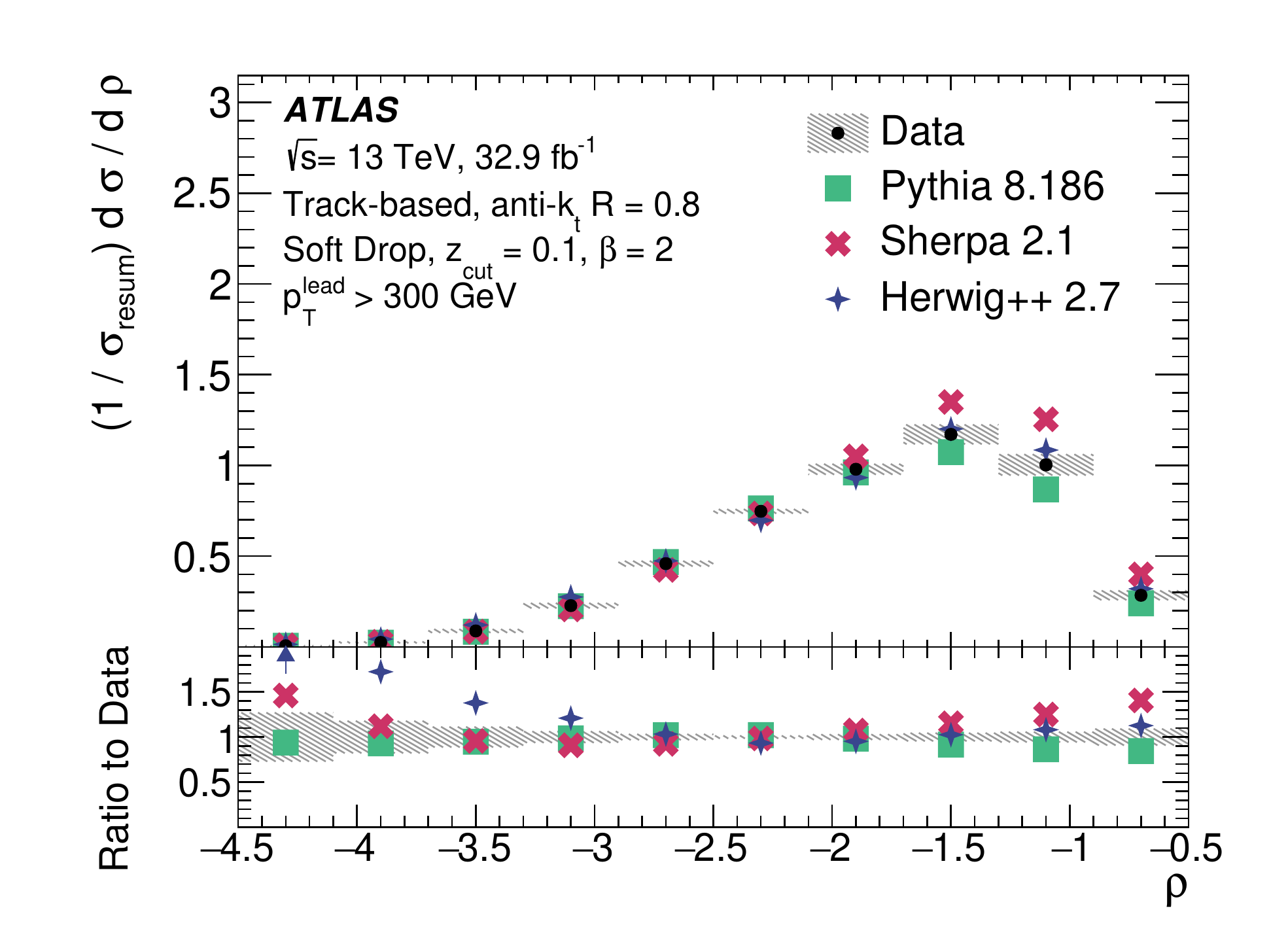}}
\end{center}
\caption{Comparison of the unfolded $\rho$ distribution with MC predictions. The uncertainty bands include all sources: data and MC statistical uncertainties, nonclosure, modeling, and cluster or tracking uncertainties where relevant.}
\label{results:unfold:rho:mc}
\end{figure}
 
\begin{figure}[hp!]
\begin{center}
\subfigure[$\beta = 0$, calorimeter-based]{\includegraphics[width=0.45\textwidth,page=1]{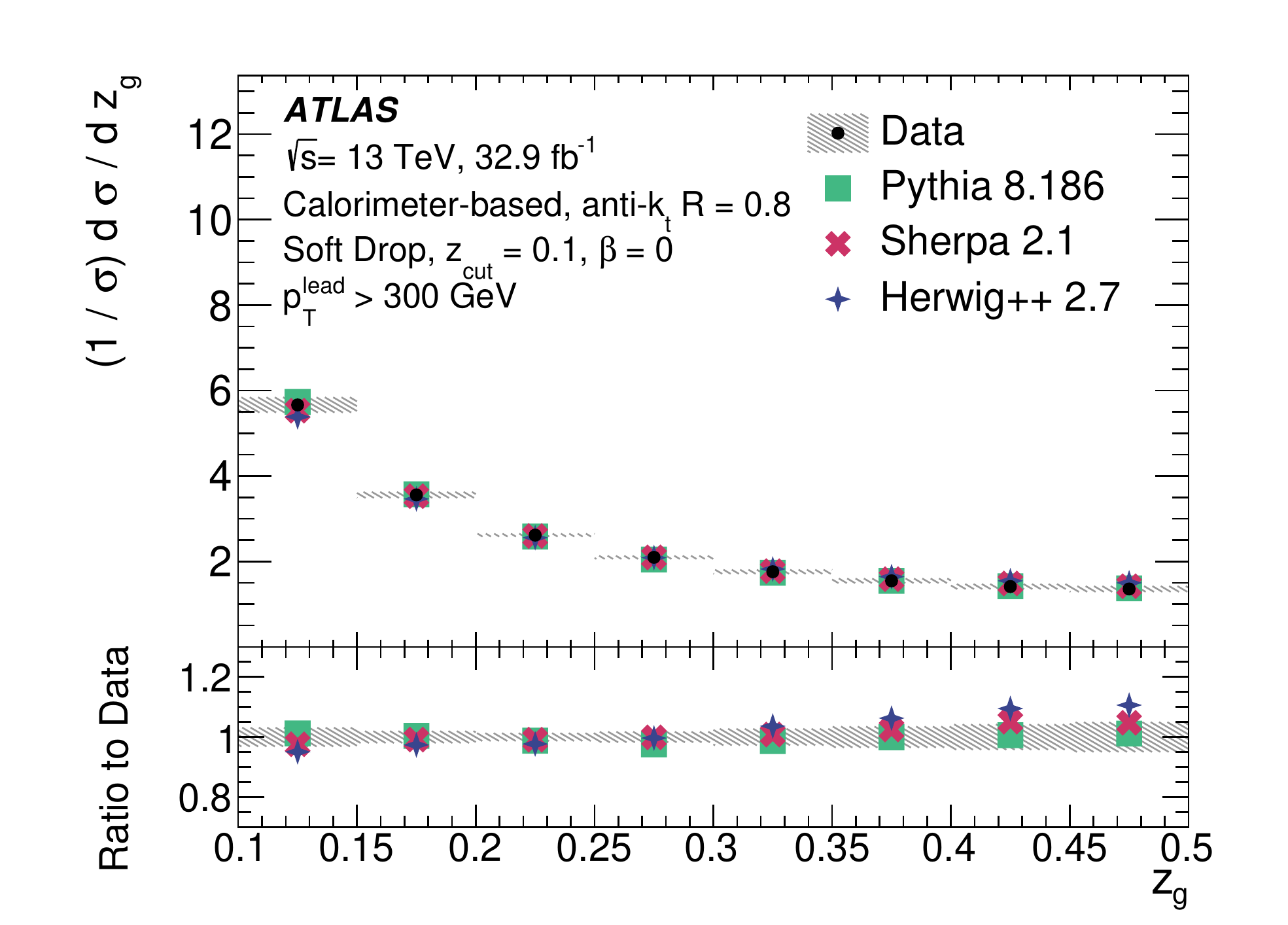}}
\subfigure[$\beta = 0$, track-based]{\includegraphics[width=0.45\textwidth,page=1]{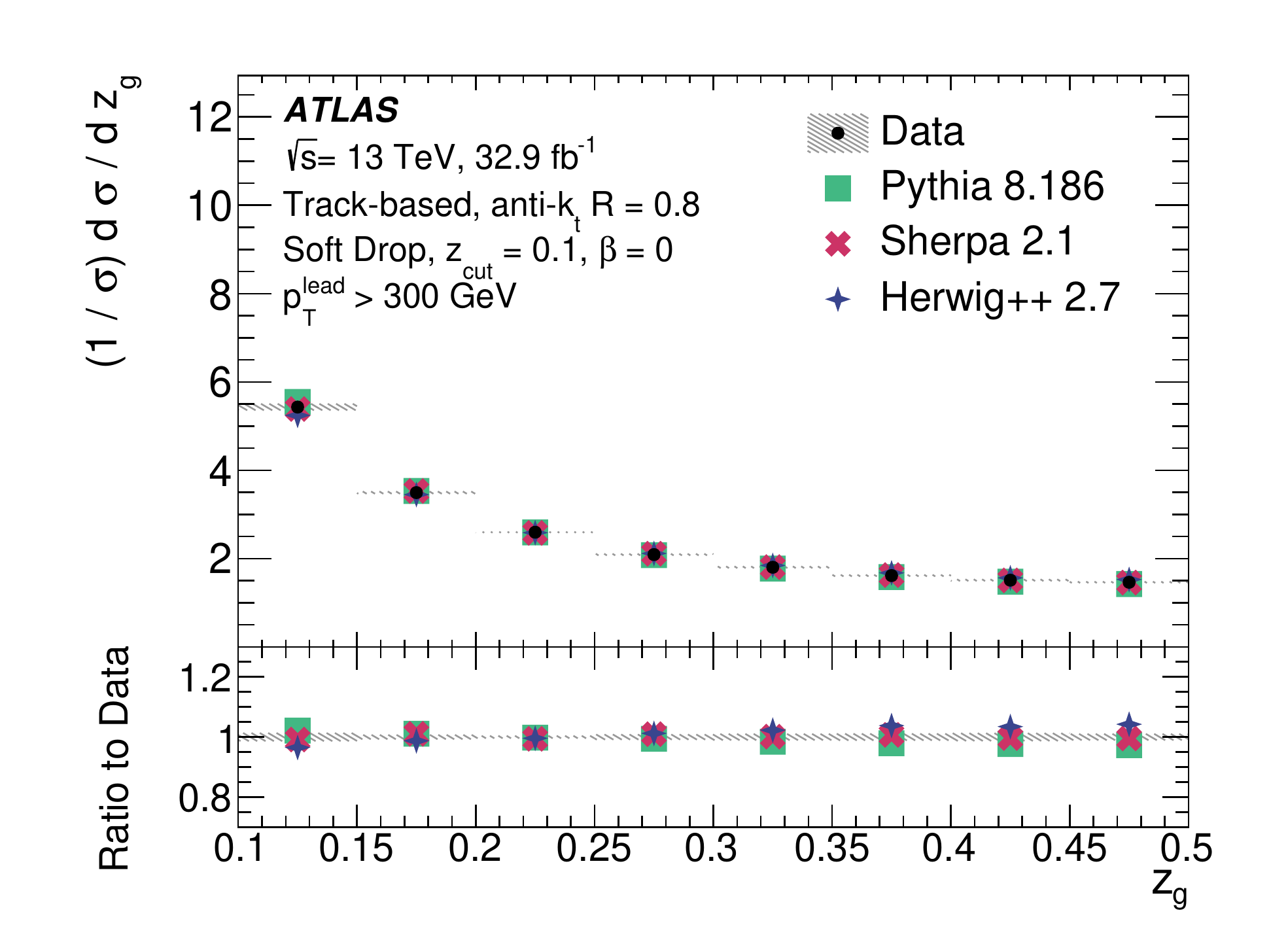}}
\subfigure[$\beta = 1$, calorimeter-based]{\includegraphics[width=0.45\textwidth,page=1]{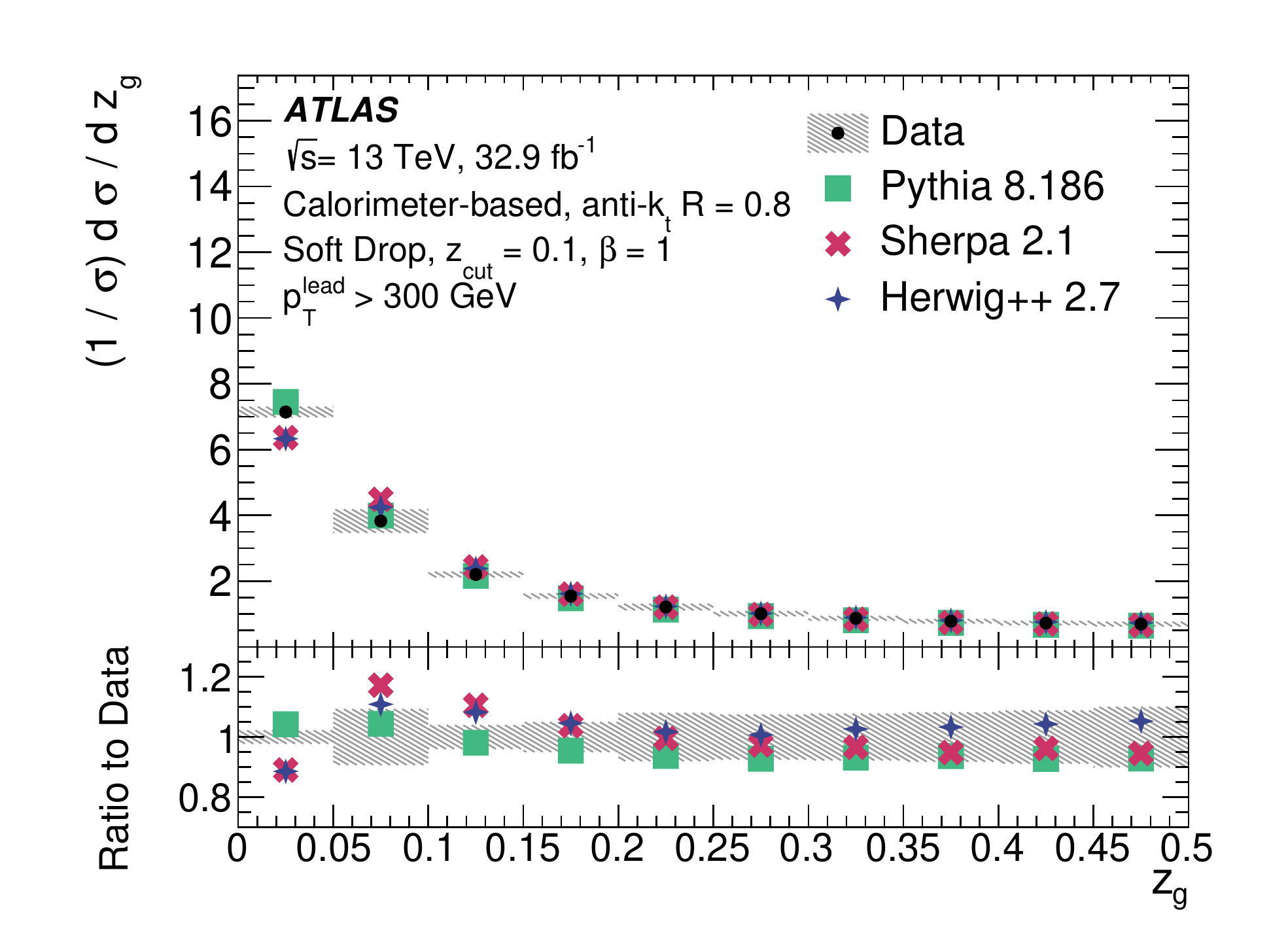}}
\subfigure[$\beta = 1$, track-based]{\includegraphics[width=0.45\textwidth,page=1]{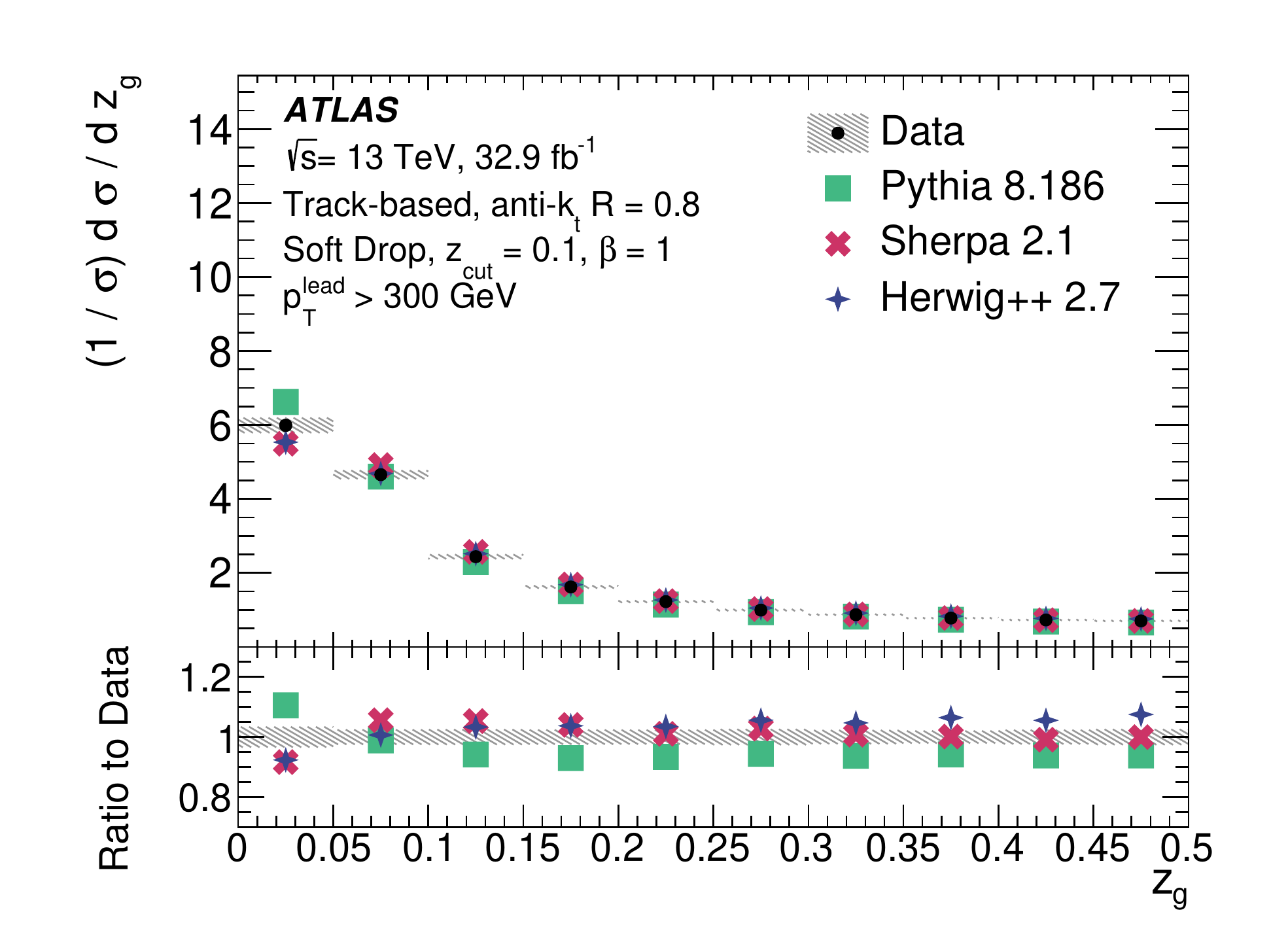}}
\subfigure[$\beta = 2$, calorimeter-based]{\includegraphics[width=0.45\textwidth,page=1]{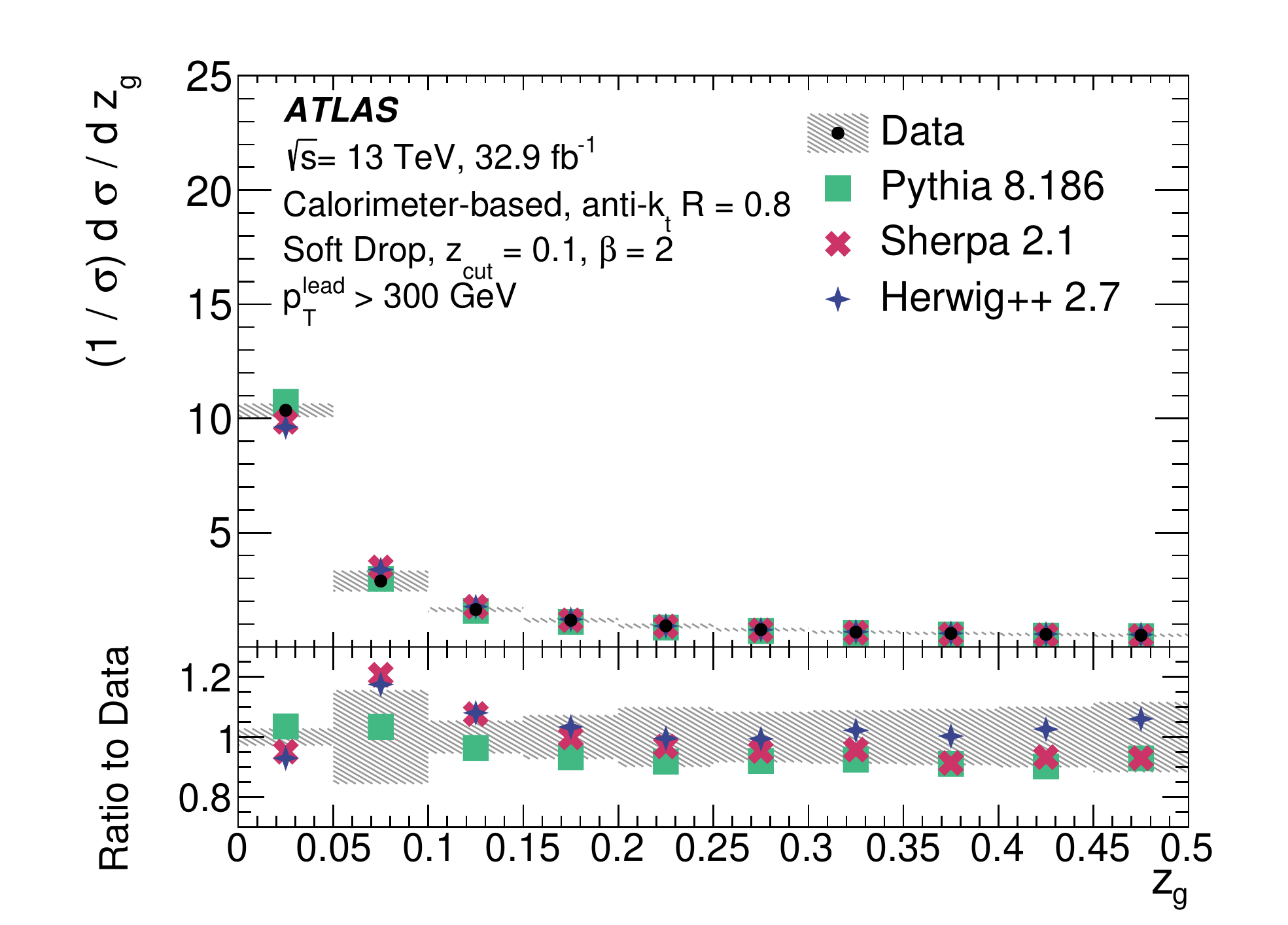}}
\subfigure[$\beta = 2$, track-based]{\includegraphics[width=0.45\textwidth,page=1]{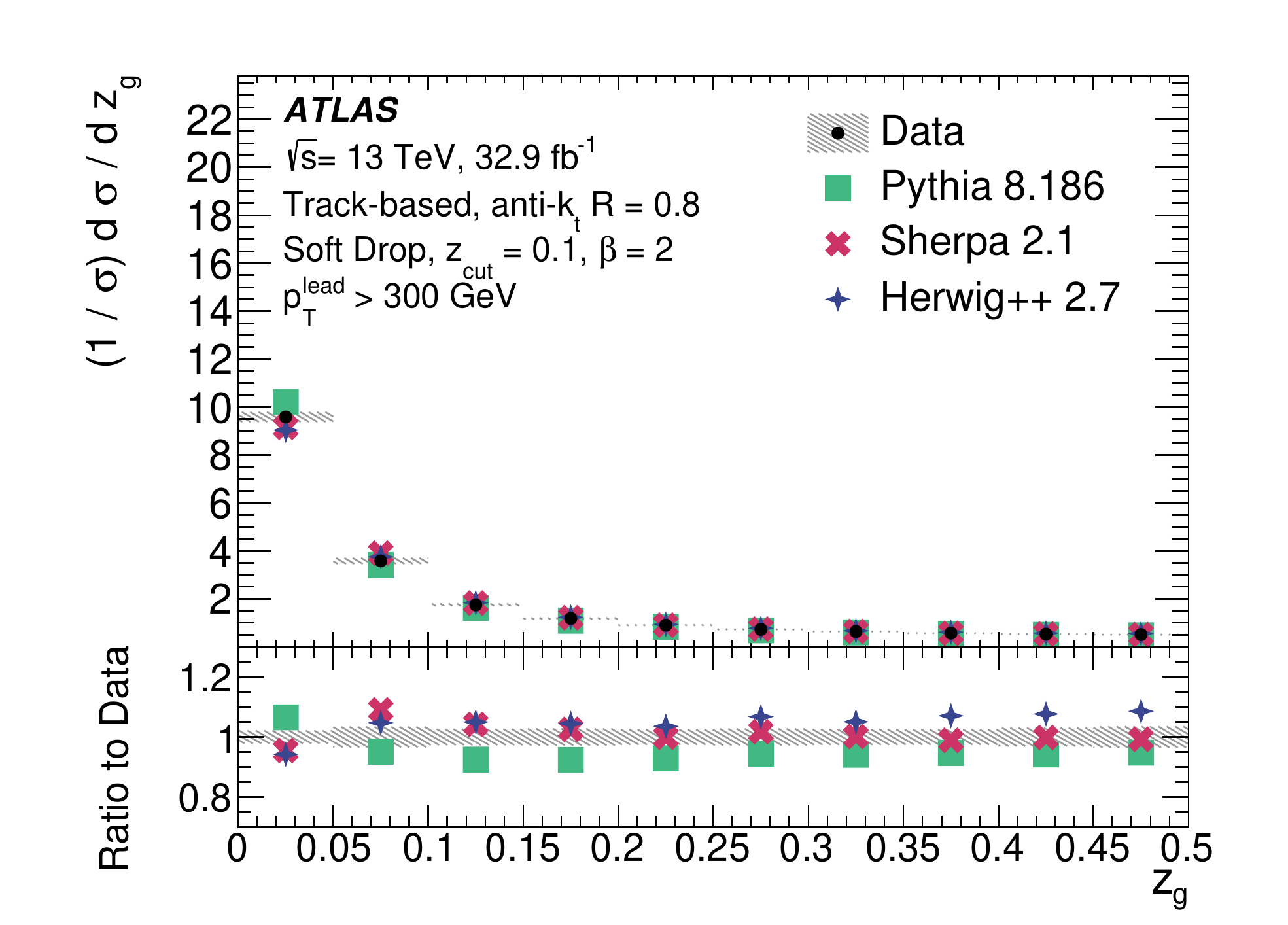}}
\end{center}
\caption{Comparison of the unfolded $z_g$ distribution with MC predictions.  The uncertainty bands include all sources: data and MC statistical uncertainties, nonclosure, modeling, and cluster or tracking uncertainties where relevant.}
\label{results:unfold:zg:mc}
\end{figure}
 
\begin{figure}[hp!]
\begin{center}
\subfigure[$\beta = 0$, calorimeter-based]{\includegraphics[width=0.48\textwidth,page=1]{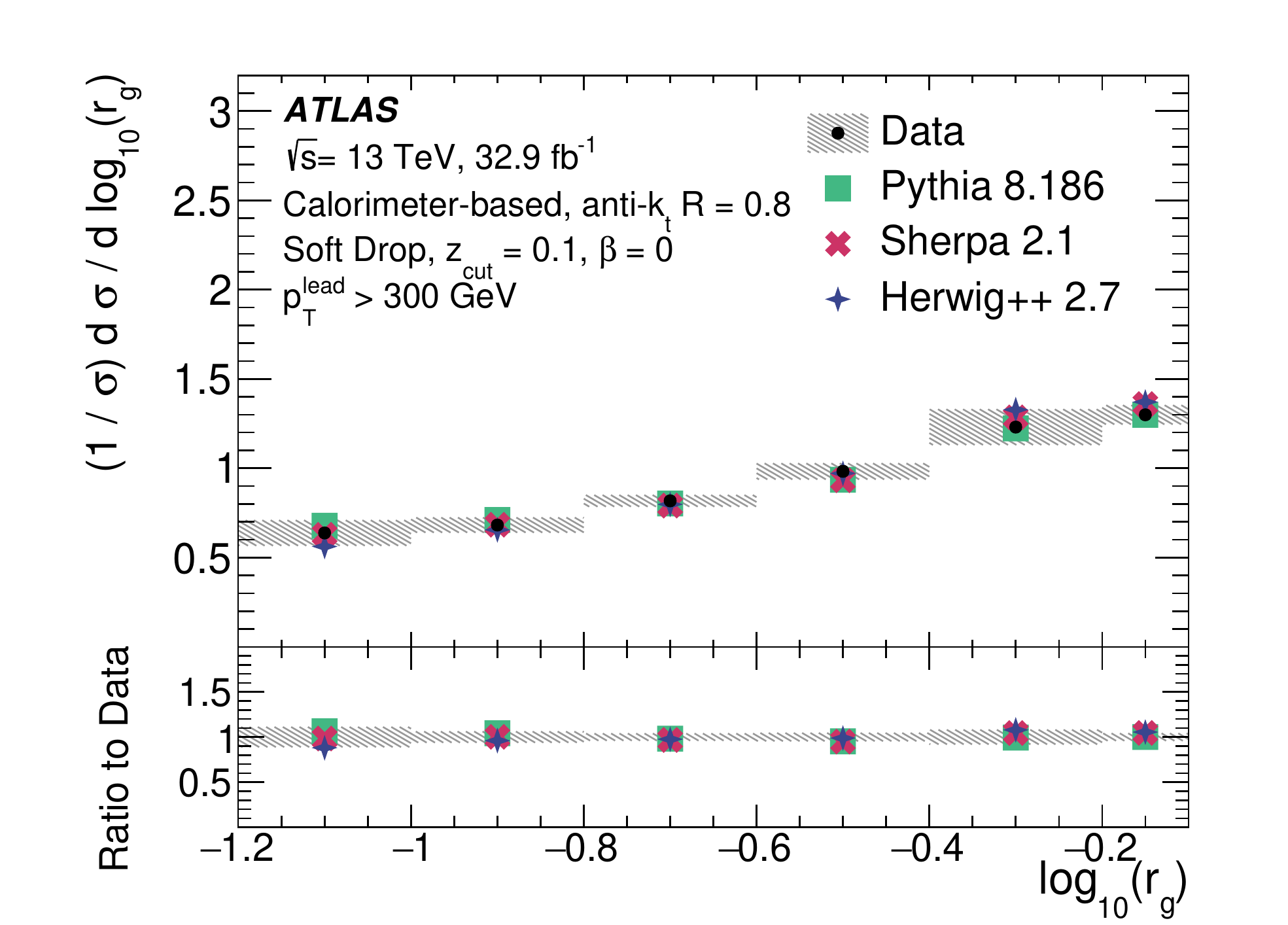}}
\subfigure[$\beta = 0$, track-based]{\includegraphics[width=0.48\textwidth,page=1]{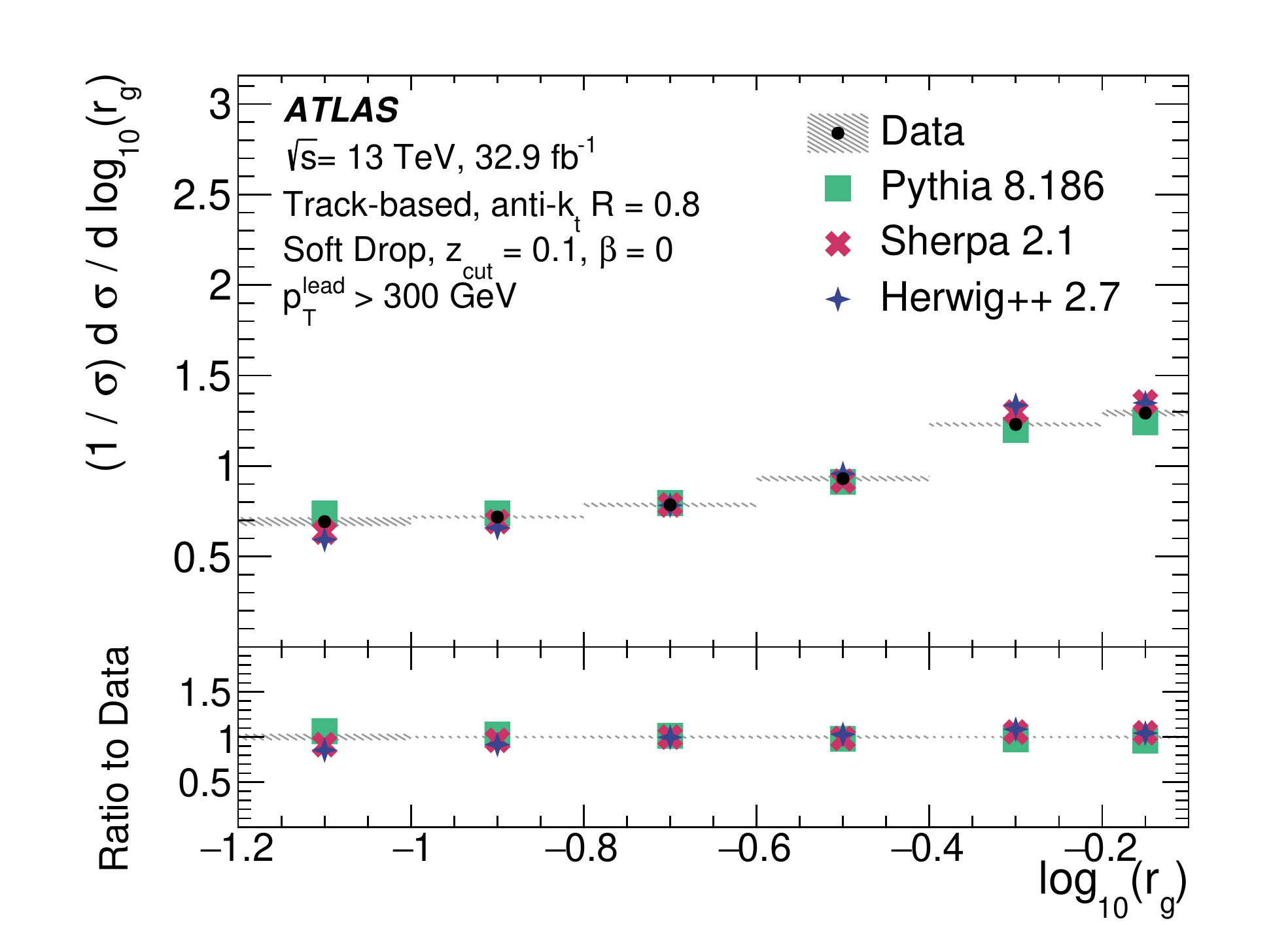}}
\subfigure[$\beta = 1$, calorimeter-based]{\includegraphics[width=0.48\textwidth,page=1]{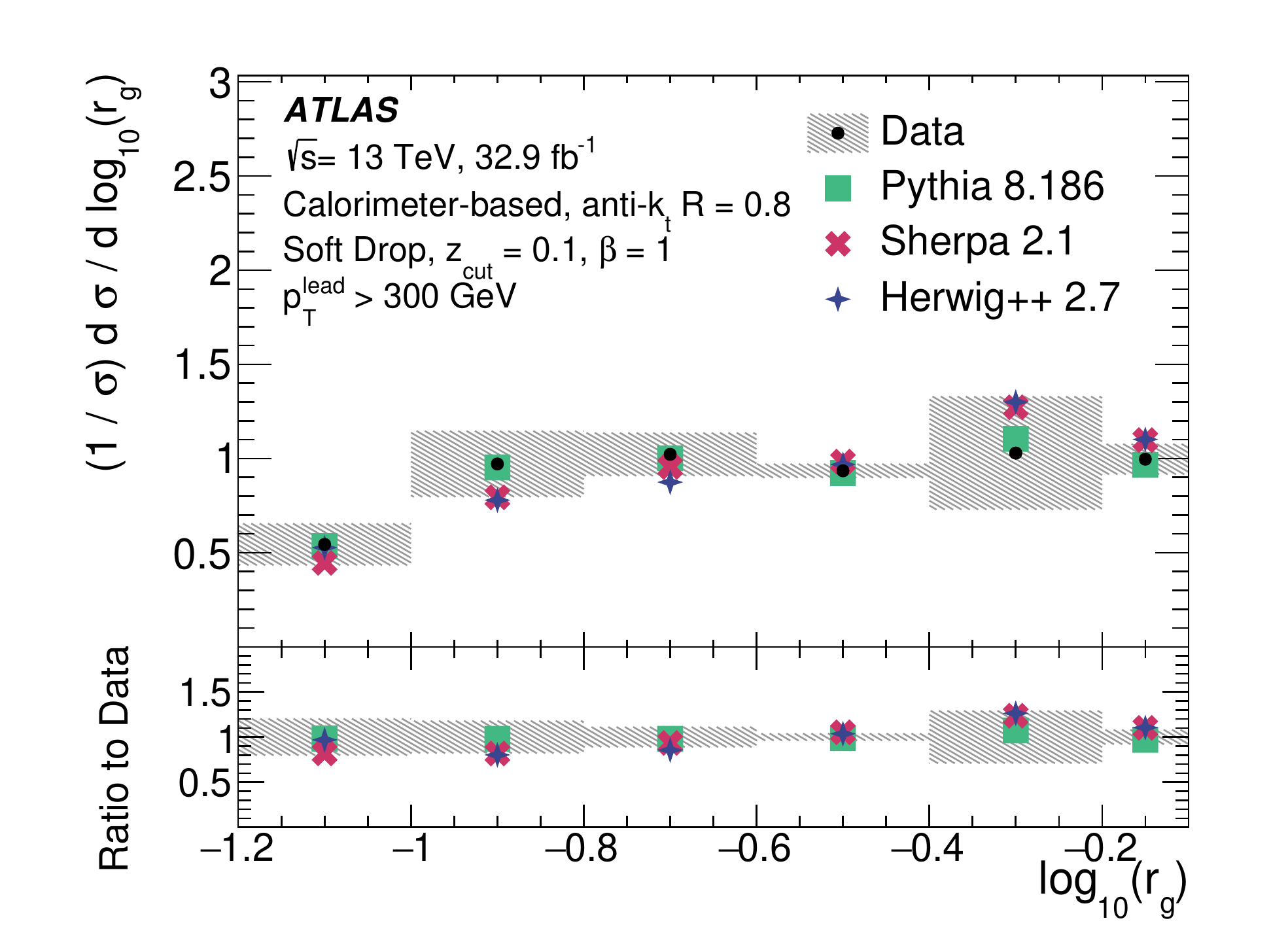}}
\subfigure[$\beta = 1$, track-based]{\includegraphics[width=0.48\textwidth,page=1]{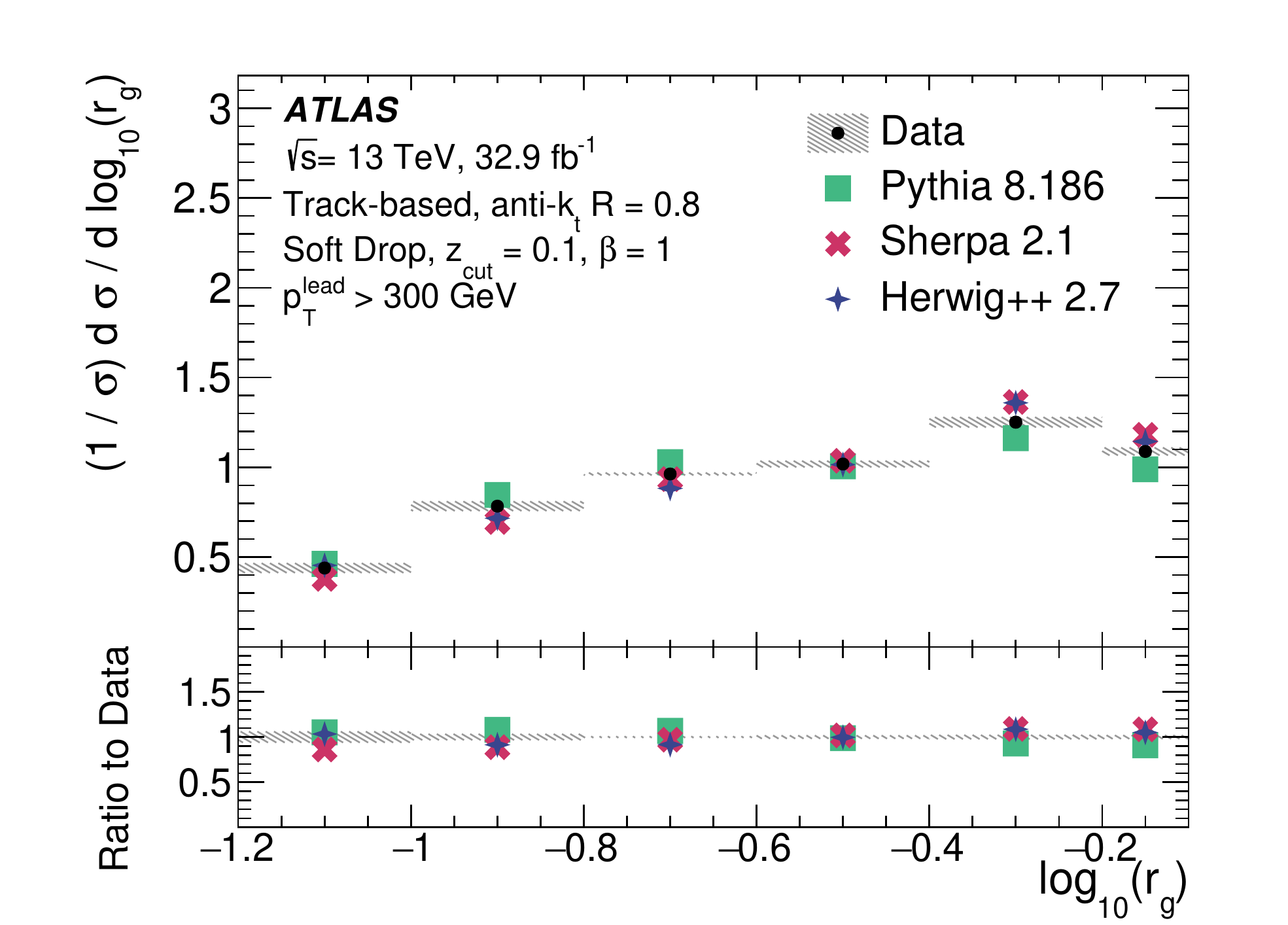}}
\subfigure[$\beta = 2$, calorimeter-based]{\includegraphics[width=0.48\textwidth,page=1]{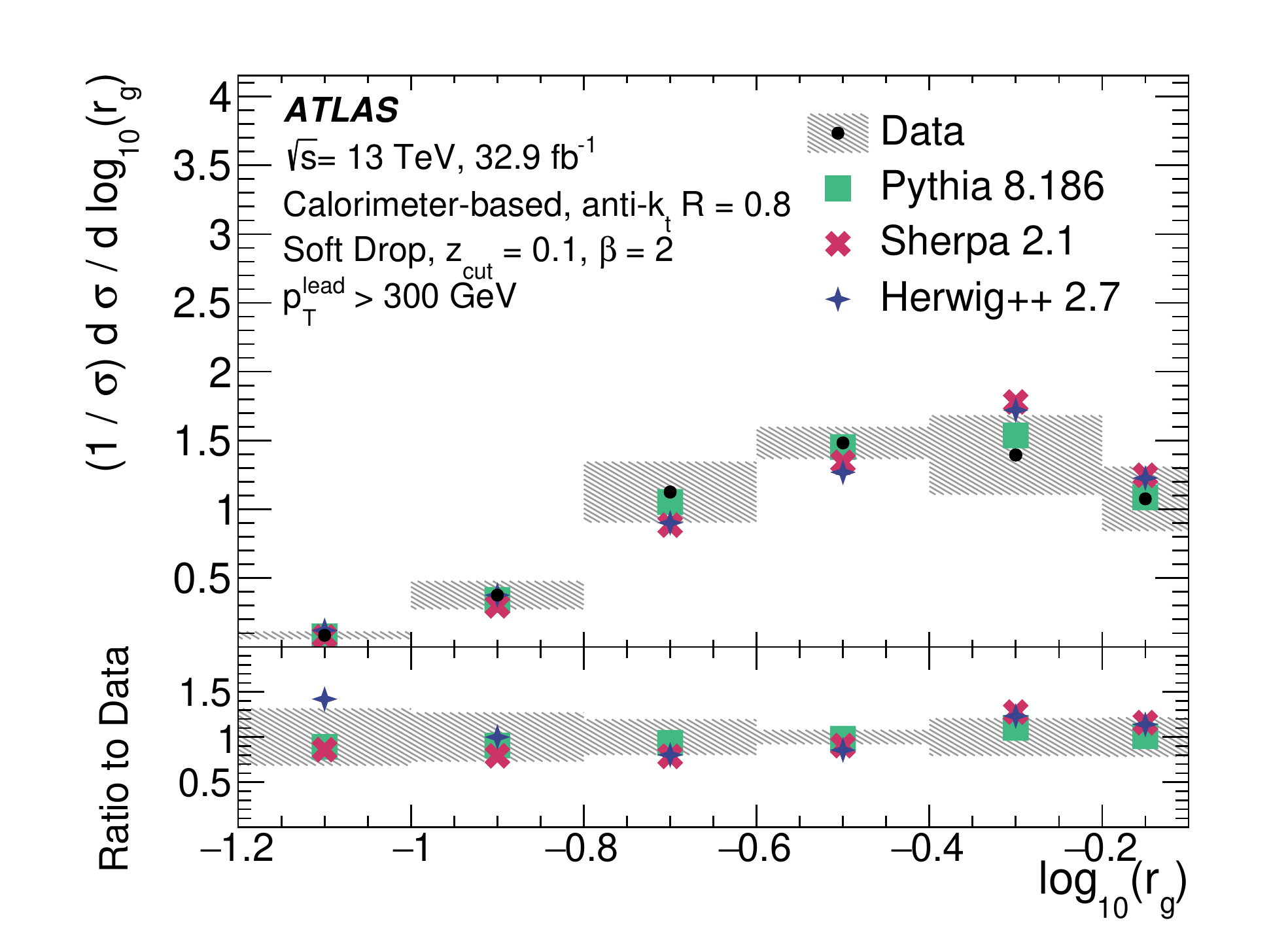}}
\subfigure[$\beta = 2$, track-based]{\includegraphics[width=0.48\textwidth,page=1]{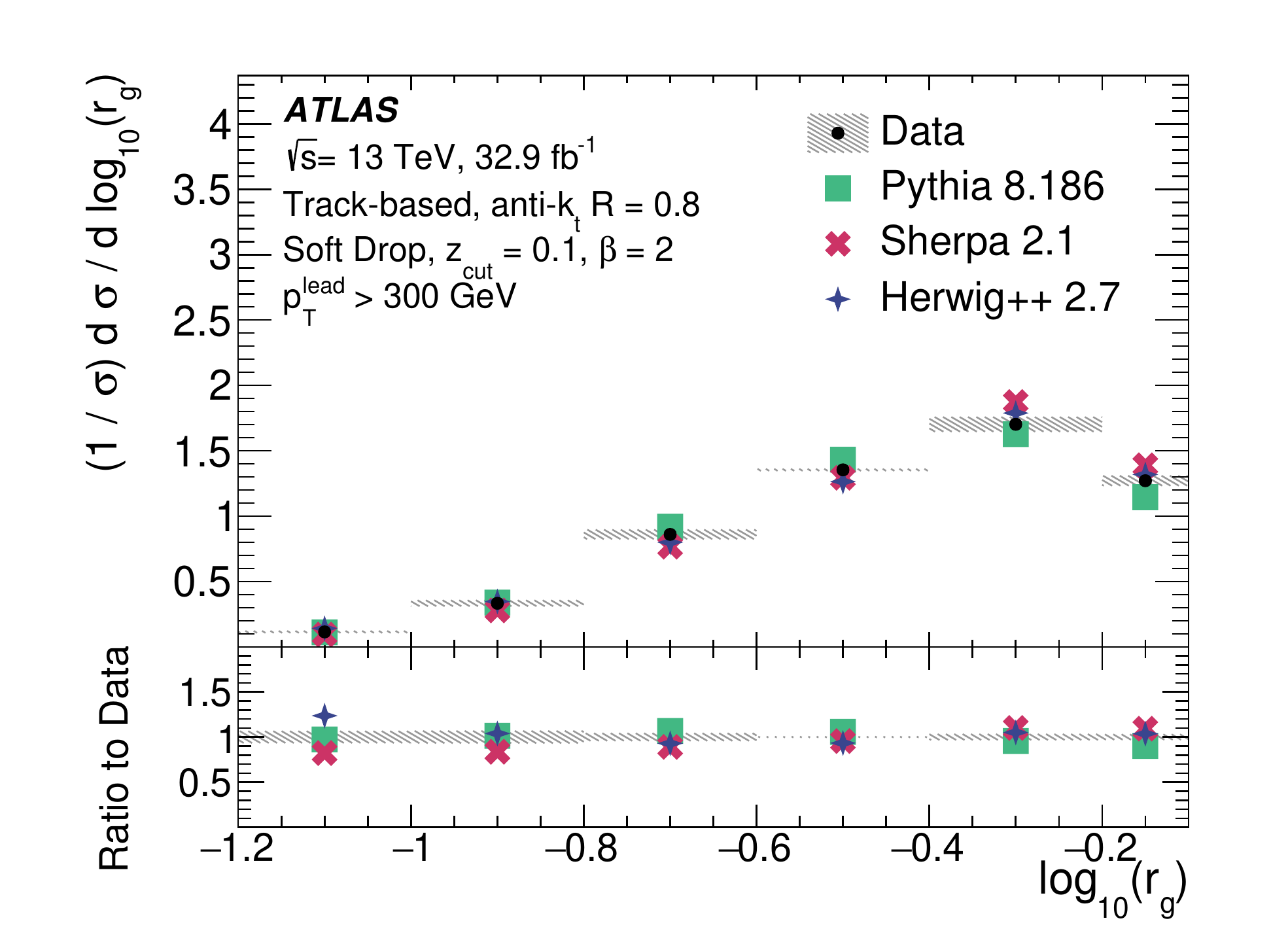}}
\end{center}
\caption{Comparison of the unfolded $r_g$ distribution with MC predictions.  The uncertainty bands include all sources: data and MC statistical uncertainties, nonclosure, modeling, and cluster or tracking uncertainties where relevant.}
\label{results:unfold:rg:mc}
\end{figure}
 
\clearpage
 
\subsection{Comparison with analytical predictions}
\label{sec:theoryComparison}
 
Currently, it is only possible to perform analytical predictions when including both charged and neutral particles, and therefore results in this section are only compared with the calorimeter-based results.  Subleading logarithms have been computed for $\rho$ and $r_g$, as described below.
Several calculations have been performed to predict the $\rho$ distribution, and these predictions are compared with the unfolded data.
In addition, only $\rho$ and $r_g$ are studied, since no predictions exist for $z_g$ beyond leading-logarithmic
accuracy.
In particular, these include the NLO+NLL prediction from Refs.~\cite{Marzani:2017mva,Marzani:2017kqd}, the LO+NNLL prediction from Refs.~\cite{Frye:2016aiz,Frye:2016okc}, and the NNLL prediction from Refs.~\cite{Kang:2018vgn,Kang:2018jwa}.
 
The LO+NNLL and NNLL calculations are based on soft collinear effective theory (SCET)~\cite{Bauer:2000yr,Bauer:2001yt}.  The former is matched to leading order using MadGraph5\_aMC@NLO~\cite{Alwall:2014hca} with the MSTW2008LO PDF.  The latter uses the CT14nlo~\cite{Dulat:2015mca} PDF set and includes finite $z_\text{cut}$ resummation as well as nonperturbative corrections based on an analytic shape function with one free parameter that is chosen based on comparisons with \PYTHIA8.  While strictly for inclusive jets, the NNLL calculation is also applicable here because at high jet $p_\text{T}$, the difference between inclusive jets and dijets is negligible.  The NLO+NLL calculation is matched to fixed order using NLOJet++~\cite{Catani:1996vz,Nagy:2003tz} with the CT14nlo PDF and includes finite $z_\text{cut}$ resummation as well as nonperturbative corrections from the envelope of parton shower MC predictions from \HERWIG 6.521~\cite{Corcella:2002jc} AUET2~\cite{ATL-PHYS-PUB-2011-008}, \PYTHIA 6.428~\cite{Sjostrand:2006za} Perugia 2011~\cite{Skands:2010ak}, \PYTHIA 6.428 Z2~\cite{Field:2010bc}, \PYTHIA 8.223~\cite{Sjostrand:2006za,Sjostrand:2007gs,Sjostrand:2014zea} 4C~\cite{Corke:2010yf}, and \PYTHIA 8.223 Monash 13~\cite{Skands:2014pea}.

These predictions are compared with the unfolded data in Figure~\ref{results:unfold:rho:theory}. Because the LO+NNLL and NLO+NLL calculations for $\rho$ are only available for $\pt>600$~\GeV, the unfolded data are shown for both a low-\pt jet selection ($\pt > 300$~\GeV) and a high-\pt jet selection (($\pt > 600$~\GeV).  The calculations are able to model the data in the resummation region (approximately  $-3\lesssim\rho\lesssim -1$) at the level of a 10\% difference.  The NLO+NLL calculation also provides an accurate model of the data at the high values of $\rho$, while the LO+NNLL and NNLL calculations do not model this region as accurately. This is the region where the fixed-order effects are dominant, and so this behavior is expected.
 
At lower values of relative mass, the nonperturbative corrections are needed to describe the data. This can be seen particularly from the low-\pt results, which show the NNLL prediction with and without the inclusion of nonperturbative effects. As expected, the inclusion of these effects brings the prediction much closer to the unfolded data distribution, although the level of agreement is still not as good as in the resummation region.
The region where nonperturbative corrections are relevant shifts to higher relative mass with increased values of $\beta$, since more soft radiation is included within the jet.
In general, similar levels of agreement are seen in the low-\pt and high-\pt cases, although it is noted that the nonperturbative region shifts to slightly lower relative mass in the high-\pt case.
 
\begin{figure}[ht!]
\begin{center}
\subfigure[$\beta = 0$, low \pt]{\includegraphics[width=0.48\textwidth,page=1]{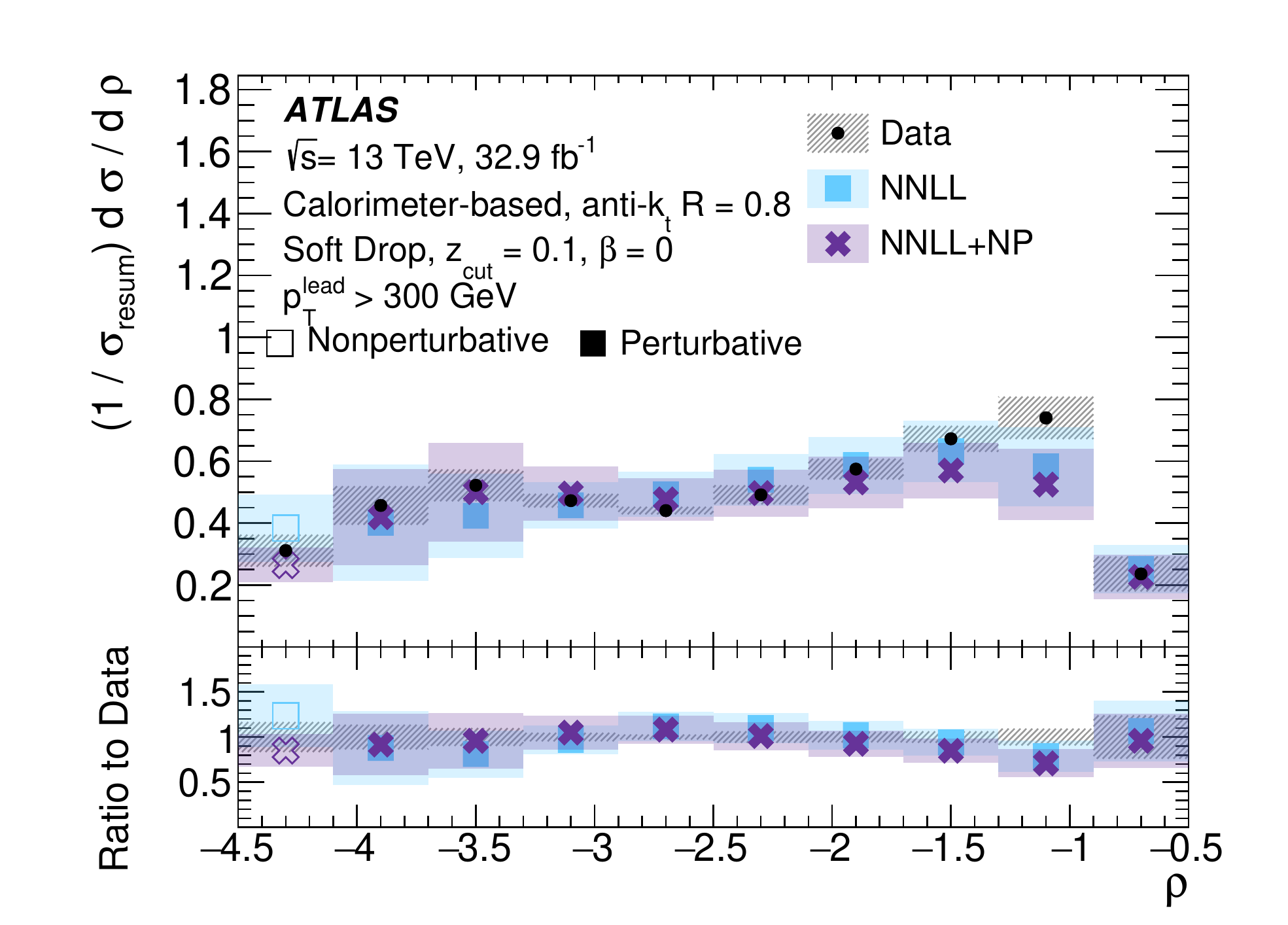}}
\subfigure[$\beta = 0$, high \pt]{\includegraphics[width=0.48\textwidth,page=1]{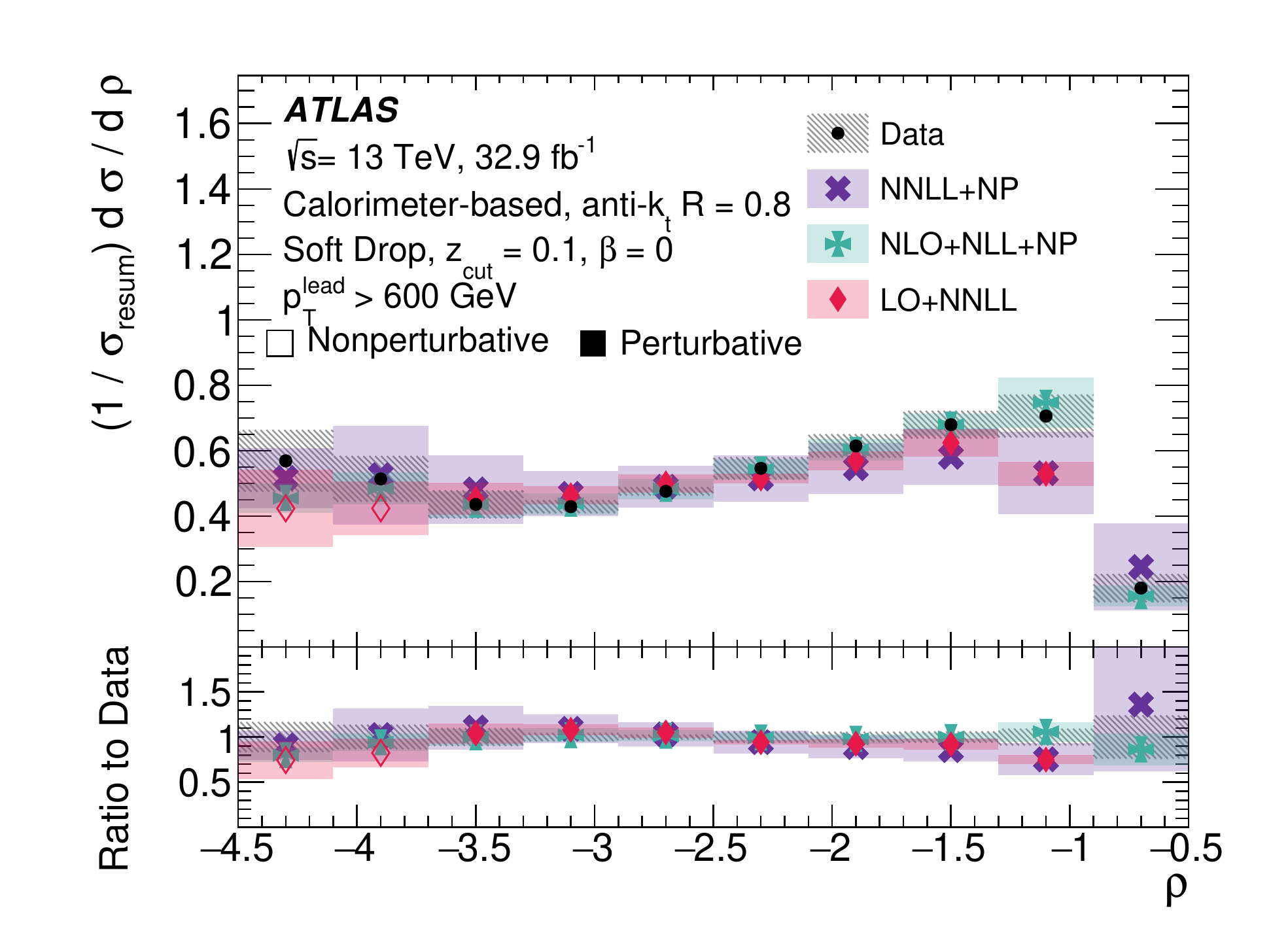}}
\subfigure[$\beta = 1$, low \pt]{\includegraphics[width=0.48\textwidth,page=1]{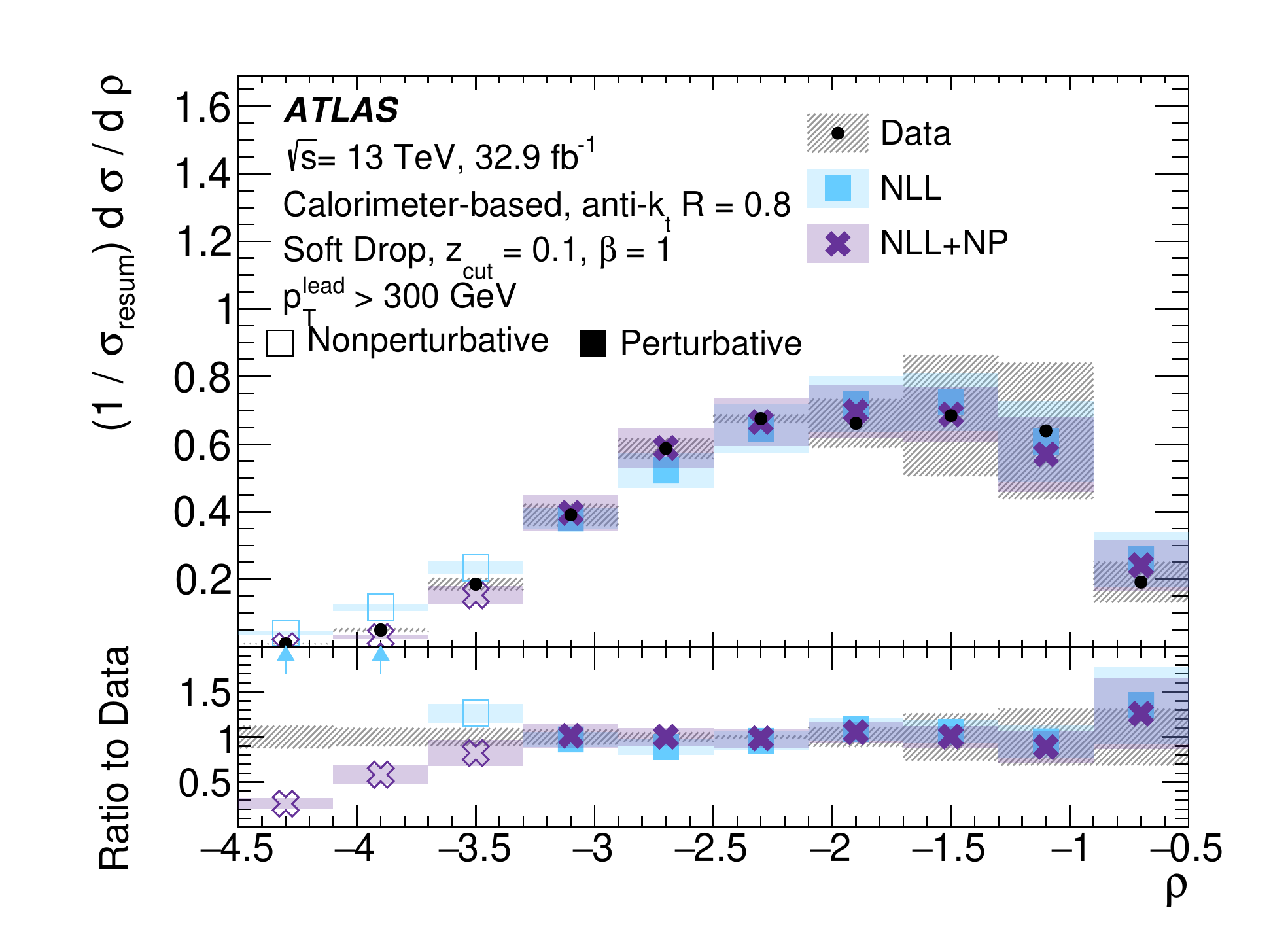}}
\subfigure[$\beta = 1$, high \pt]{\includegraphics[width=0.48\textwidth,page=1]{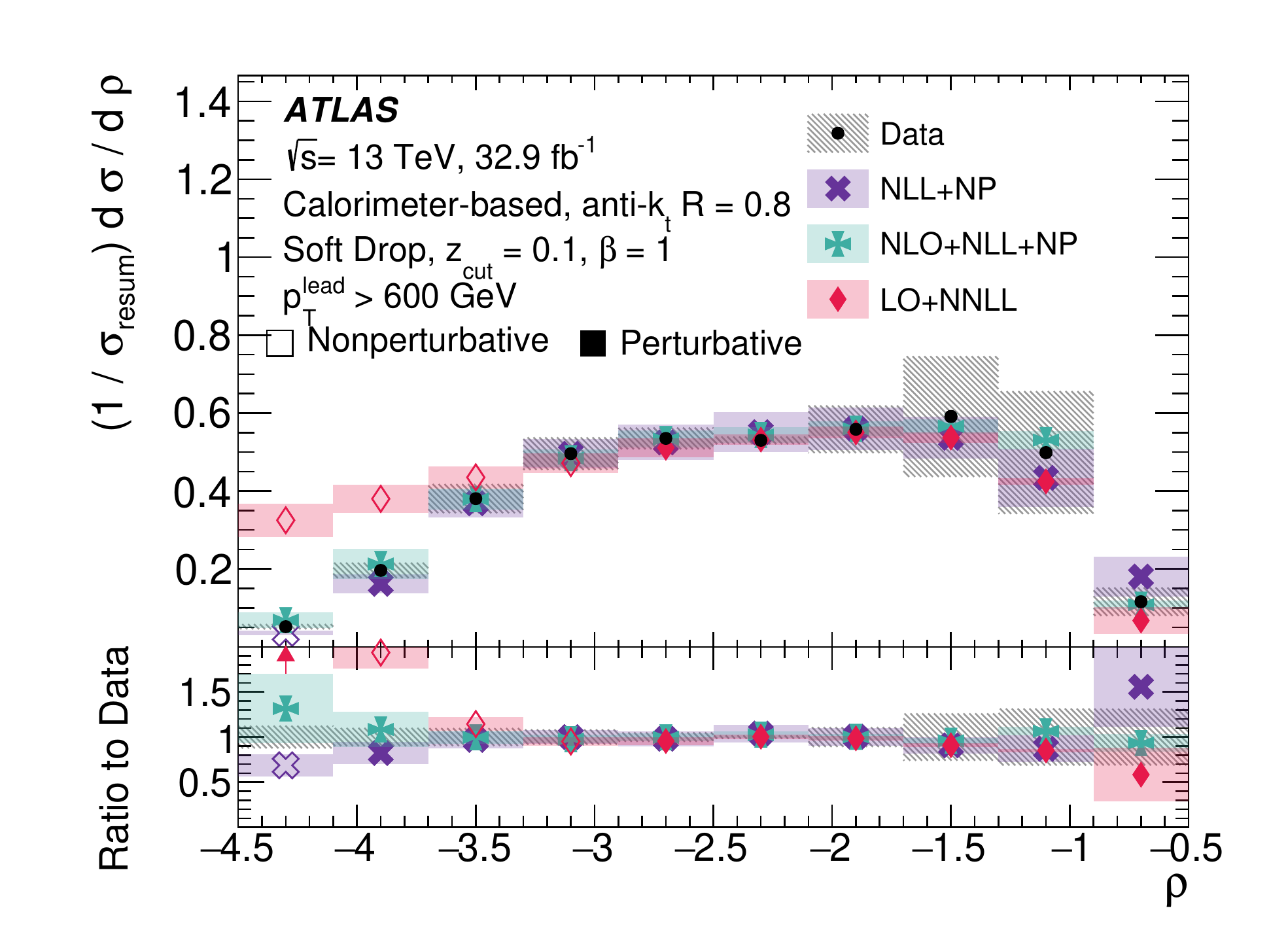}}
\subfigure[$\beta = 2$, low \pt]{\includegraphics[width=0.48\textwidth,page=1]{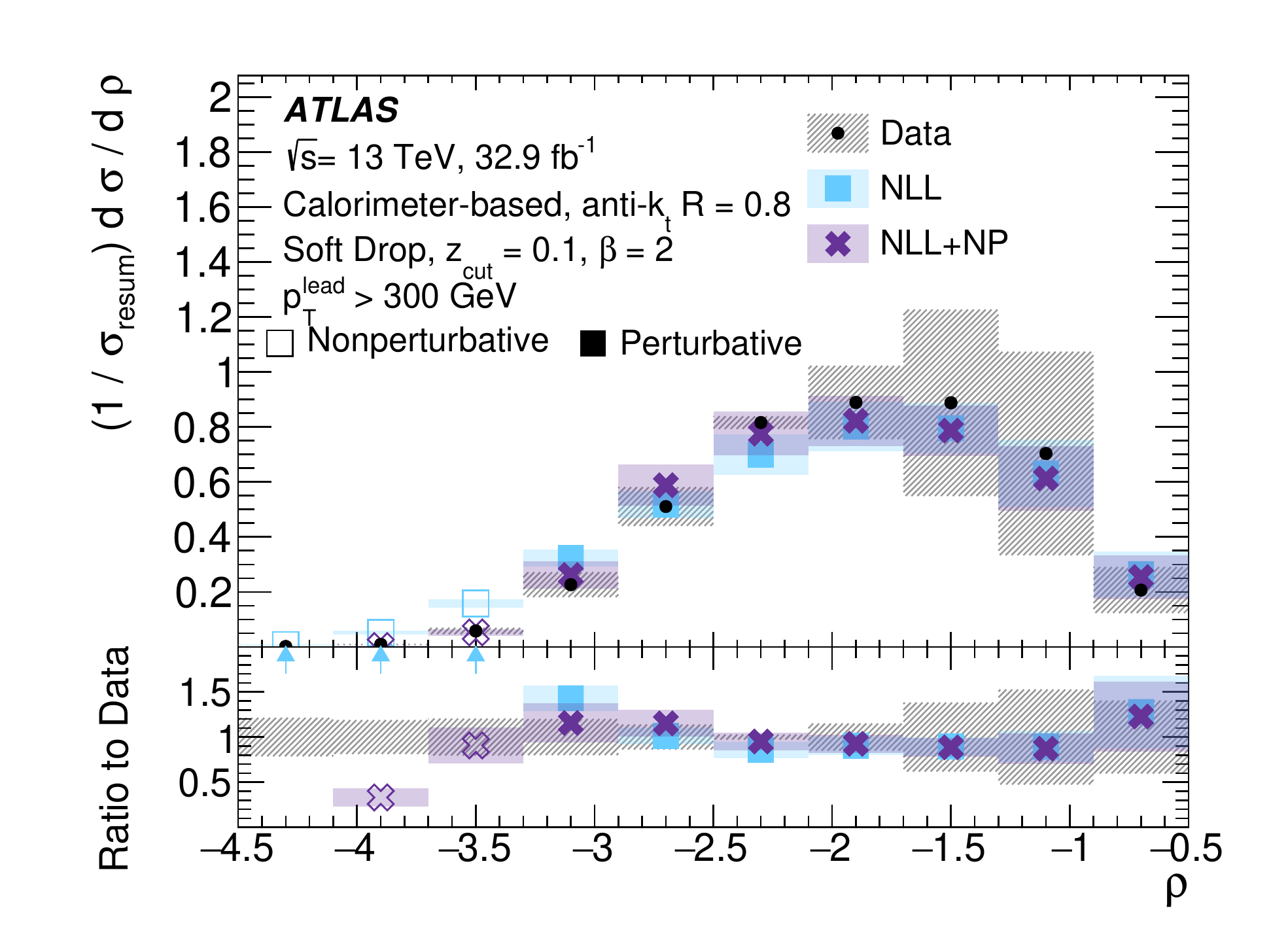}}
\subfigure[$\beta = 2$, high \pt]{\includegraphics[width=0.48\textwidth,page=1]{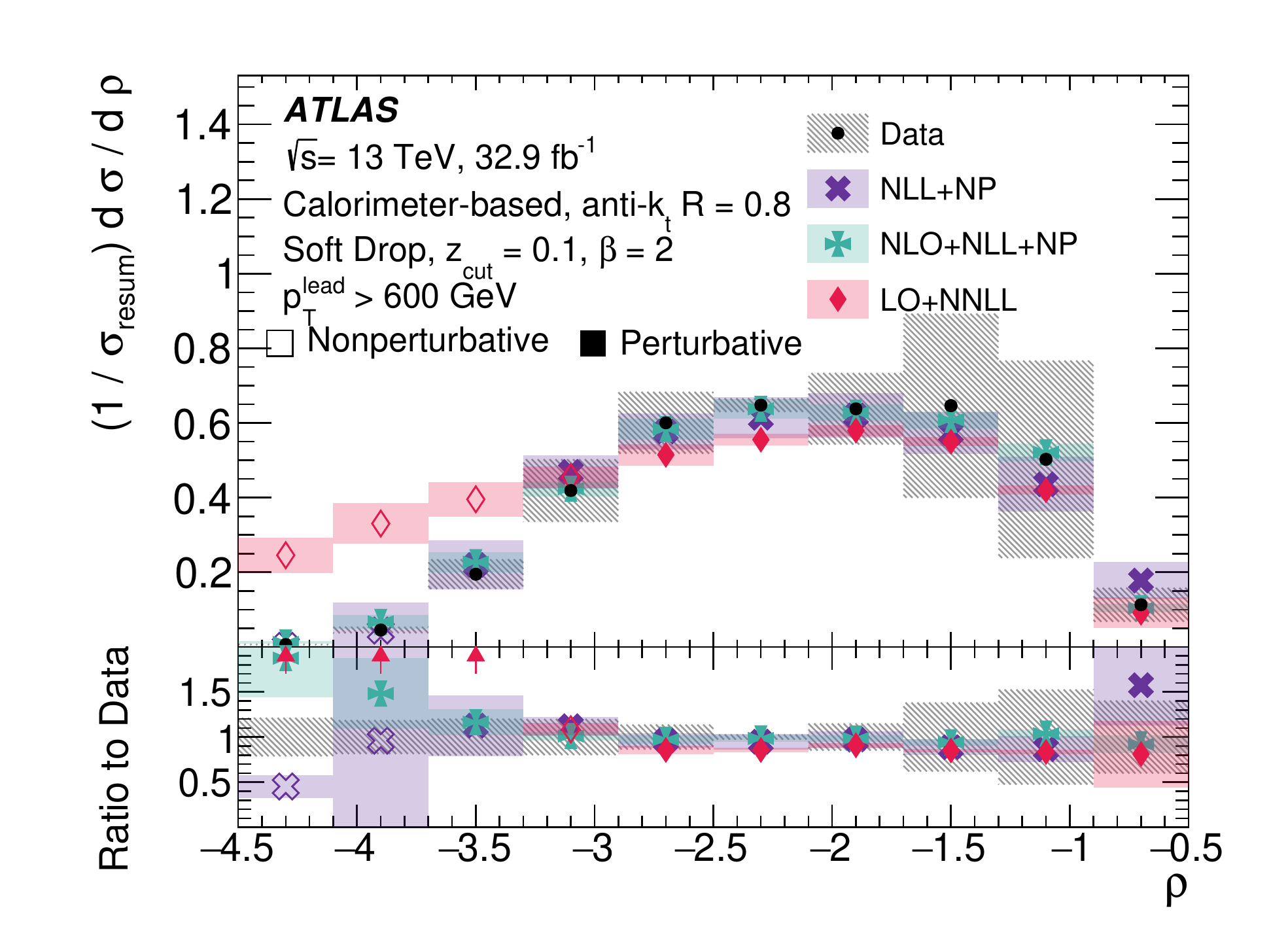}}
\end{center}
\caption{Comparison of the unfolded $\rho$ distribution with the theory predictions. For the (N)NLL, (N)NLL+NP, and LO+NNLL predictions, the open marker style indicates that
nonperturbative effects on the calculation are expected to be large.  `NP' indicates that nonperturbative corrections have been applied.  The experimental uncertainty bands include all sources: data and MC statistical uncertainties, nonclosure, modeling, and cluster or tracking uncertainties where relevant.  The theory error bands include perturbative scale variations as well as nonperturbative model variations (NLO+NLL only).}
\label{results:unfold:rho:theory}
\end{figure}
 
An NLL calculation of $r_g$ has been performed recently~\cite{Kang:2019prh}, and the results of this calculation are compared with the unfolded data distribution in Figure~\ref{results:unfold:rg:theory}.  Unlike the jet mass case, nonglobal logarithms are not absent ($\beta=0)$ or power suppressed ($\beta>0)$.
The calculation includes both the nonglobal and clustering logarithms to achieve full NLL accuracy.
In general, in the region where nonperturbative effects are expected to be small, the prediction agrees with the data within uncertainties, while in the regions
where nonperturbative effects are large, the prediction is systematically higher than the data.
 
\begin{figure}[ht!]
\begin{center}
\subfigure[$\beta = 0$, low \pt]{\includegraphics[width=0.48\textwidth,page=1]{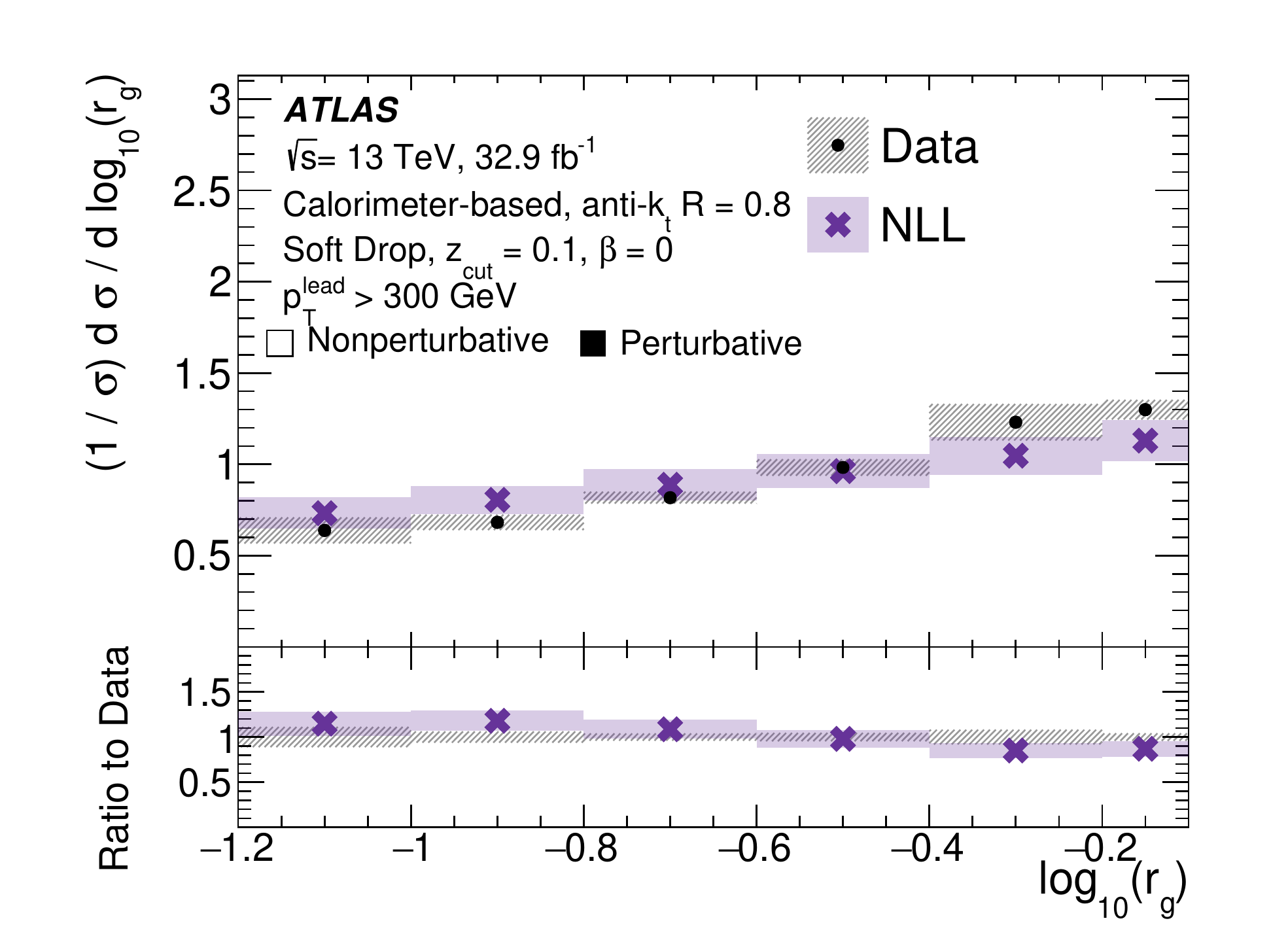}}
\subfigure[$\beta = 0$, high \pt]{\includegraphics[width=0.48\textwidth,page=1]{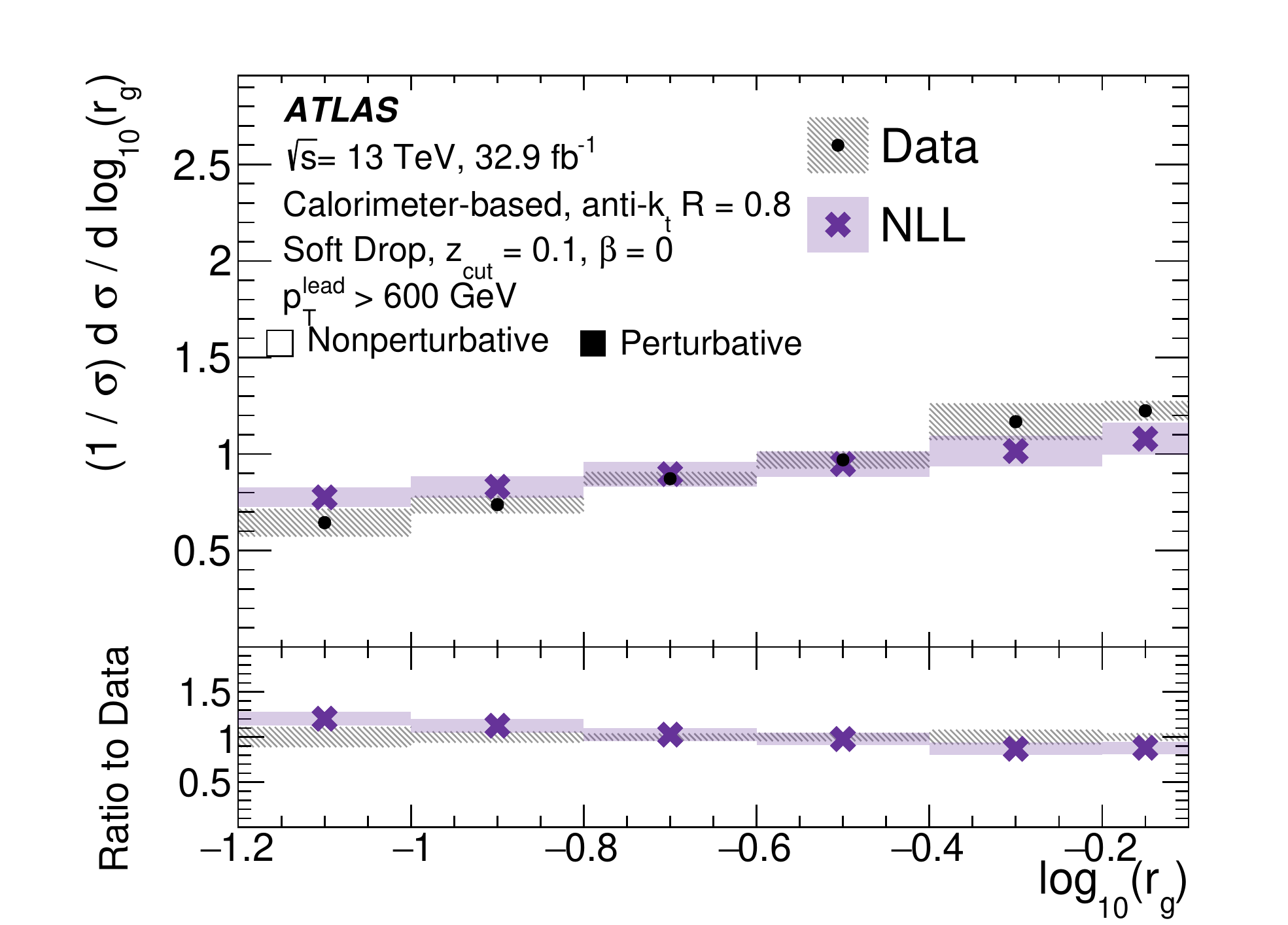}}
\subfigure[$\beta = 1$, low \pt]{\includegraphics[width=0.48\textwidth,page=1]{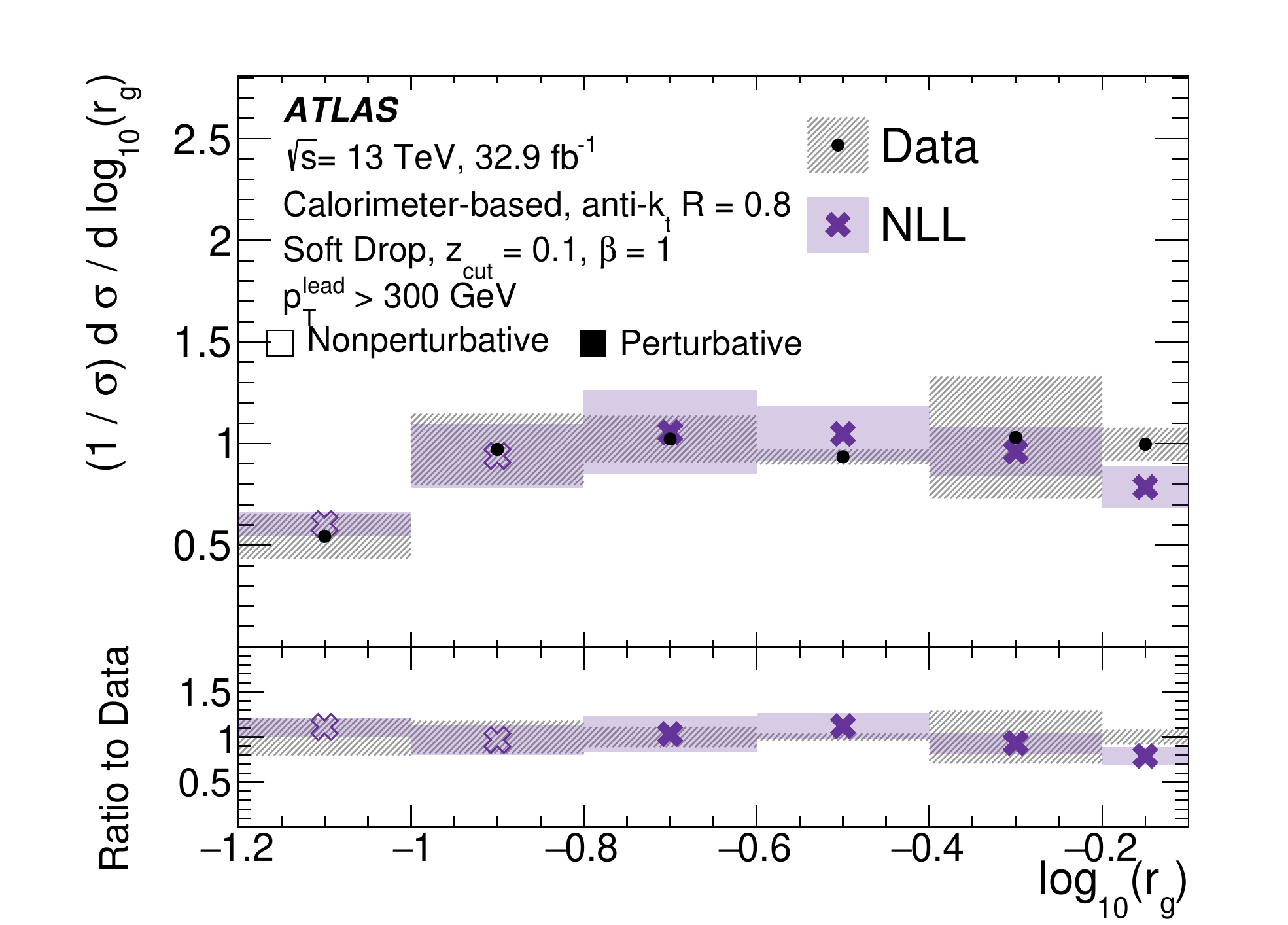}}
\subfigure[$\beta = 1$, high \pt]{\includegraphics[width=0.48\textwidth,page=1]{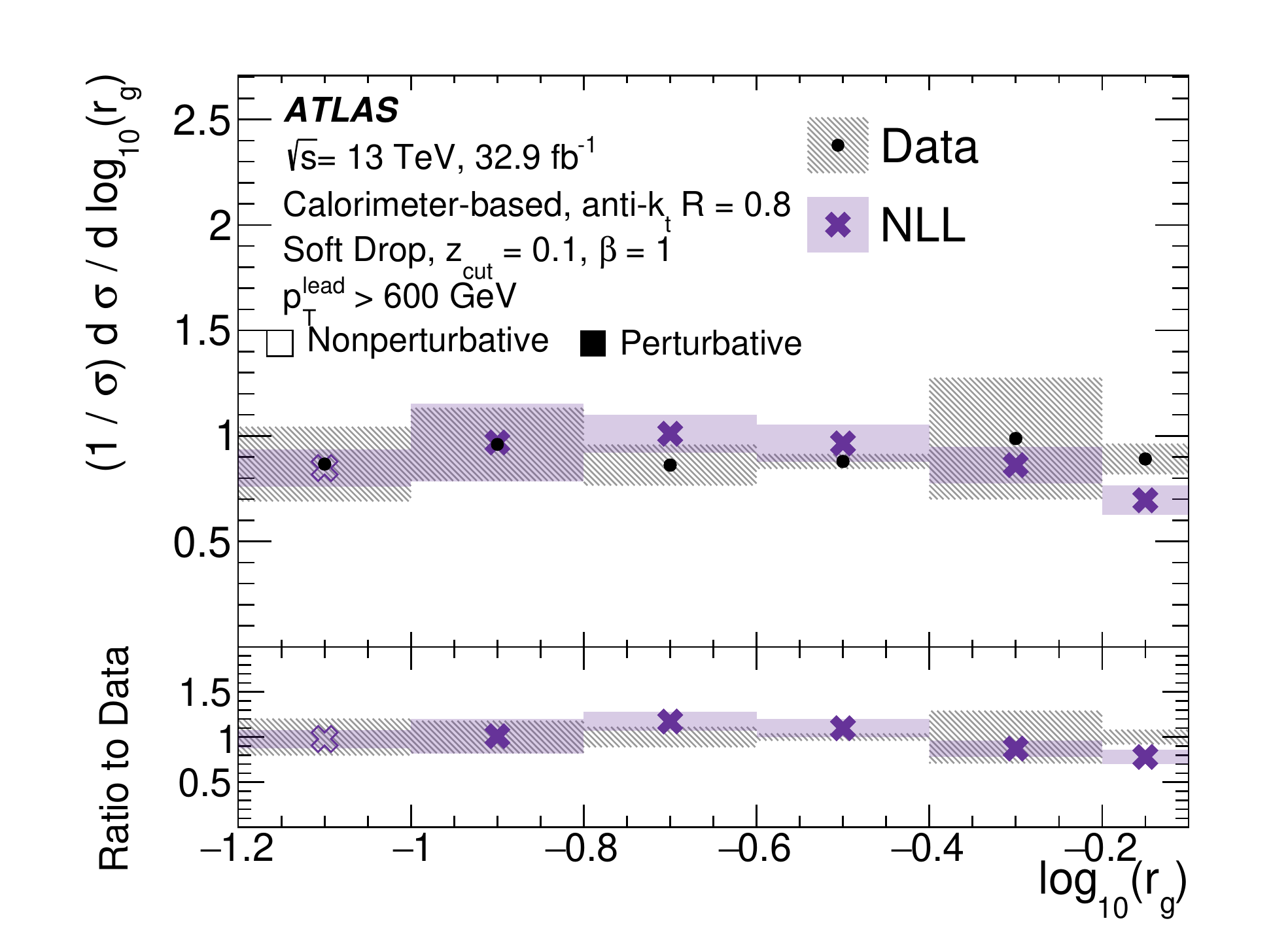}}
\subfigure[$\beta = 2$, low \pt]{\includegraphics[width=0.48\textwidth,page=1]{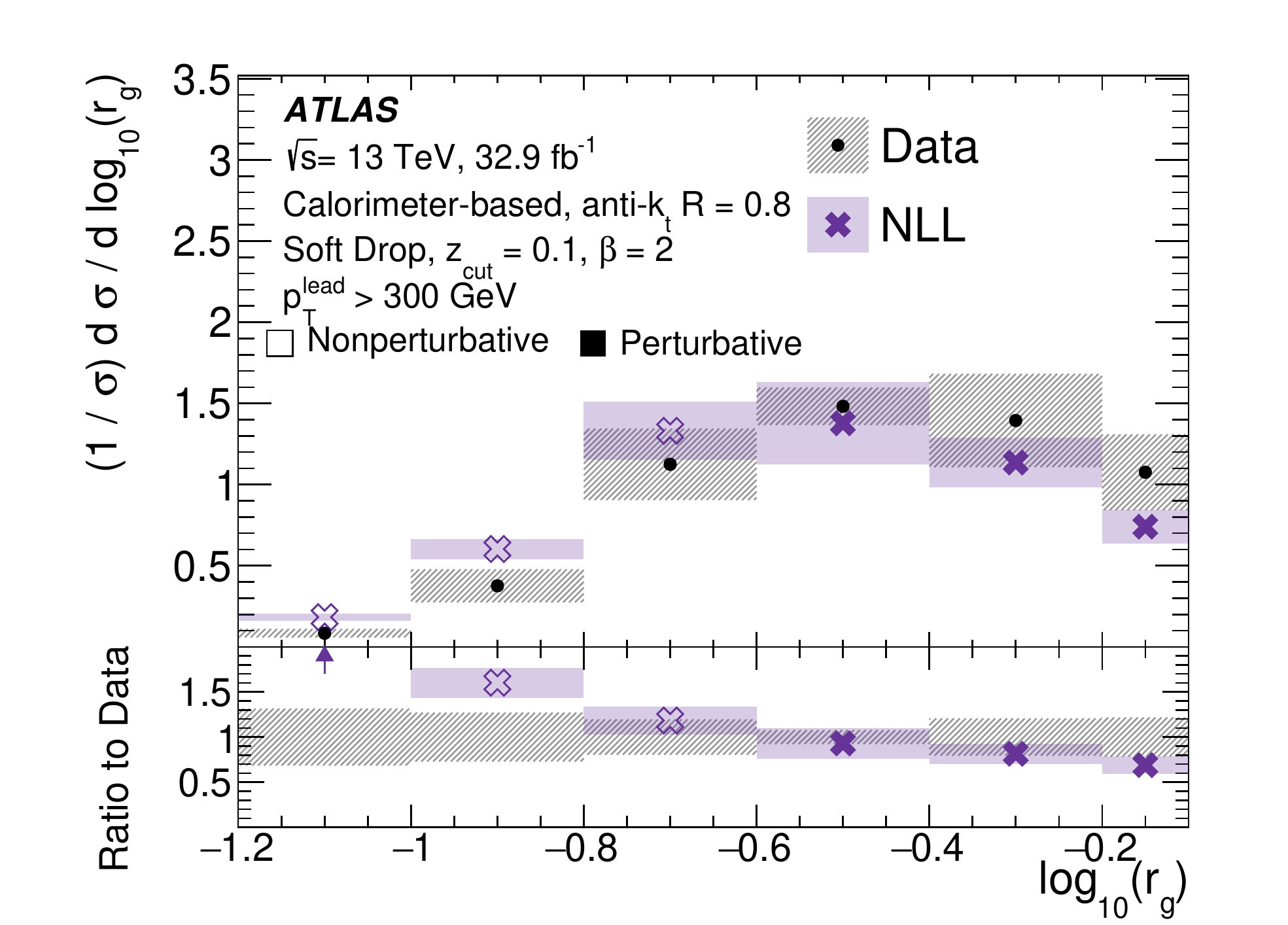}}
\subfigure[$\beta = 2$, high \pt]{\includegraphics[width=0.48\textwidth,page=1]{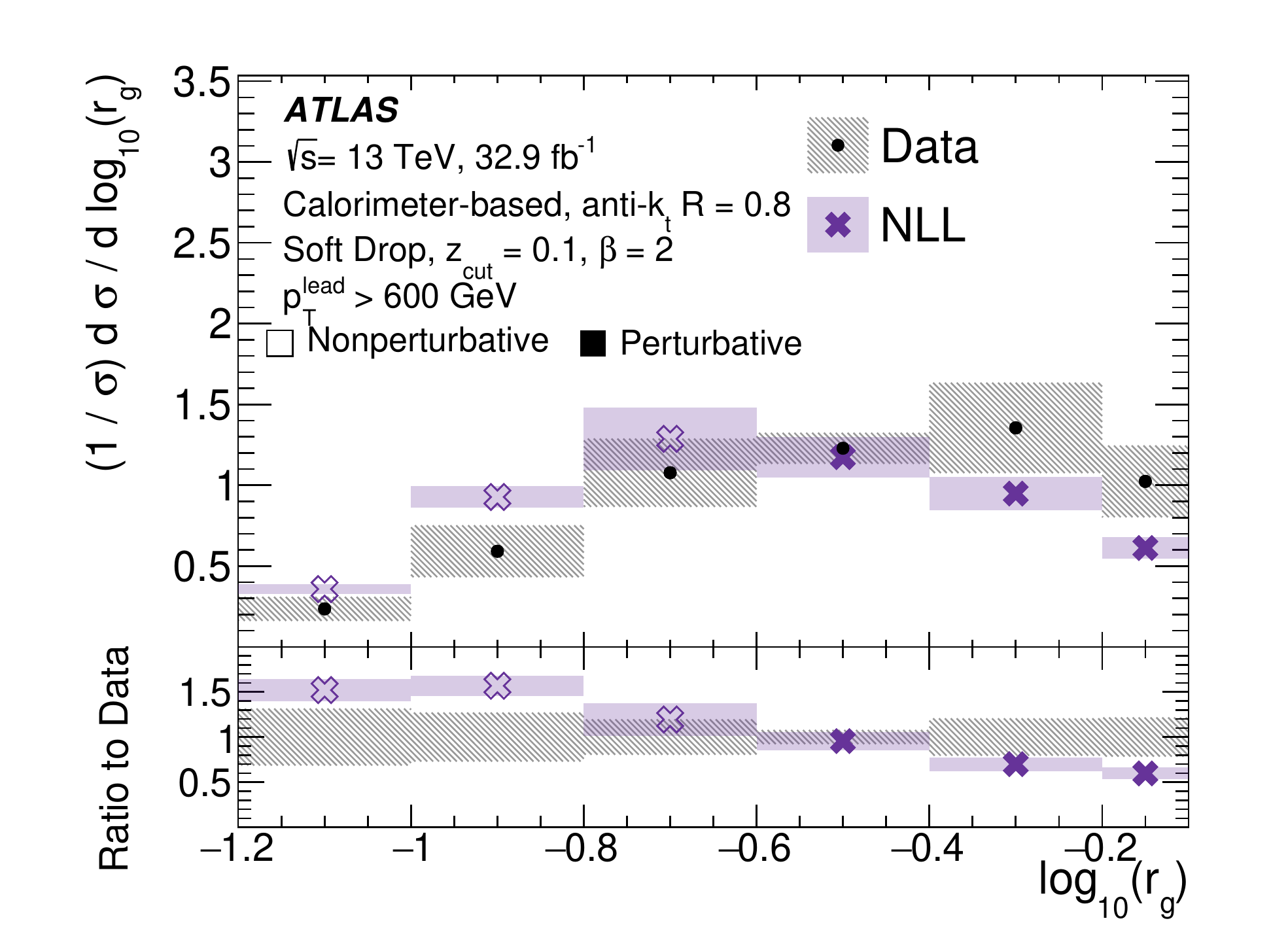}}
\end{center}
\caption{Comparison of the unfolded $r_g$ distribution with the theory predictions. For the NLL predictions, the open marker style indicates that
nonperturbative effects on the calculation are expected to be large.  The experimental uncertainty bands include all sources: data and MC statistical uncertainties, nonclosure, modeling, and cluster or tracking uncertainties where relevant. The theory error bands include perturbative scale variations.}
\label{results:unfold:rg:theory}
\end{figure}
 
\FloatBarrier
 
\subsection{Comparison of track-based and calorimeter-based measurements}
\label{sec:trackCalo}
 
On a jet-by-jet basis, the value of the all-particles and charged-particles jet substructure observables are largely uncorrelated.  However, due to isospin symmetry, the probability distributions for all-particles and charged-particles distributions are nearly identical.
This is studied by comparing the unfolded distributions for the cluster-based and track-based measurements,
which are shown in Figures~\ref{results:unfold:rho:trackCalo_inclPt}--\ref{results:unfold:rg:trackCalo_inclPt} for the region
which includes both jets in the dijet system.
The results generally agree in the perturbative regions at high values of $\rho$ and $r_g$, and there is disagreement in the low-relative-mass regions.
There is also some disagreement for low values of $z_g$ for $\beta > 0$.
 
These studies also enable a comparison of the sizes of the uncertainties for calorimeter-based and track-based observables.
For all of these observables, the uncertainties for the track-based observables are significantly smaller than those for the calorimeter-based observables,
particularly for higher values of $\beta$, where more soft radiation is included within the jet.
However, since no track-based calculations exist at the present time, calorimeter-based measurements are still useful for precision QCD studies.
 
\begin{figure}[ht!]
\begin{center}
\subfigure[$\rho$ distribution, $\beta = 0$]{\includegraphics[width=0.48\textwidth,page=1]{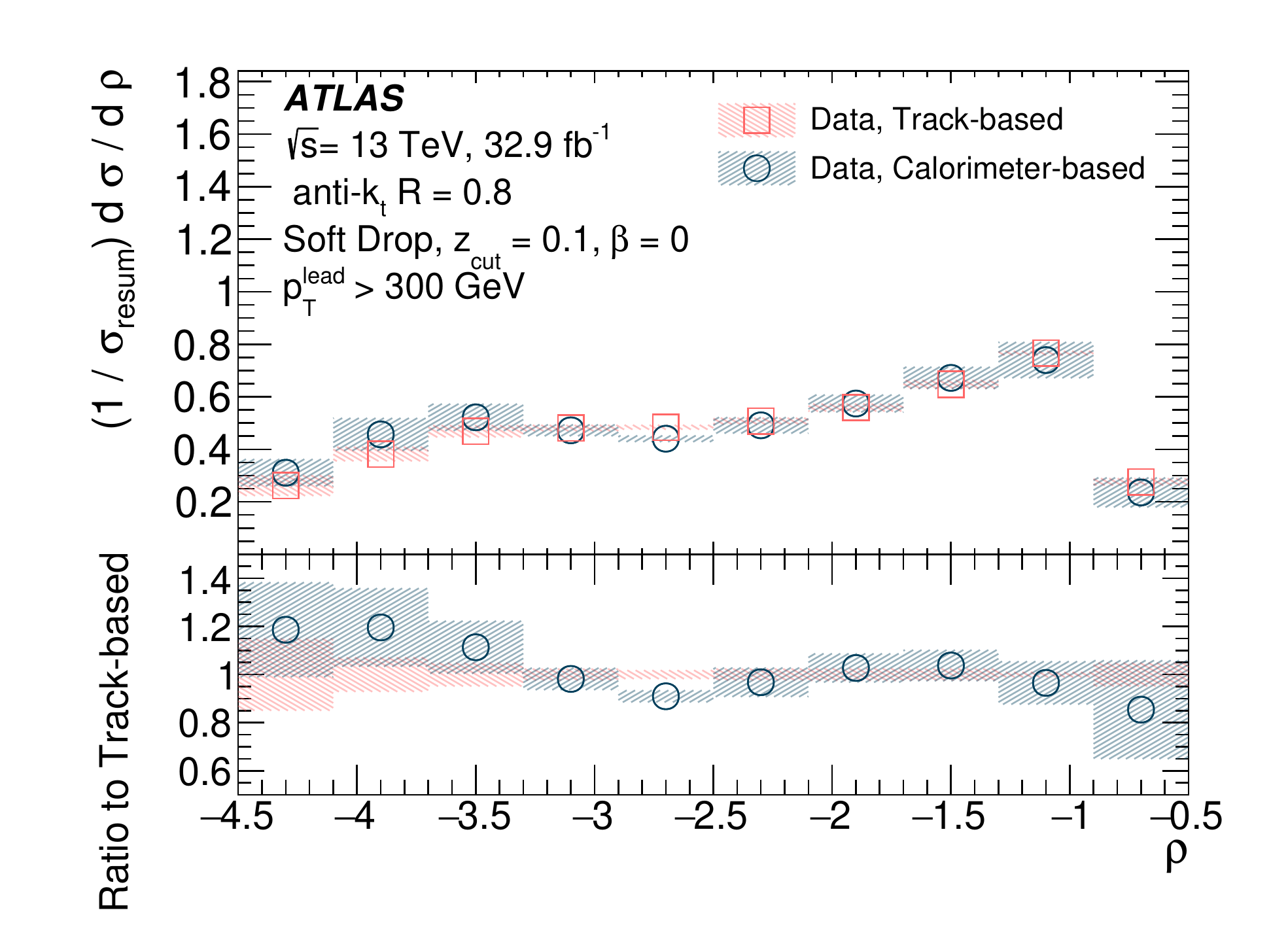}}
\subfigure[$\rho$ distribution, $\beta = 1$]{\includegraphics[width=0.48\textwidth,page=1]{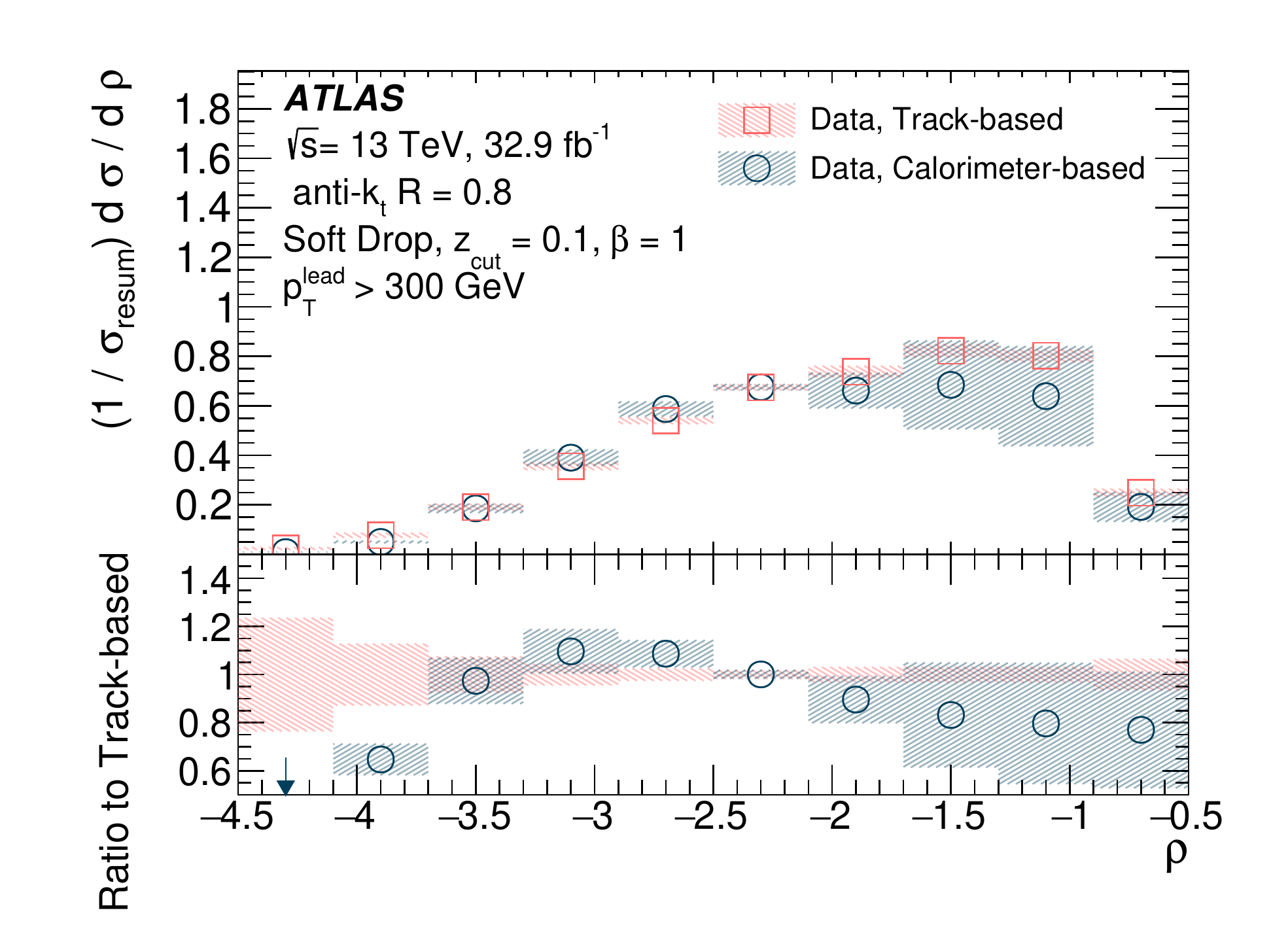}}
\subfigure[$\rho$ distribution, $\beta = 2$]{\includegraphics[width=0.48\textwidth,page=1]{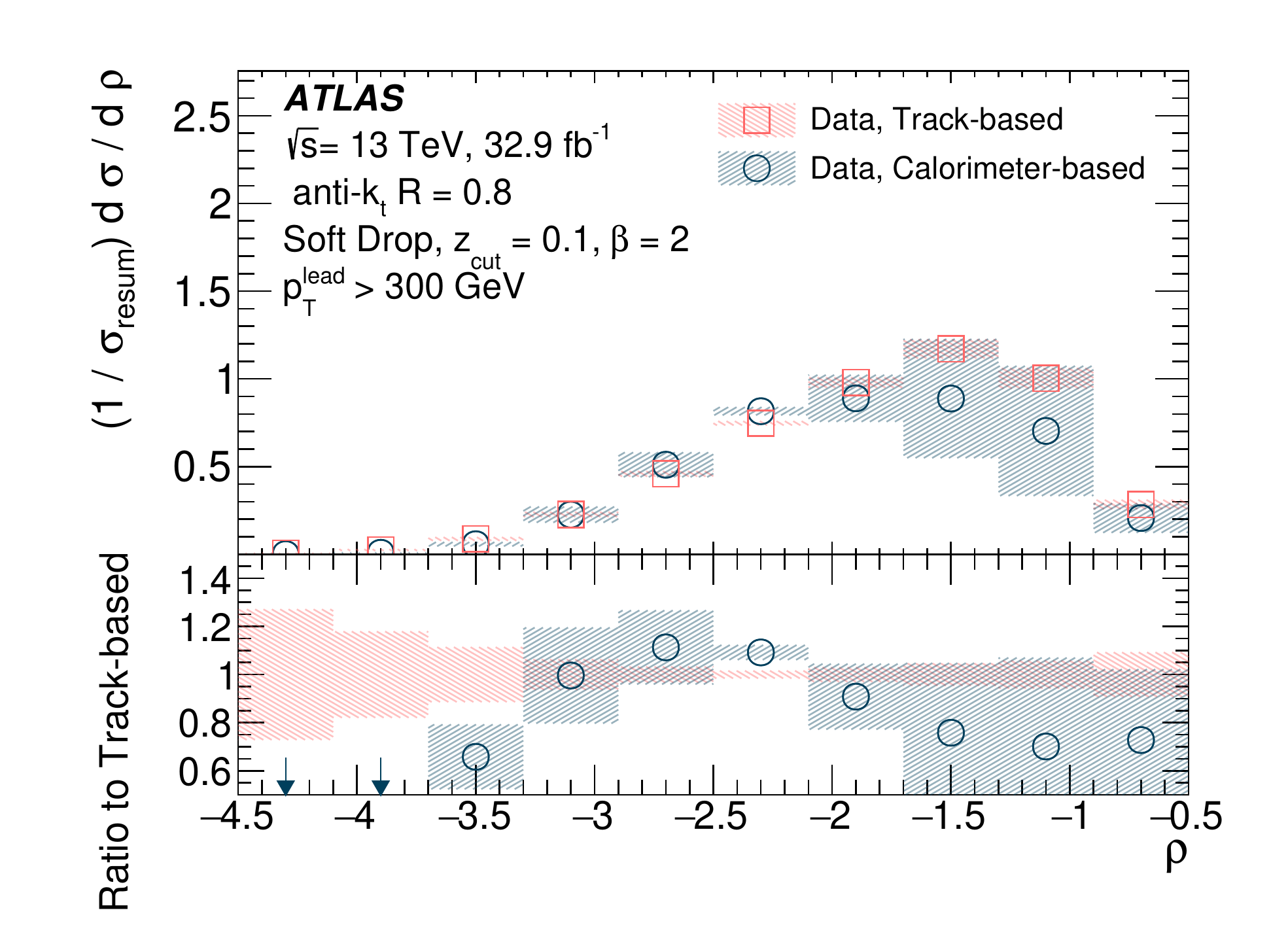}}
\end{center}
\caption{Unfolded $\rho$ distribution, for calorimeter- and track-based jets. The uncertainty bands include all sources: data and MC statistical uncertainties, nonclosure, modeling, and cluster or tracking uncertainties where relevant.}
\label{results:unfold:rho:trackCalo_inclPt}
\end{figure}
 
\begin{figure}[ht!]
\begin{center}
\subfigure[$z_g$ distribution, $\beta = 0$]{\includegraphics[width=0.48\textwidth,page=1]{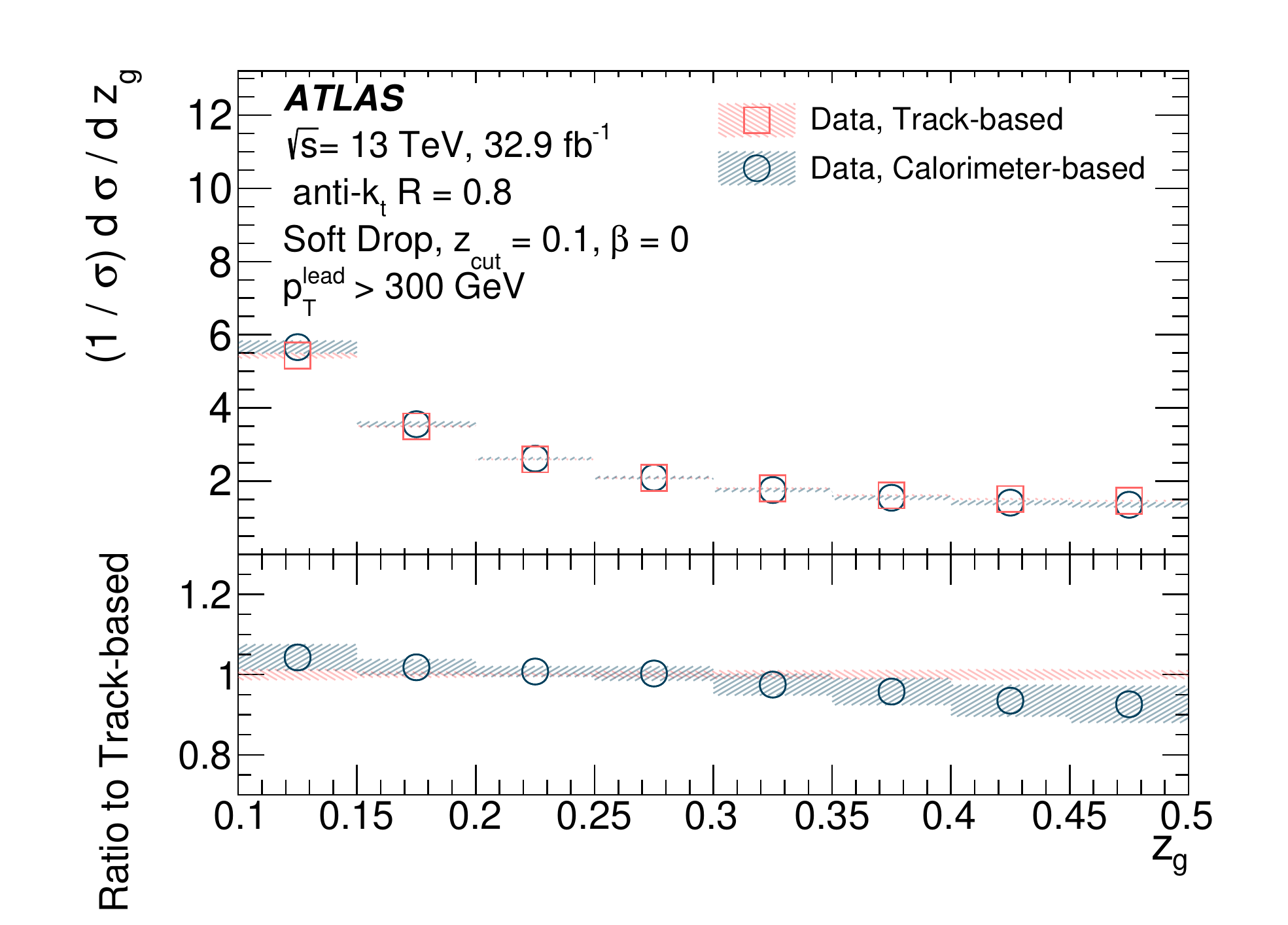}}
\subfigure[$z_g$ distribution, $\beta = 1$]{\includegraphics[width=0.48\textwidth,page=1]{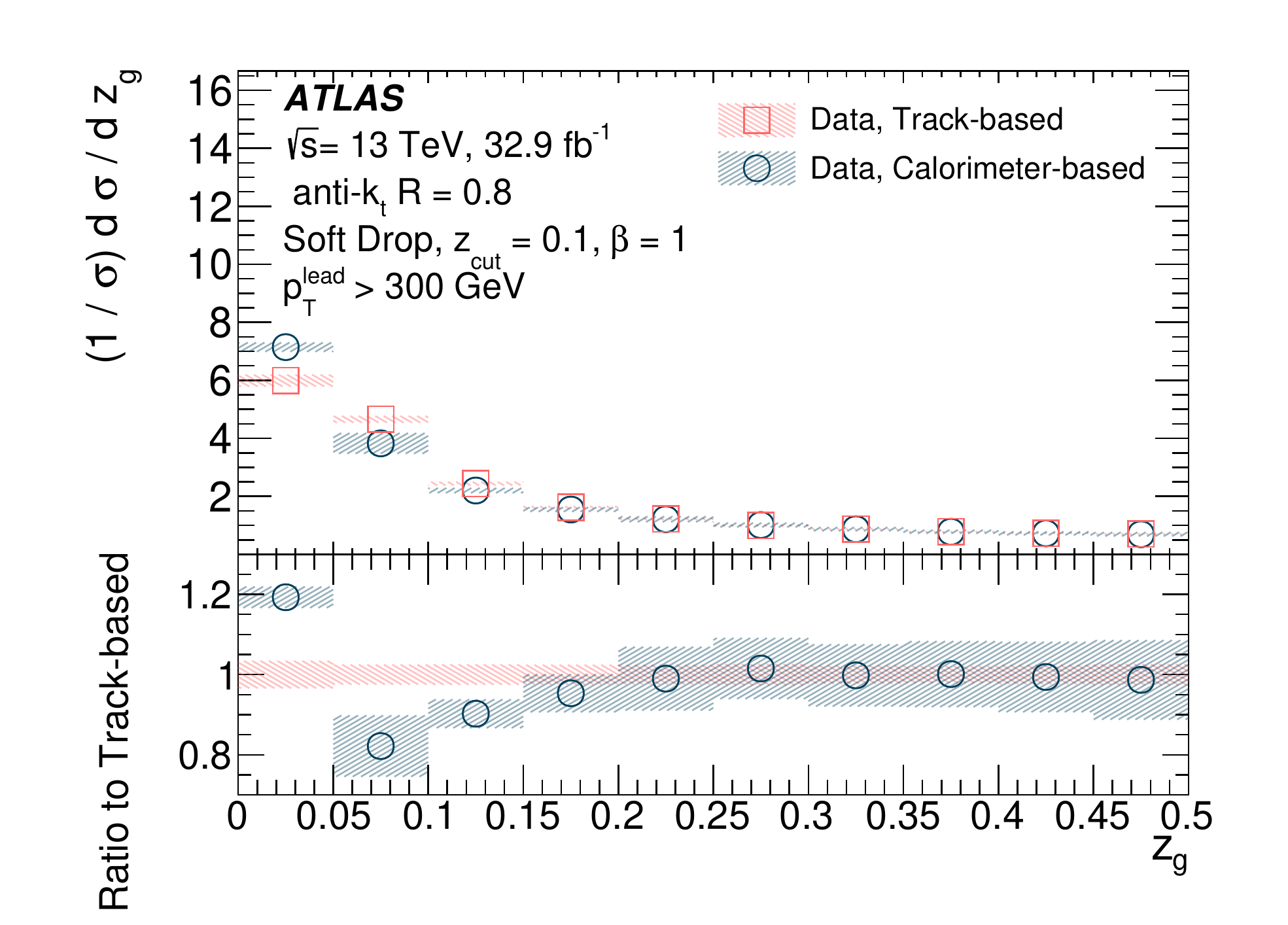}}
\subfigure[$z_g$ distribution, $\beta = 2$]{\includegraphics[width=0.48\textwidth,page=1]{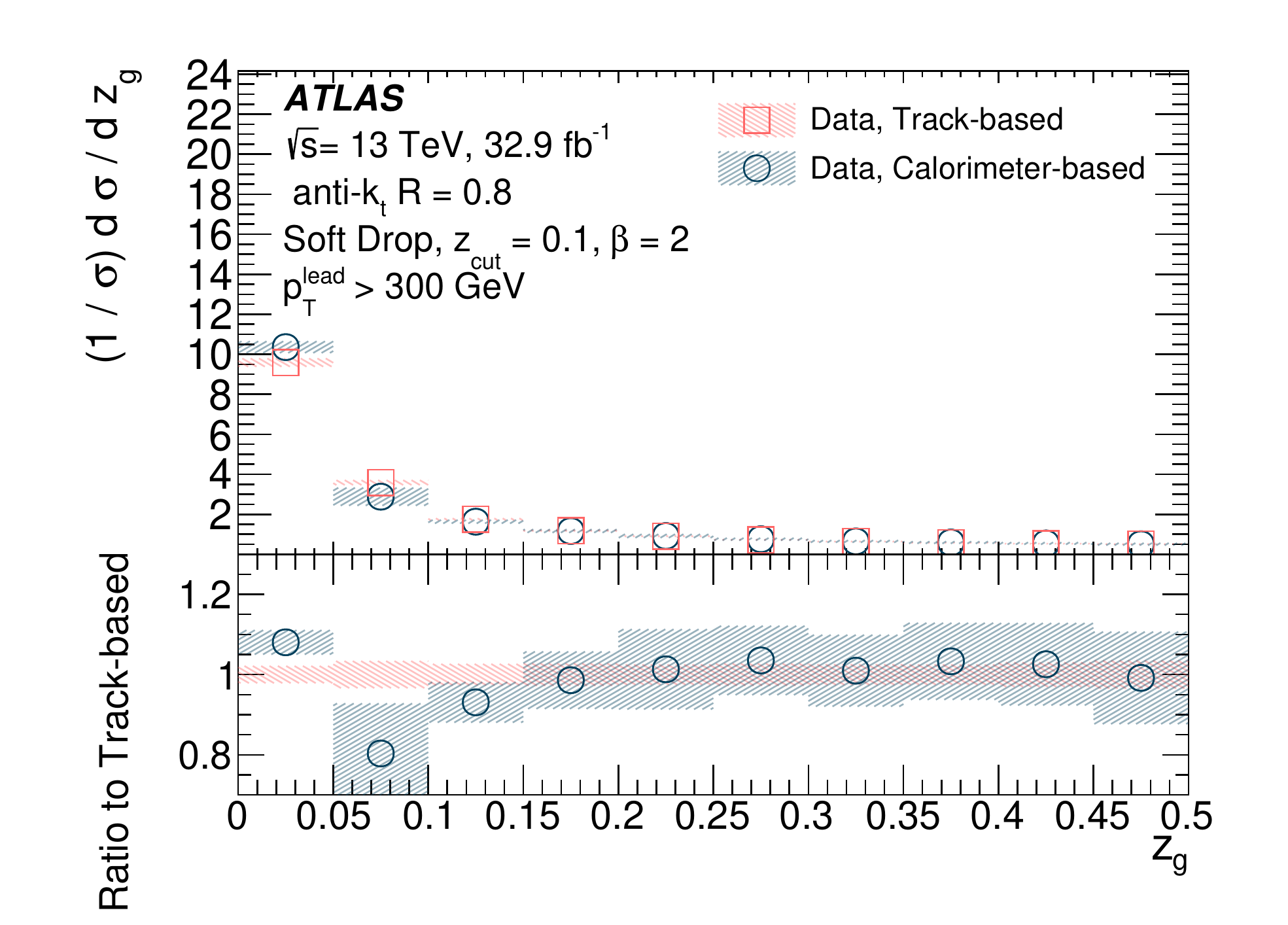}}
\end{center}
\caption{Unfolded $z_g$ distribution, for calorimeter- and track-based jets. The uncertainty bands include all sources: data and MC statistical uncertainties, nonclosure, modeling, and cluster or tracking uncertainties where relevant.}
\end{figure}
 
\begin{figure}[ht!]
\begin{center}
\subfigure[$r_g$ distribution, $\beta = 0$]{\includegraphics[width=0.48\textwidth,page=1]{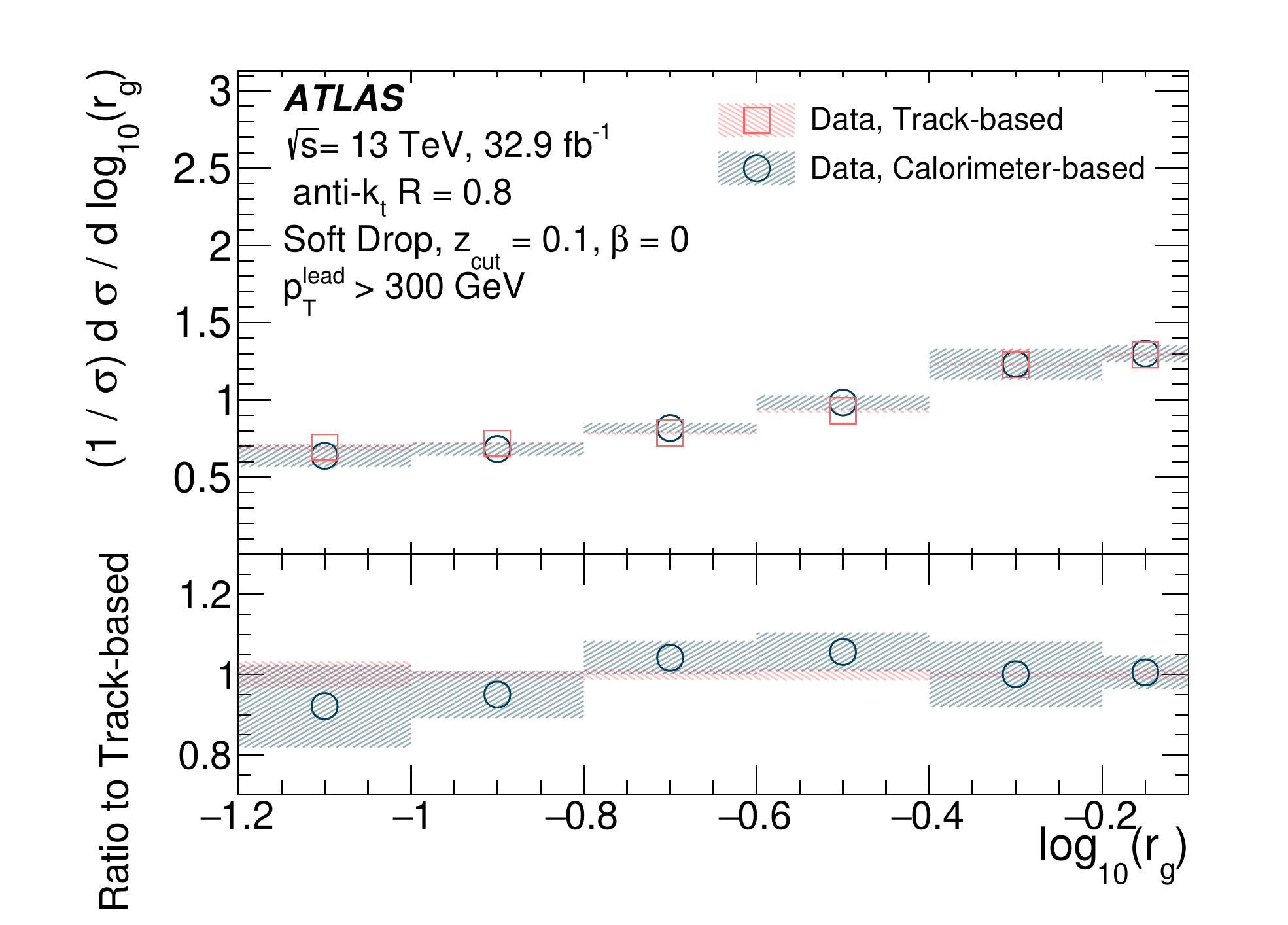}}
\subfigure[$r_g$ distribution, $\beta = 1$]{\includegraphics[width=0.48\textwidth,page=1]{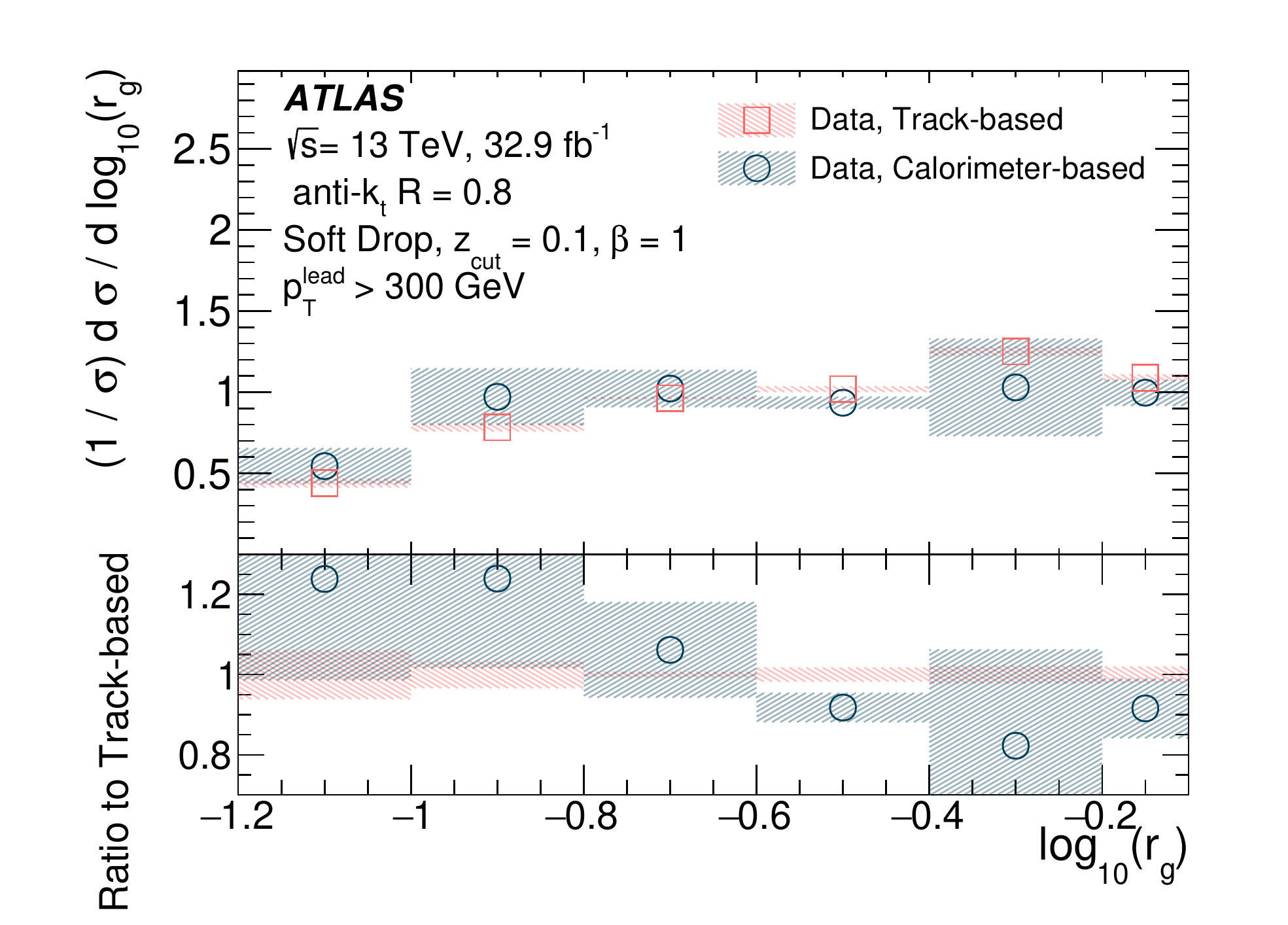}}
\subfigure[$r_g$ distribution, $\beta = 2$]{\includegraphics[width=0.48\textwidth,page=1]{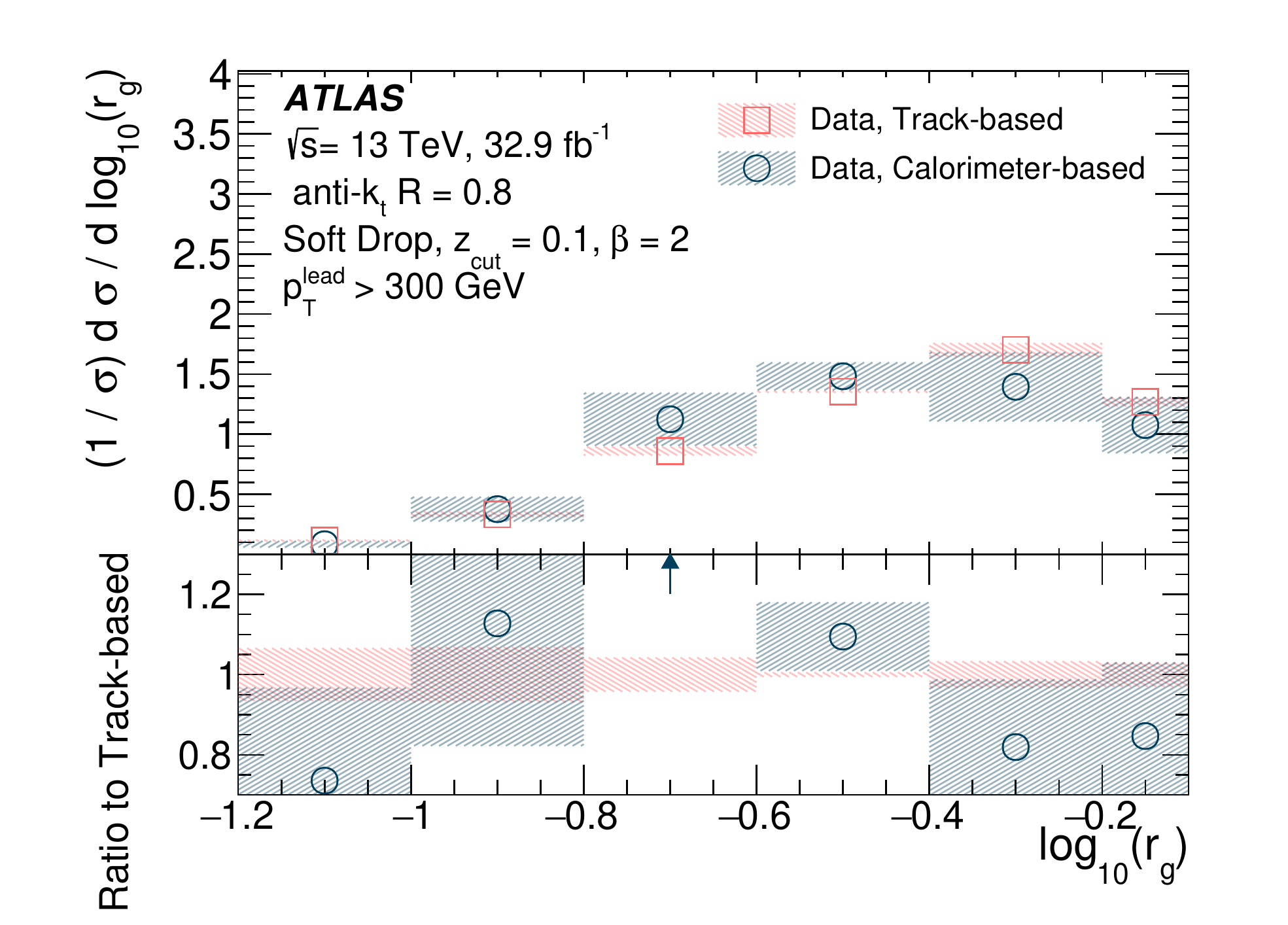}}
\end{center}
\caption{Unfolded $r_g$ distribution, for calorimeter- and track-based jets. The uncertainty bands include all sources: data and MC statistical uncertainties, nonclosure, modeling, and cluster or tracking uncertainties where relevant.}
\label{results:unfold:rg:trackCalo_inclPt}
\end{figure}

\subsection{Comparison of forward and central measurements}
 
\label{sec:forwardCentral}
 
The distribution of the substructure observables at a given \pt is a function of the composition of the initiating parton type, and should not be affected by where the jet
is produced within the detector. Therefore, any differences seen between the distribution of the observable in different regions of the detector are related
to the quark--gluon composition of the events produced there. Since this measurement was done separately for the more forward and more central jet in the
dijet samples, it is possible to compare these distributions to see if these effects are visible.
The fraction of central and forward jets originating from gluons, $f_G$,
in \PYTHIA8 multijet events is shown in Table~\ref{table:qgFraction}, where the jet flavor is determined by the highest-energy parton inside the jet cone.\footnote{Various definitions were studied in Ref.~\cite{STDM-2015-12} and found to have a small effect on quark/gluon extractions.  Furthermore, the universality of this definition was studied in Ref.~\cite{Bright-Thonney:2018mxq}.}
This shows that the gluon fractions in the forward and central regions differ by about 5--10\%.
For each of the three observables, Figure~\ref{results:unfold:rho:forwardCentral_inclPt} compares the unfolded distributions for the jets in the forward
region with those for jets in the central region.
As expected, since the forward region is quark-enhanced, it has more jets at lower relative masses.
These differences are numerically small because the gluon fractions are similar for the forward and central jets.
 
\begin{table}[ht!]
\caption{The gluon fractions predicted by the \PYTHIA8 multijet simulation. }
\centering
\begin{tabular}{|c|c|c|}
\hline & \multicolumn{2}{c|}{Gluon Fraction [\%]} \\ \hline
\hline & Central Region & Forward Region \\ \hline
300 ~\GeV< \pt < 400 ~\GeV  & 75.1                                                               & 69.5                                                                         \\
400 ~\GeV< \pt < 600 ~\GeV  & 71.7                                                               & 64.4                                                                         \\
600 ~\GeV< \pt < 800 ~\GeV  & 66.2                                                               & 56.9                                                                         \\
~~800 ~\GeV< \pt < 1000 ~\GeV & 61.0                                                               & 50.5                                                                         \\
1000 ~\GeV< \pt < 2000 ~\GeV& 54.4                                                               & 43.3                                                                         \\
\hline
\end{tabular}
\label{table:qgFraction}
\end{table}

\begin{figure}[ht!]
\begin{center}
\subfigure[$\rho$ distribution, $\beta = 0$, calorimeter-based]{\includegraphics[width=0.48\textwidth,page=1]{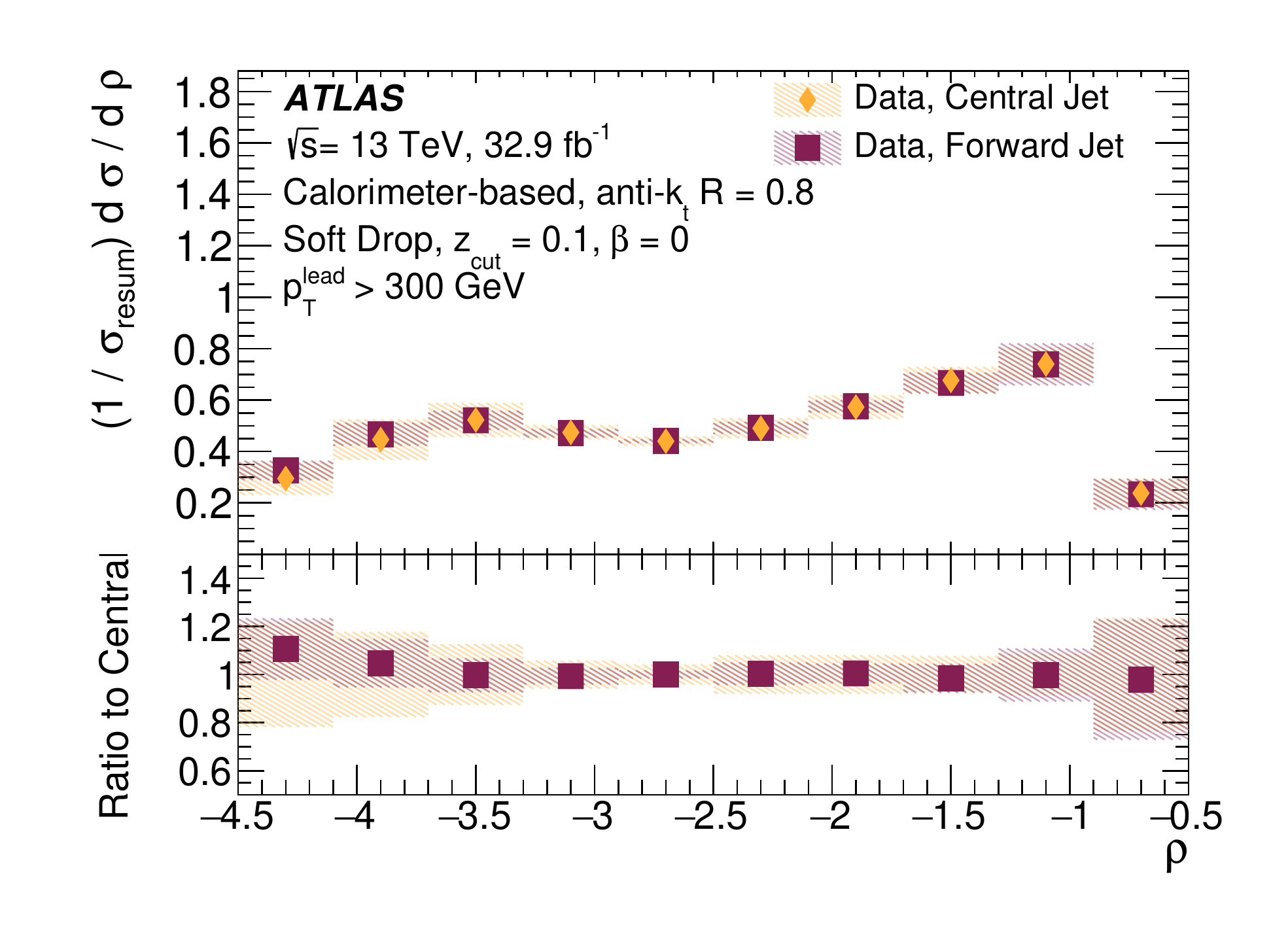}}
\subfigure[$\rho$ distribution, $\beta = 0$, track-based]{\includegraphics[width=0.48\textwidth,page=1]{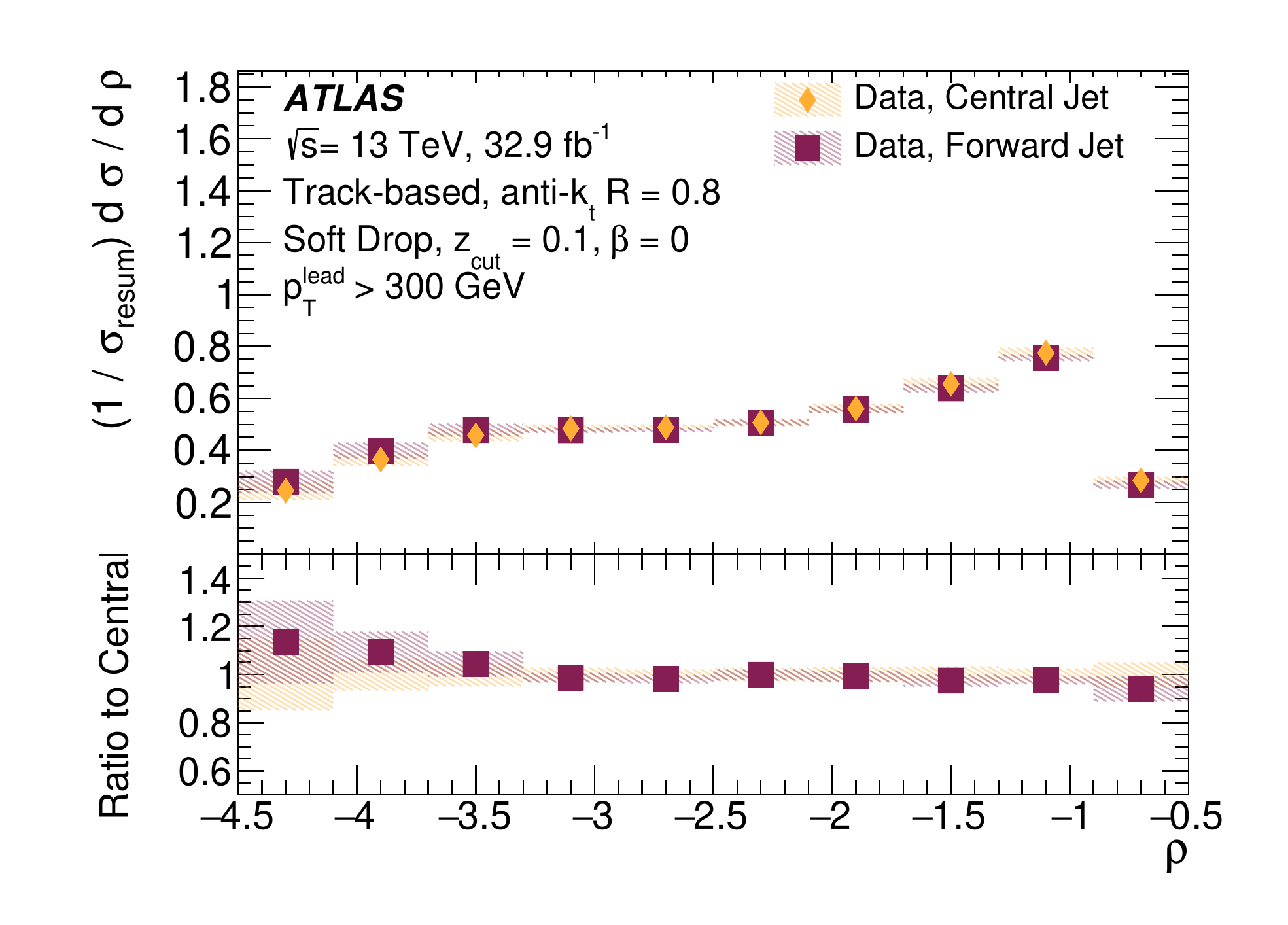}}
\subfigure[$z_g$ distribution, $\beta = 0$, calorimeter-based]{\includegraphics[width=0.48\textwidth,page=1]{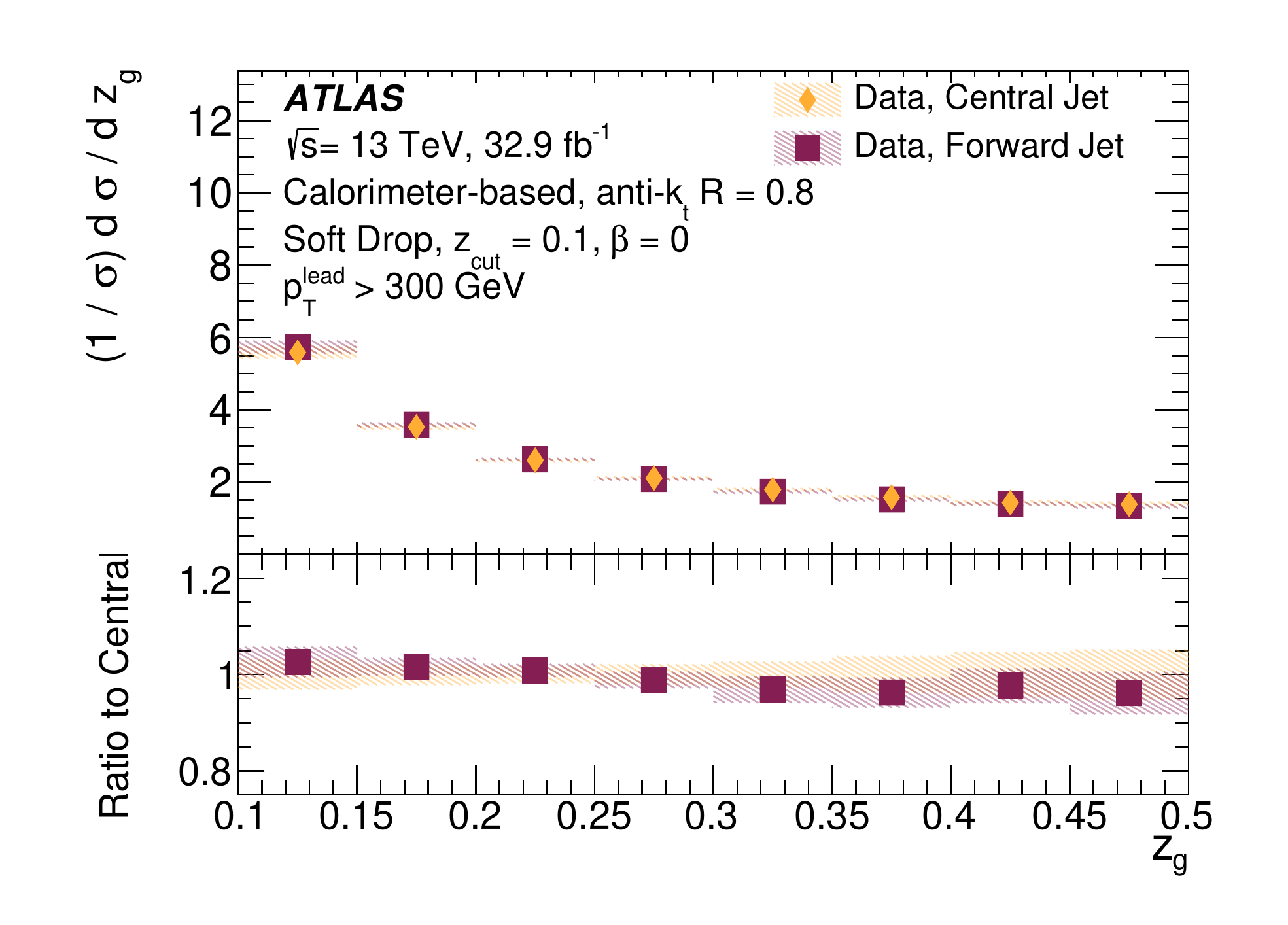}}
\subfigure[$z_g$ distribution, $\beta = 0$, track-based]{\includegraphics[width=0.48\textwidth,page=1]{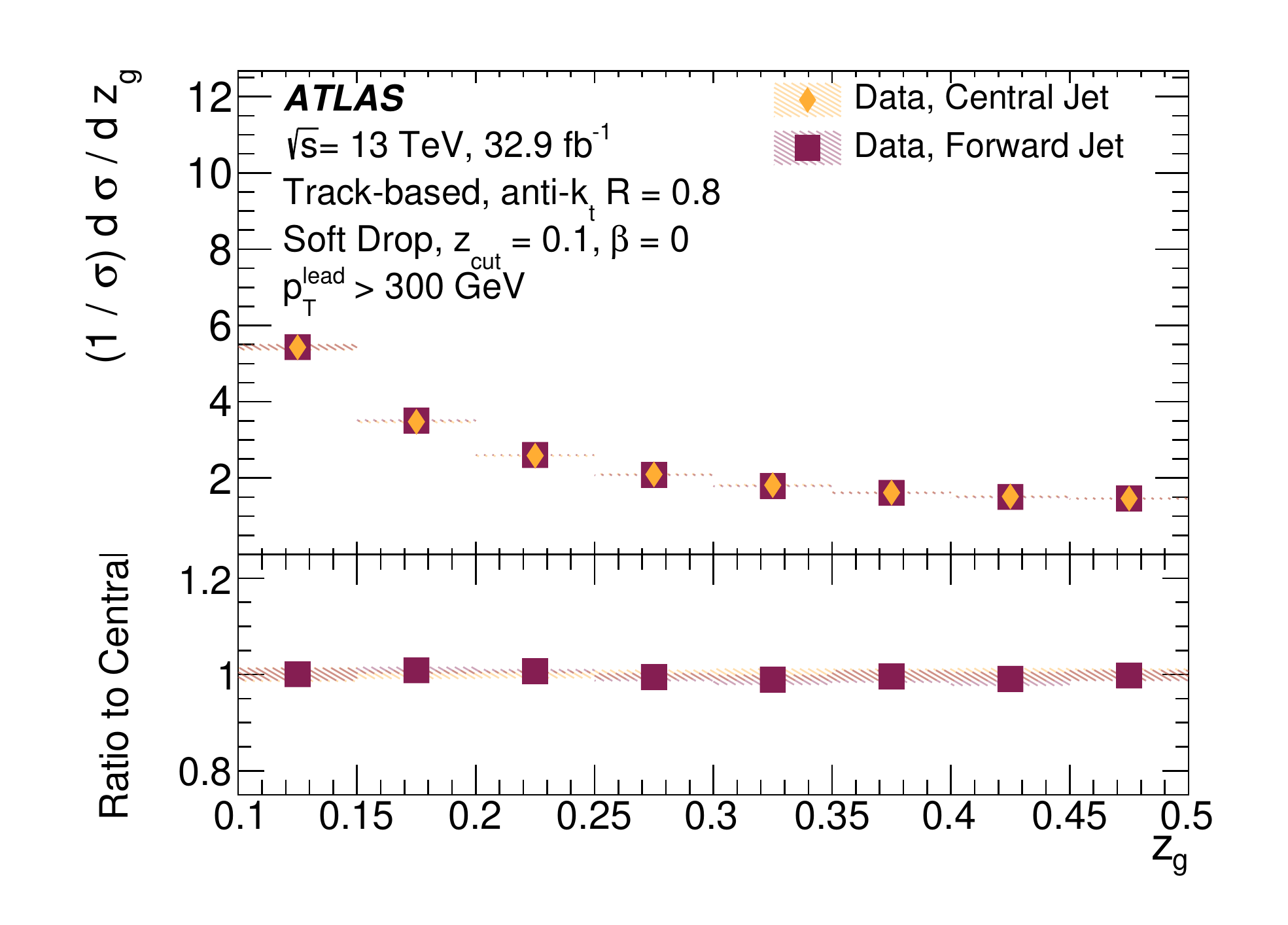}}
\subfigure[$r_g$ distribution, $\beta = 0$, calorimeter-based]{\includegraphics[width=0.48\textwidth,page=1]{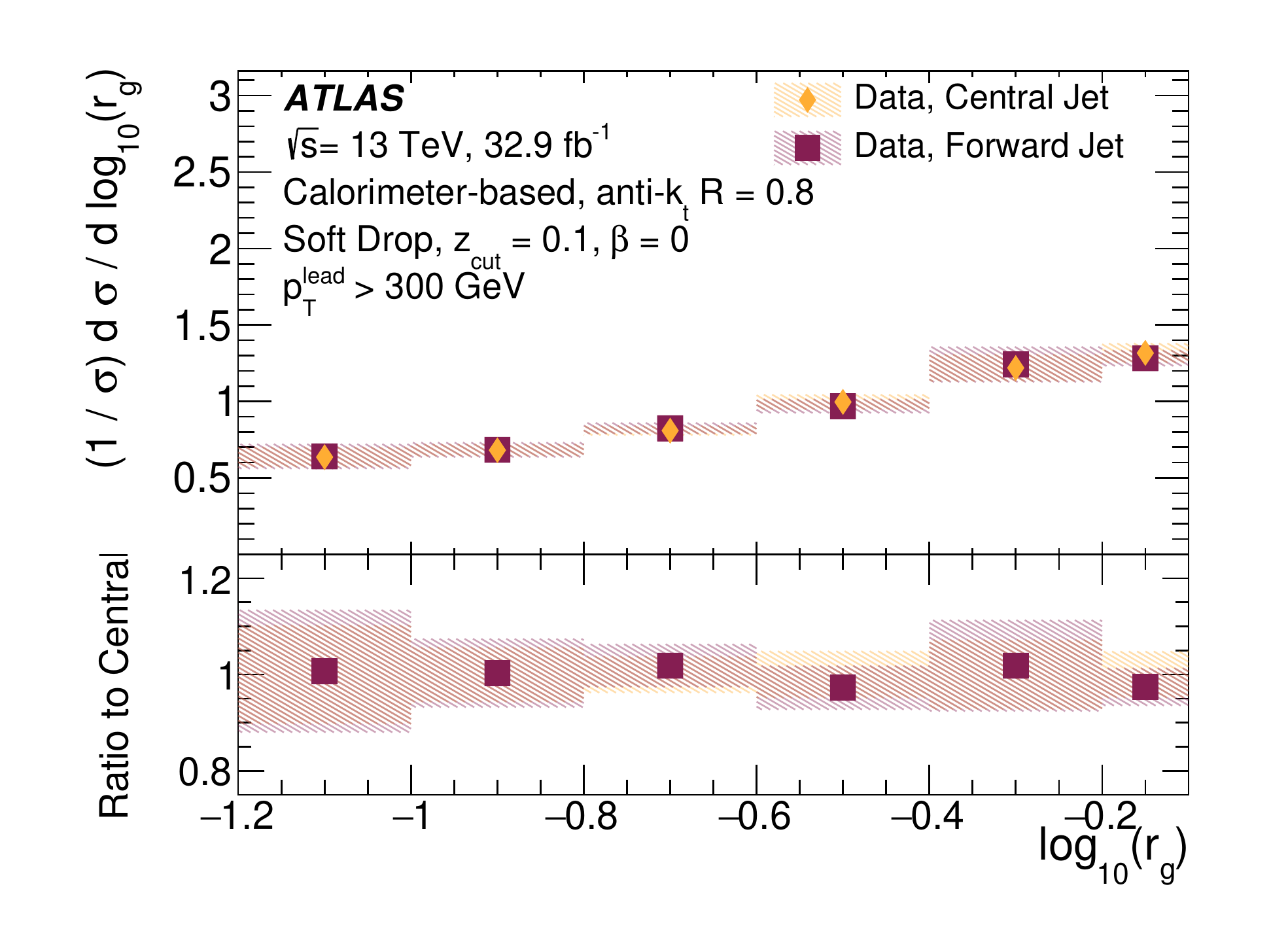}}
\subfigure[$r_g$ distribution, $\beta = 0$, track-based]{\includegraphics[width=0.48\textwidth,page=1]{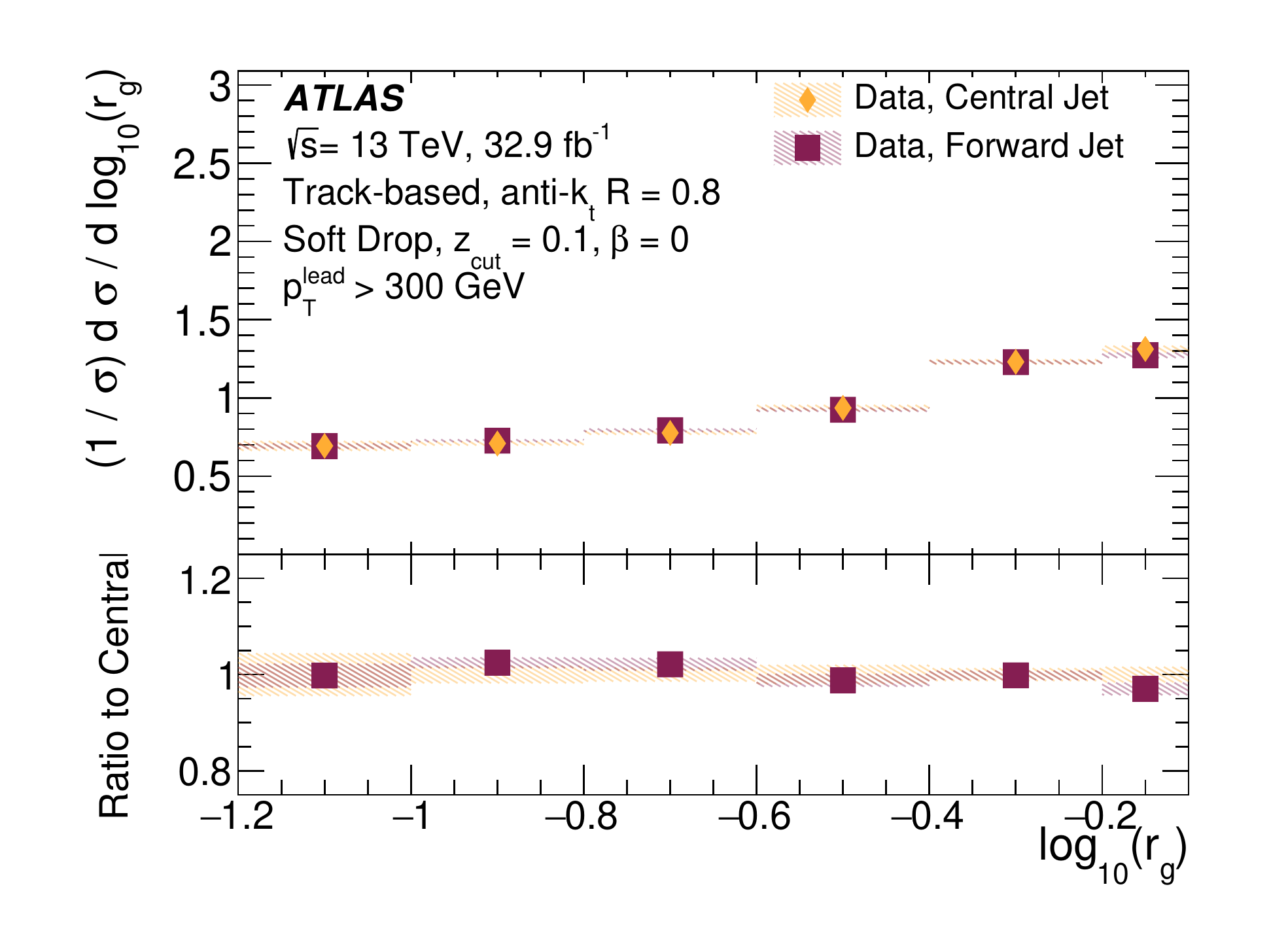}}
\end{center}
\caption{Comparison of the forward and central unfolded distributions for $\beta=0$. The uncertainty bands include all sources: data and MC statistical uncertainties, cluster uncertainties, nonclosure, and modeling.  See Section~\ref{sec:systematics} for details.}
\label{results:unfold:rho:forwardCentral_inclPt}
\end{figure}

\FloatBarrier
 
\subsection{Quark--gluon Extraction of the observables}
\label{sec:quarkGluon}

Since the shape of the $\rho$, $r_g$, and $z_g$ distributions at a given jet \pt only depends on the flavor of the initiating parton and not on the rapidity, the quark and gluon distributions
may be extracted from the measurements of the central and forward distributions if the quark--gluon fraction is known for each region.
In particular, the central and forward distributions for these observables may be described as the sum of the quark and gluon distributions, weighted by the
quark and gluon composition of the sample:
\begin{equation}
\begin{aligned}
h_{i}^{F} &= f_{Q}^{F} h_{i}^{Q} + f_{G}^{F} h_{i}^{G}, \\
h_{i}^{C} &= f_{Q}^{C} h_{i}^{Q} + f_{G}^{C} h_{i}^{G},
\end{aligned}
\label{qgEquations}
\end{equation}
where $h_i$ is a bin of a histogram for an observable, $F$ and $C$ represent the forward and central regions, and $Q$ and $G$ represent
quark or gluon. The quark and gluon fractions ($f_Q$ and $f_{G}$) for the more forward and more central jets are determined from the nominal \PYTHIA8 MC event sample,
where the quark fraction $f_Q$ is given by $1-f_G$.
This extraction is model-dependent, but the more forward and more central distributions are made public for reinterpretation using any model.
Table~\ref{table:qgFraction} shows these values for each \pt bin.
Equations (~\ref{qgEquations}) may then be solved for $h_{i}^{G}$ and $h_{i}^{Q}$
to extract these distributions from the forward and central distributions.
 
The extracted quark and gluon distributions are shown in Figures~\ref{results:unfold:rho:quarkgluon_inclPt}--~\ref{results:unfold:rg:quarkgluon_inclPt} for track-based observables.
Cluster-based observables are not shown, but exhibit similar behavior overall.
For these results, the PDF uncertainties and the uncertainties in the jet inputs are taken to be fully correlated between the more forward and more central jets, while all other uncertainties are considered fully uncorrelated.
In addition, to account for the uncertainty in the composition of the sample, the difference between the extracted distributions using the \PYTHIA8 and \SHERPA compositions
is taken as an uncertainty.
A few observations can be made about the differences between the quark and gluon distributions.
For the jet mass, the gluon distribution tends towards higher values of the mass, which is expected due to the larger color factor associated with gluons.
These differences become more apparent at larger values of $\beta$.
For $\beta=0$, $z_g$ is independent of $\alphas$ to leading order, and the distributions are very similar,
while for $\beta > 0$, some differences begin to appear.
Finally, for $r_g$, the gluon distribution tends towards a larger splitting, which is similarly more
apparent at larger values of $\beta$.

\begin{figure}[ht!]
\begin{center}
\subfigure[$\rho$ distribution, $\beta = 0$, track-based]{\includegraphics[width=0.48\textwidth,page=1]{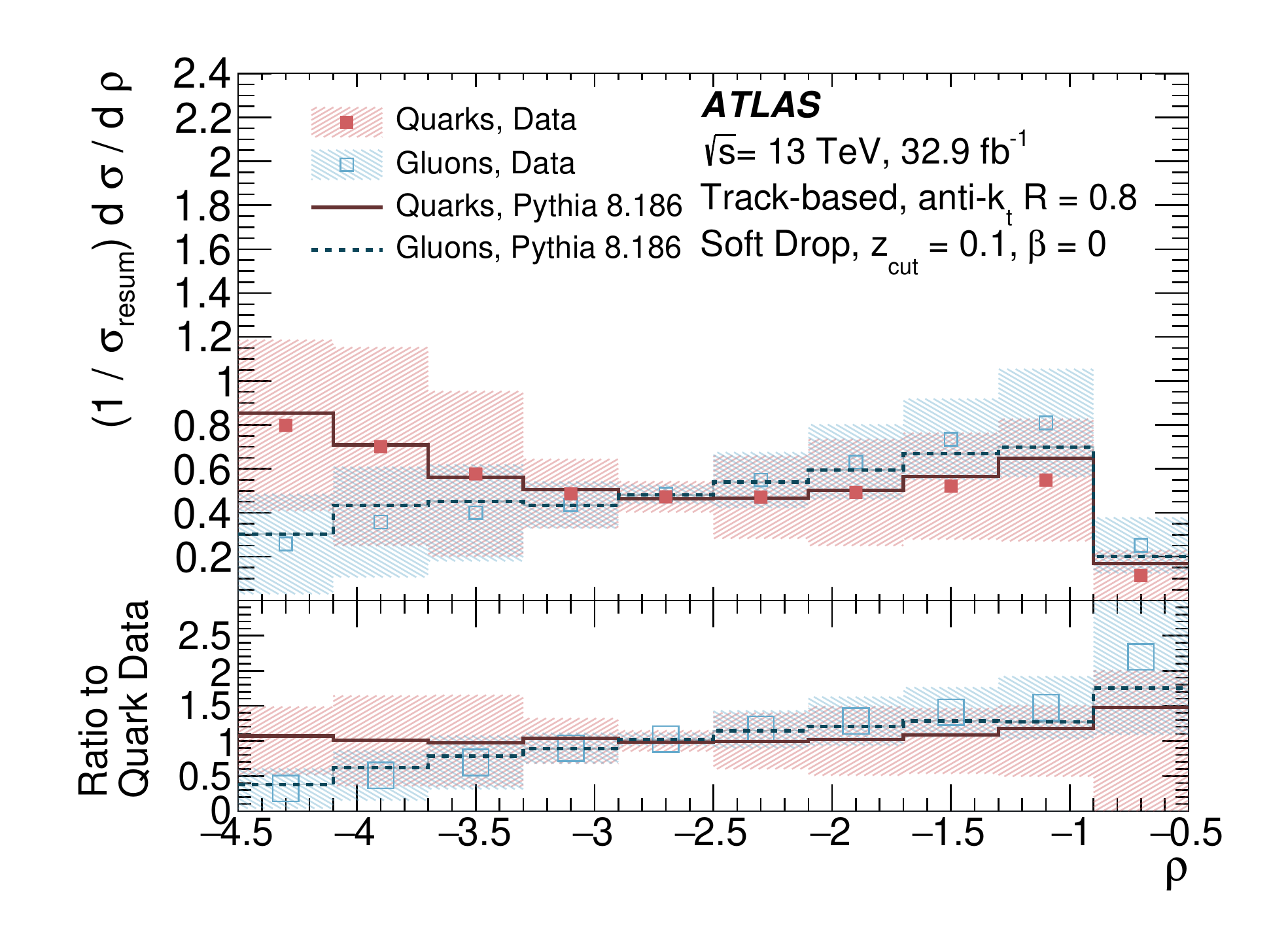}}
\subfigure[$\rho$ distribution, $\beta = 1$, track-based]{\includegraphics[width=0.48\textwidth,page=1]{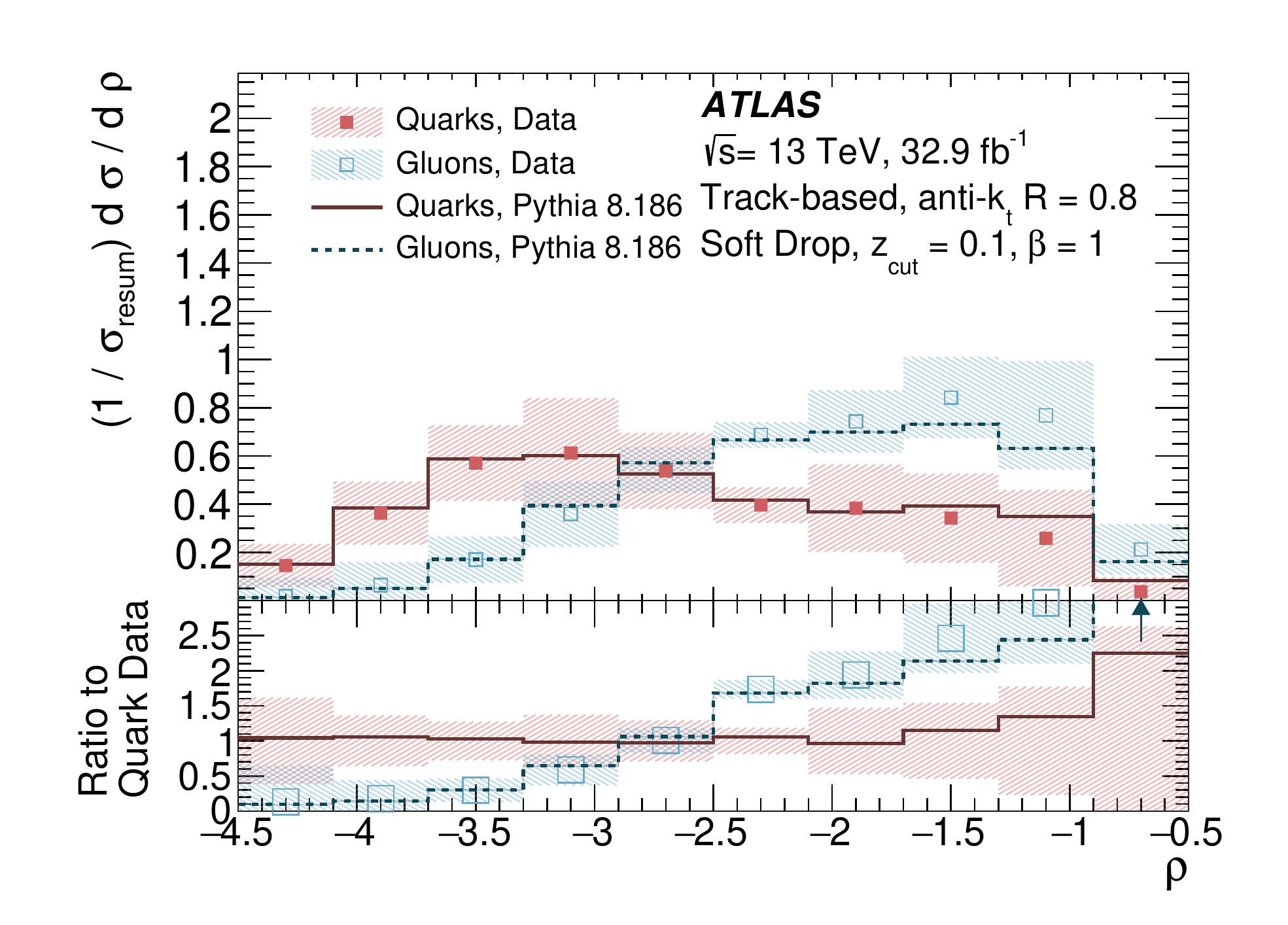}}
\subfigure[$\rho$ distribution, $\beta = 2$, track-based]{\includegraphics[width=0.48\textwidth,page=1]{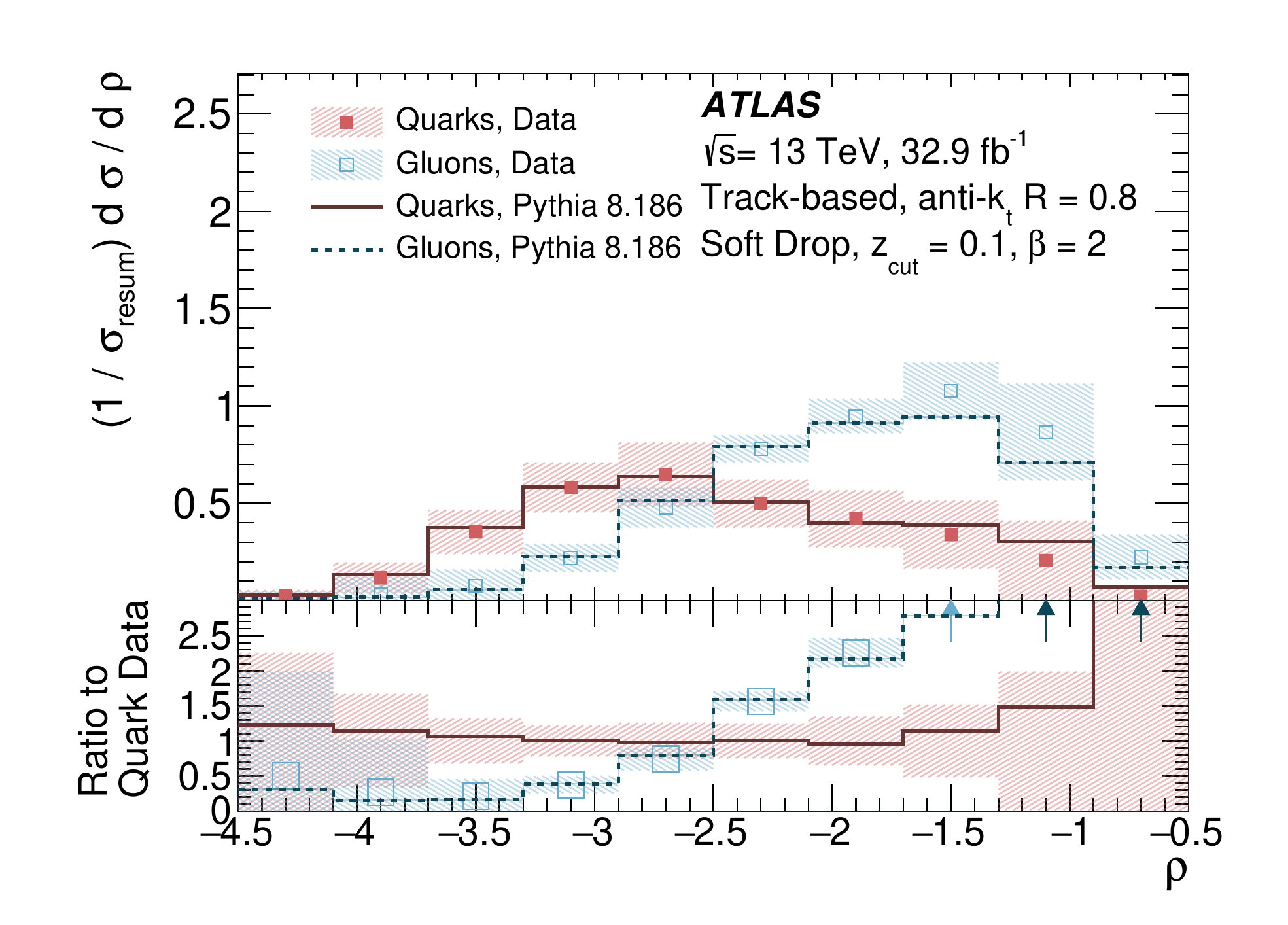}}
\end{center}
\caption{Comparison of the quark and gluon unfolded $\rho$ distributions for the track-based measurement. The uncertainty bands include all sources: data and MC statistical uncertainties, tracking uncertainties, nonclosure, and modeling. }
\label{results:unfold:rho:quarkgluon_inclPt}
\end{figure}
 
\begin{figure}[ht!]
\begin{center}
\subfigure[$z_g$ distribution, $\beta = 0$, track-based]{\includegraphics[width=0.48\textwidth,page=1]{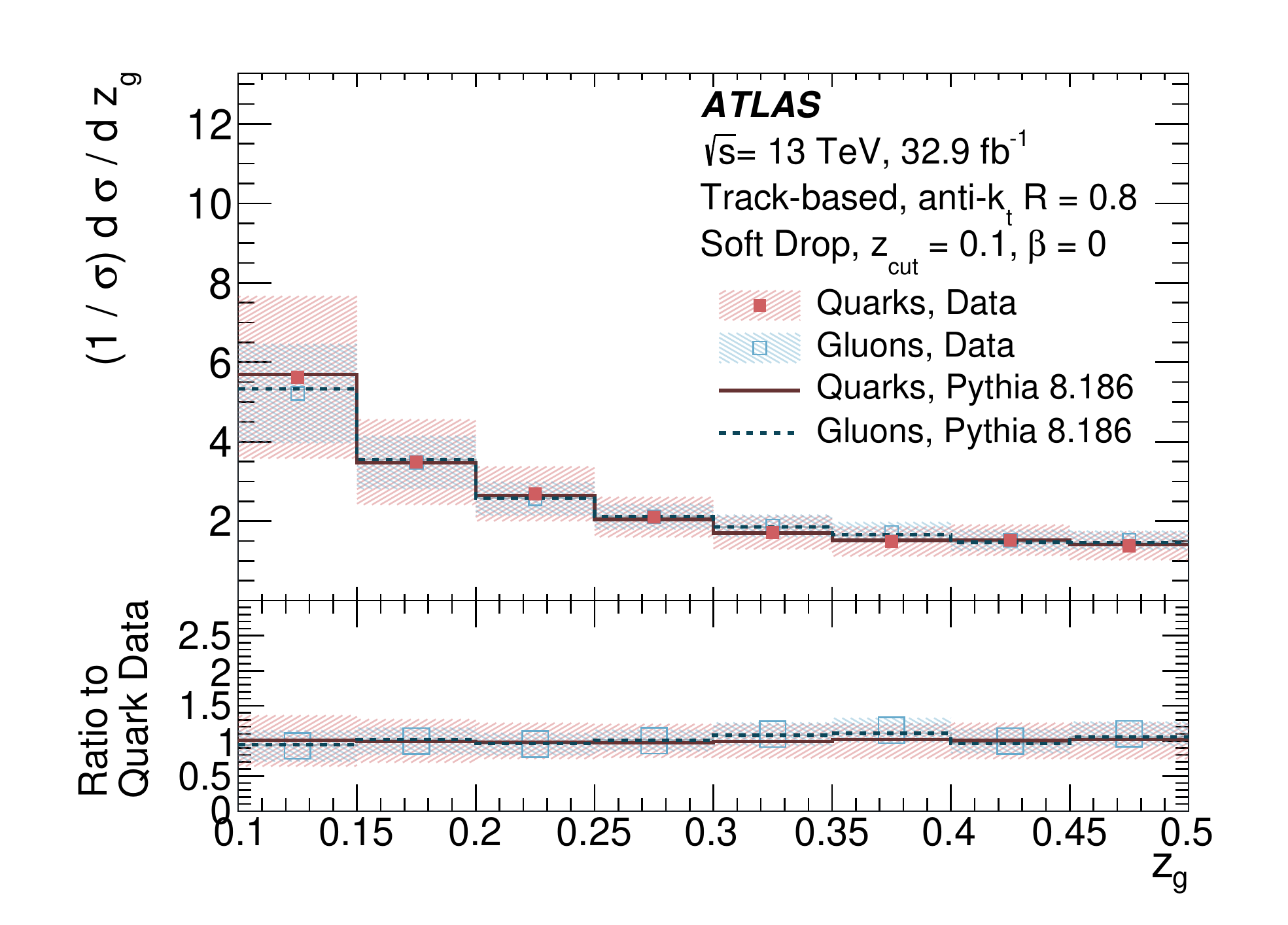}}
\subfigure[$z_g$ distribution, $\beta = 1$, track-based]{\includegraphics[width=0.48\textwidth,page=1]{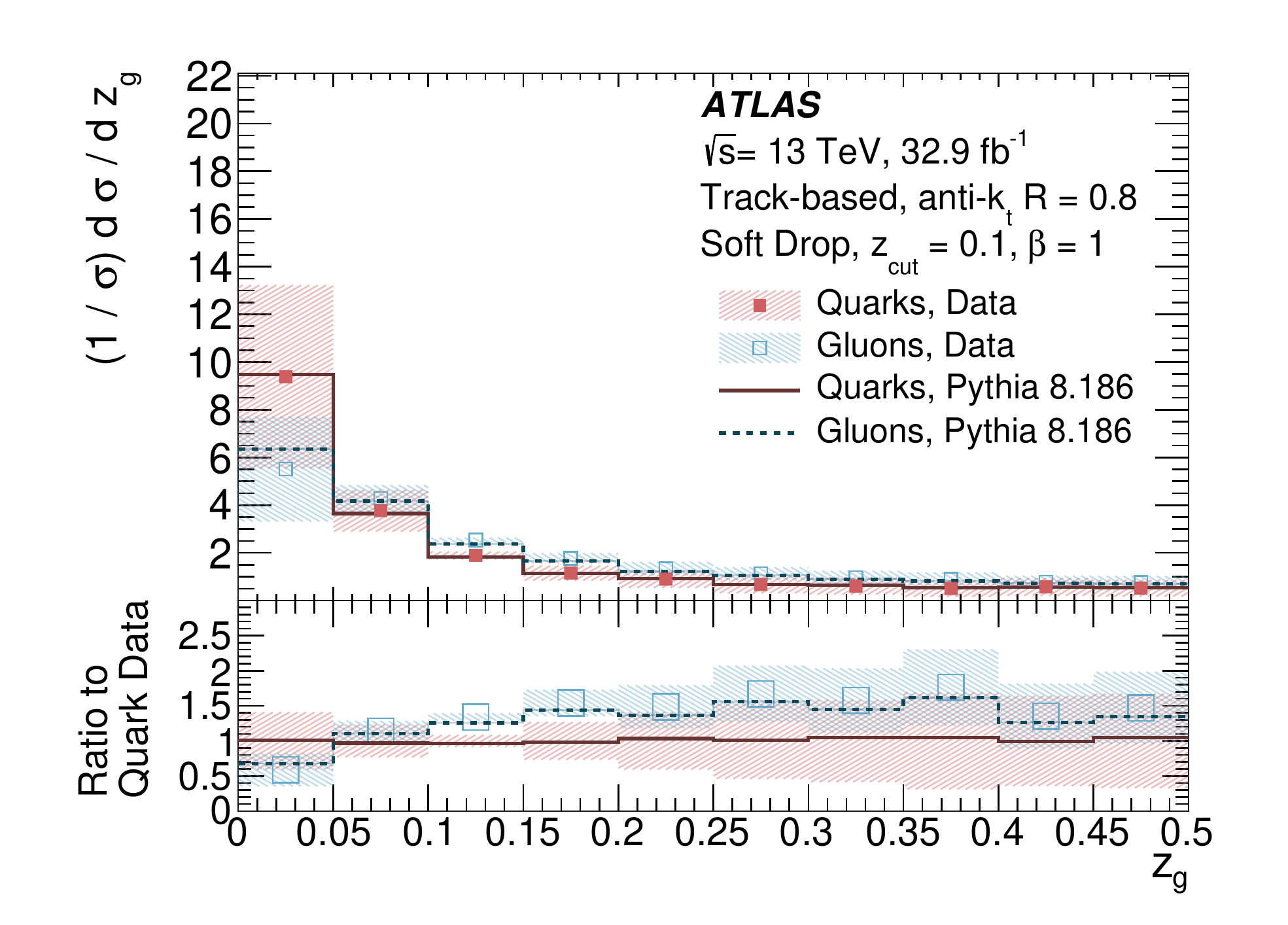}}
\subfigure[$z_g$ distribution, $\beta = 2$, track-based]{\includegraphics[width=0.48\textwidth,page=1]{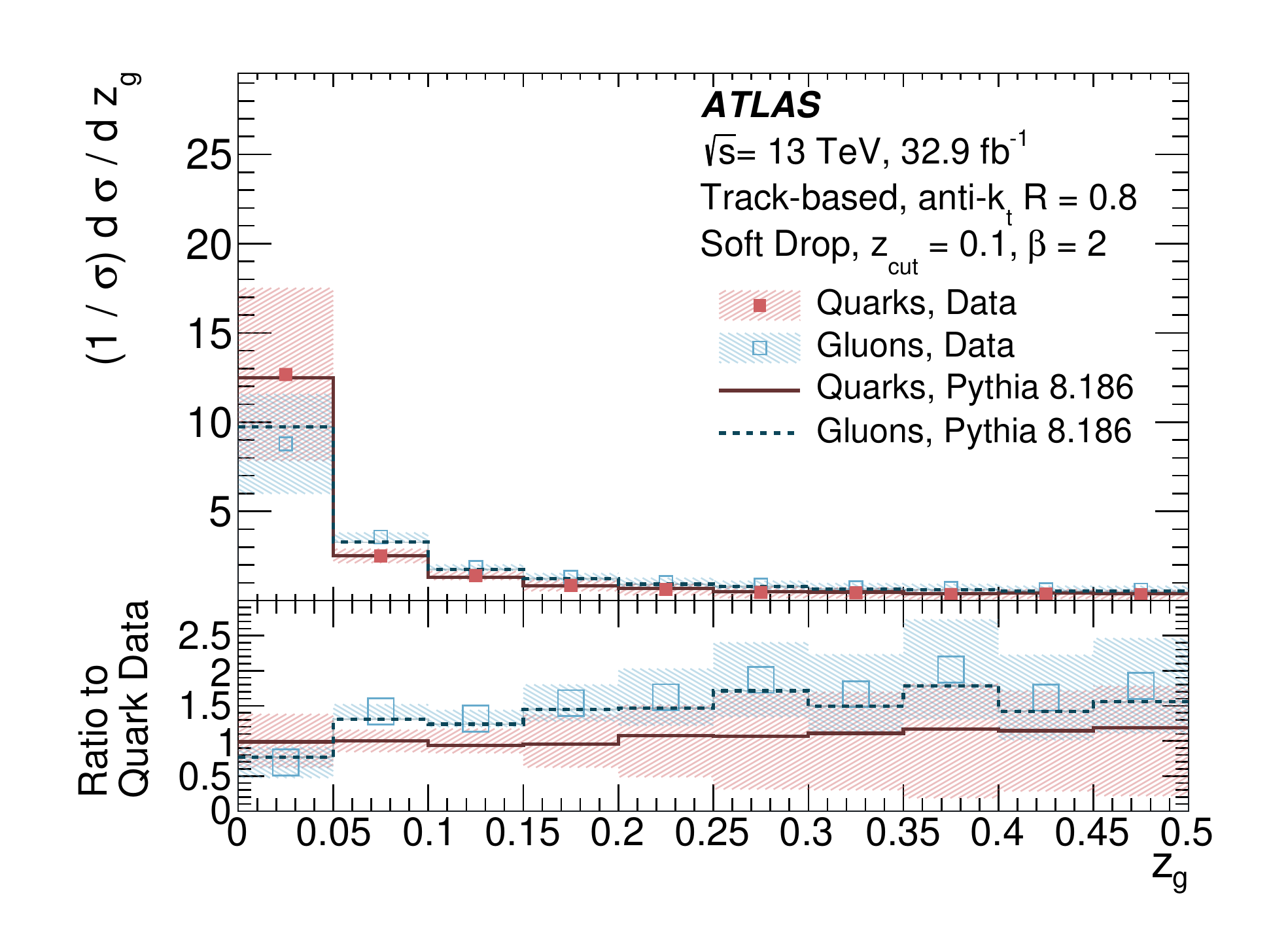}}
\end{center}
\caption{Comparison of the quark and gluon unfolded $z_g$ distribution for the track-based measurement. The uncertainty bands include all sources: data and MC statistical uncertainties, nonclosure, modeling, and tracking uncertainties where relevant.}
\end{figure}
 
\begin{figure}[ht!]
\begin{center}
\subfigure[$r_g$ distribution, $\beta = 0$, track-based]{\includegraphics[width=0.48\textwidth,page=1]{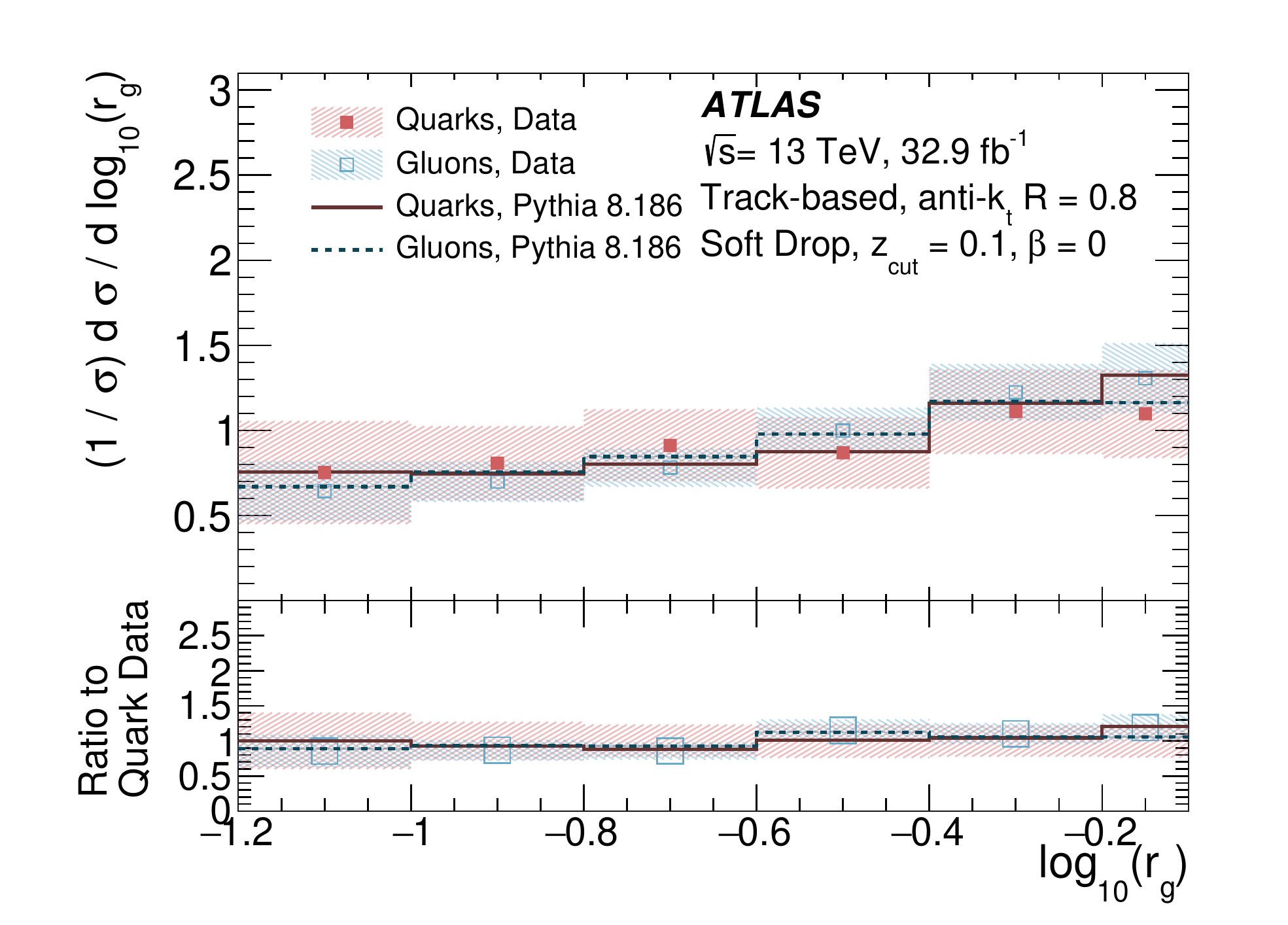}}
\subfigure[$r_g$ distribution, $\beta = 1$, track-based]{\includegraphics[width=0.48\textwidth,page=1]{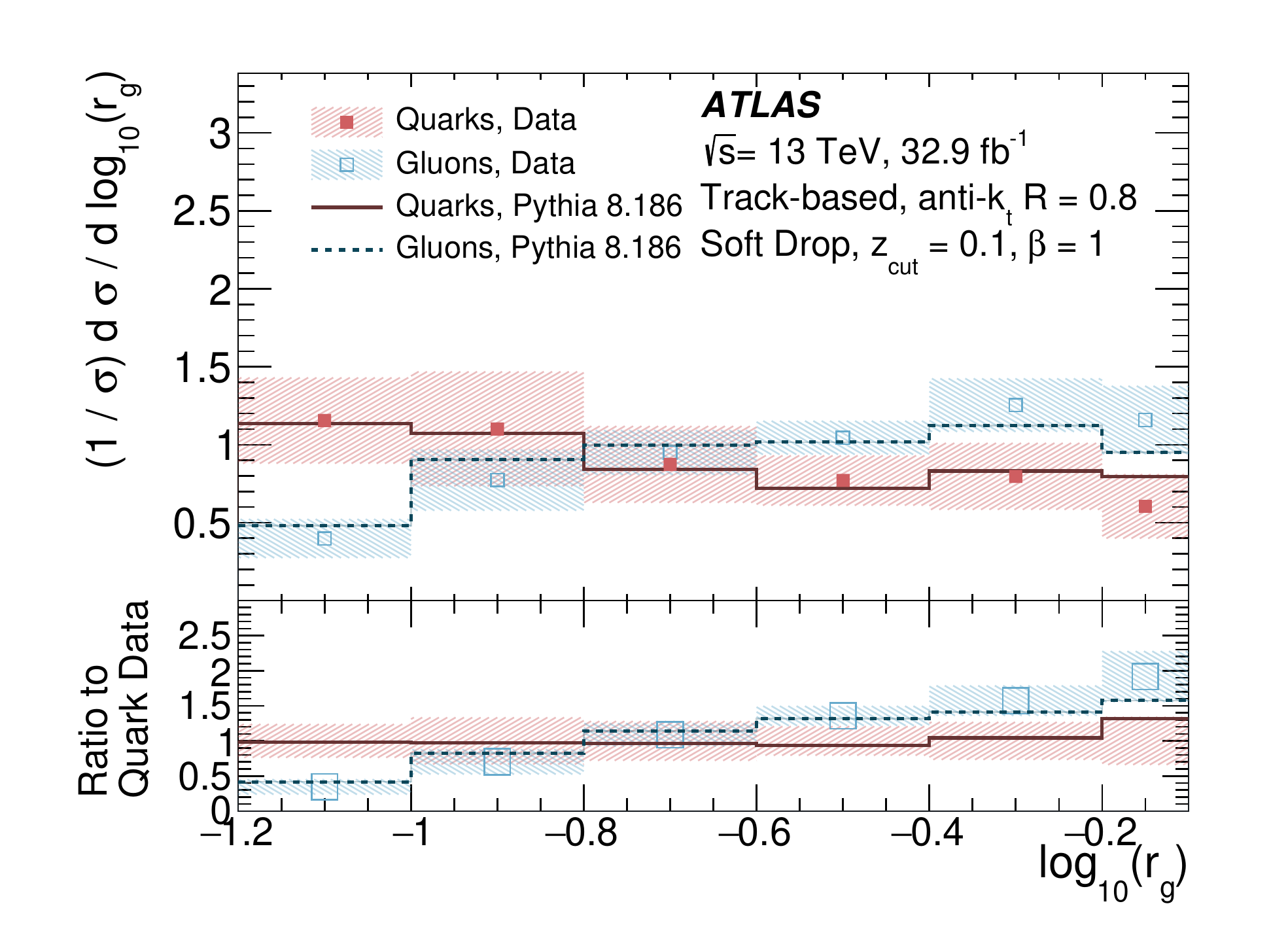}}
\subfigure[$r_g$ distribution, $\beta = 2$, track-based]{\includegraphics[width=0.48\textwidth,page=1]{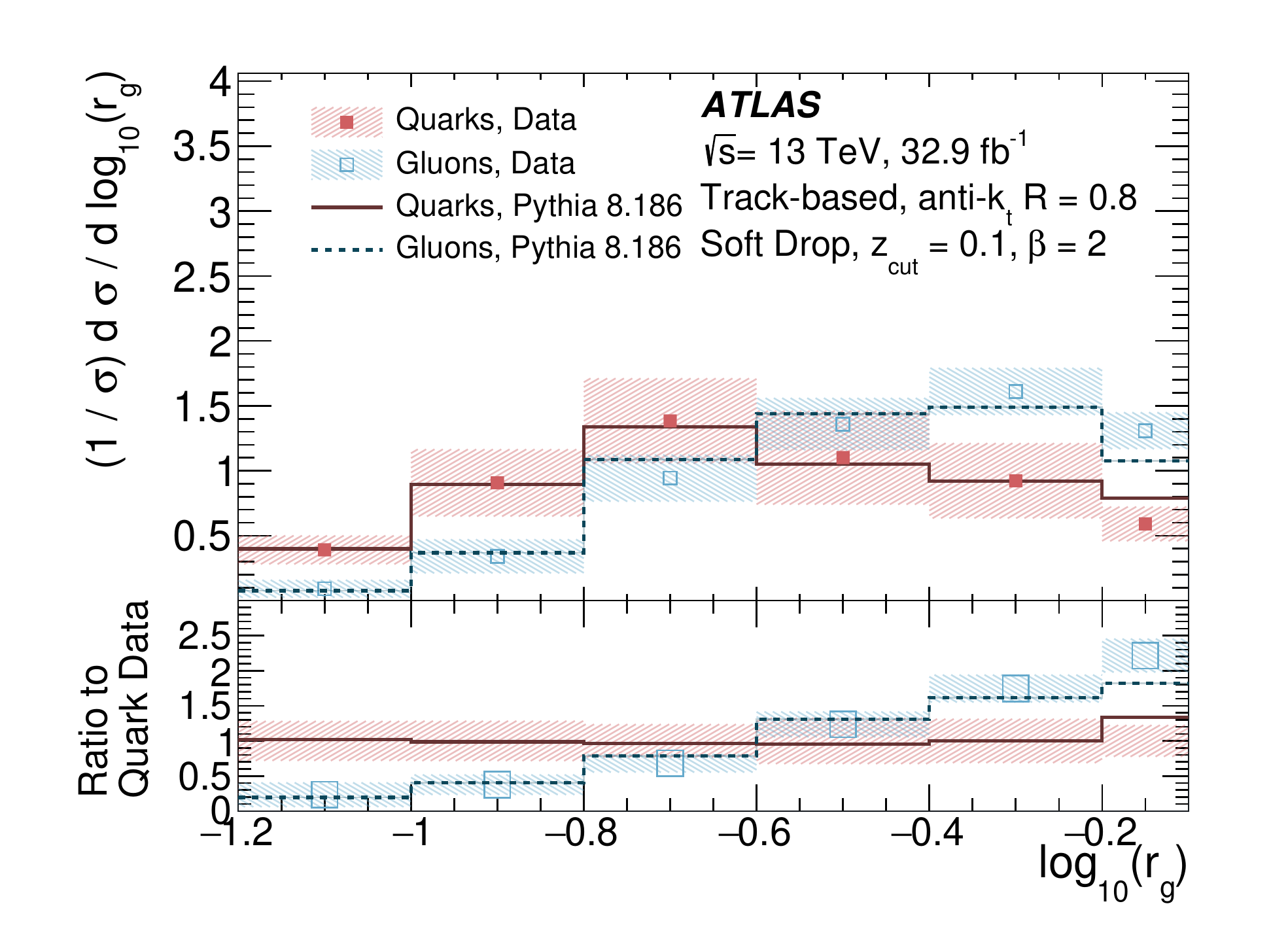}}
\end{center}
\caption{Comparison of the quark and gluon unfolded $r_g$ distribution for the track-based measurement. The uncertainty bands include all sources: data and MC statistical uncertainties, nonclosure, modeling, and tracking uncertainties where relevant.}
\label{results:unfold:rg:quarkgluon_inclPt}
\end{figure}

% End of text imported from the .//sections/results.tex input file
 
\FloatBarrier

\section{Conclusion}
 
This paper presents a measurement of soft-drop jet substructure observables in dijet events in $pp$ collisions at $\sqrt{s} = 13$ ~\TeV~using a dataset corresponding
to an integrated luminosity of 32.9 $\mathrm{fb}^{-1}$ collected with the ATLAS detector at the LHC.
Unfolded measurements of three substructure observables are shown for both the calorimeter-based observables unfolded to the all-particle level and
track-based observables unfolded to the charged-particles level.
These two types of measurements allow a direct comparison of how the different object definitions affect on the measured observables.
The calorimeter-based measurements for the relative jet mass and $r_g$ are compared with analytical predictions and are shown to be in good agreement in the perturbative region.
In particular, this provides the first comparison between an analytical prediction and an unfolded measurement of $r_g$.
Particularly for observables which are sensitive to the angular distribution of radiation within a jet, track-based observables are shown to be more precise
than calorimeter-based observables, due to the better angular resolution of tracks.
Since analytical predictions of track-based observables are not currently available, cluster-based observables are still relevant for probing the perturbative region.
The forward and central jets are measured separately, which enables an extraction of the quark- and gluon-jet distributions using input from simulation.
The extractions demonstrate differences between the observables in their sensitivity to the quark and gluon composition of the sample, which are most pronounced for the least amount of grooming.

\section*{Acknowledgements}
 
% The next lines are included from the .//acknowledgements/Acknowledgements.tex input file

We thank CERN for the very successful operation of the LHC, as well as the
support staff from our institutions without whom ATLAS could not be
operated efficiently.
 
We acknowledge the support of ANPCyT, Argentina; YerPhI, Armenia; ARC, Australia; BMWFW and FWF, Austria; ANAS, Azerbaijan; SSTC, Belarus; CNPq and FAPESP, Brazil; NSERC, NRC and CFI, Canada; CERN; CONICYT, Chile; CAS, MOST and NSFC, China; COLCIENCIAS, Colombia; MSMT CR, MPO CR and VSC CR, Czech Republic; DNRF and DNSRC, Denmark; IN2P3-CNRS and CEA-DRF/IRFU, France; SRNSFG, Georgia; BMBF, HGF and MPG, Germany; GSRT, Greece; RGC and Hong Kong SAR, China; ISF and Benoziyo Center, Israel; INFN, Italy; MEXT and JSPS, Japan; CNRST, Morocco; NWO, Netherlands; RCN, Norway; MNiSW and NCN, Poland; FCT, Portugal; MNE/IFA, Romania; MES of Russia and NRC KI, Russia Federation; JINR; MESTD, Serbia; MSSR, Slovakia; ARRS and MIZ\v{S}, Slovenia; DST/NRF, South Africa; MINECO, Spain; SRC and Wallenberg Foundation, Sweden; SERI, SNSF and Cantons of Bern and Geneva, Switzerland; MOST, Taiwan; TAEK, Turkey; STFC, United Kingdom; DOE and NSF, United States of America. In addition, individual groups and members have received support from BCKDF, CANARIE, Compute Canada and CRC, Canada; ERC, ERDF, Horizon 2020, Marie Sk{\l}odowska-Curie Actions and COST, European Union; Investissements d'Avenir Labex, Investissements d'Avenir Idex and ANR, France; DFG and AvH Foundation, Germany; Herakleitos, Thales and Aristeia programmes co-financed by EU-ESF and the Greek NSRF, Greece; BSF-NSF and GIF, Israel; CERCA Programme Generalitat de Catalunya and PROMETEO Programme Generalitat Valenciana, Spain; G\"{o}ran Gustafssons Stiftelse, Sweden; The Royal Society and Leverhulme Trust, United Kingdom.
 
The crucial computing support from all WLCG partners is acknowledged gratefully, in particular from CERN, the ATLAS Tier-1 facilities at TRIUMF (Canada), NDGF (Denmark, Norway, Sweden), CC-IN2P3 (France), KIT/GridKA (Germany), INFN-CNAF (Italy), NL-T1 (Netherlands), PIC (Spain), ASGC (Taiwan), RAL (UK) and BNL (USA), the Tier-2 facilities worldwide and large non-WLCG resource providers. Major contributors of computing resources are listed in Ref.~\cite{ATL-GEN-PUB-2016-002}.
 
% End of text imported from the .//acknowledgements/Acknowledgements.tex input file

\clearpage

\printbibliography

\clearpage
\input{atlas_authlist.tex}

\end{document}

%% file: atlas_authlist.tex
% ATLAS Collaboration author list
% Reference date of STDM-2017-33 is 2019-08-28
% Author list last updated on date 10-FEB-20
% Data extracted on 10-Feb-2020 for paper reference STDM-2017-33
% at 10:16am
 
\begin{flushleft}
{\Large The ATLAS Collaboration}

\bigskip

G.~Aad$^\textrm{\scriptsize 102}$,    
B.~Abbott$^\textrm{\scriptsize 129}$,    
D.C.~Abbott$^\textrm{\scriptsize 103}$,    
A.~Abed~Abud$^\textrm{\scriptsize 36}$,    
K.~Abeling$^\textrm{\scriptsize 53}$,    
D.K.~Abhayasinghe$^\textrm{\scriptsize 94}$,    
S.H.~Abidi$^\textrm{\scriptsize 167}$,    
O.S.~AbouZeid$^\textrm{\scriptsize 40}$,    
N.L.~Abraham$^\textrm{\scriptsize 156}$,    
H.~Abramowicz$^\textrm{\scriptsize 161}$
% ...
\bigskip
\\

% ...
$^{36}$CERN, Geneva; Switzerland.\\
% ...
$^{40}$Niels Bohr Institute, University of Copenhagen, Copenhagen; Denmark.\\
% ...
$^{53}$II. Physikalisches Institut, Georg-August-Universit\"{a}t G\"ottingen, G\"ottingen; Germany.\\
% ...
$^{94}$Department of Physics, Royal Holloway University of London, Egham; United Kingdom.\\
% ...
$^{102}$CPPM, Aix-Marseille Universit\'e, CNRS/IN2P3, Marseille; France.\\
$^{103}$Department of Physics, University of Massachusetts, Amherst MA; United States of America.\\
% ...
$^{129}$Homer L. Dodge Department of Physics and Astronomy, University of Oklahoma, Norman OK; United States of America.\\
% ...
$^{156}$Department of Physics and Astronomy, University of Sussex, Brighton; United Kingdom.\\
% ...
$^{161}$Raymond and Beverly Sackler School of Physics and Astronomy, Tel Aviv University, Tel Aviv; Israel.\\
% ...
$^{167}$Department of Physics, University of Toronto, Toronto ON; Canada.\\
% ...

\end{flushleft}

% Created with Glance <Atlas.Glance@cern.ch>